\documentclass[12pt]{iopart} 

\usepackage{iopams}
\usepackage{graphicx}
\usepackage{color}

\begin{document}

\topical{Spintronic oxides grown by laser-MBE}

\author{Matthias Opel}
\address{Walther-Meissner-Institut,
         Bayerische Akademie der Wissenschaften,
         85748 Garching,
         GERMANY}
\ead{Matthias.Opel@wmi.badw.de}

\begin{abstract} 
The recent study of oxides led to the discovery of several new fascinating physical phenomena. High-temperature superconductivity, colossal magnetoresistance, dilute magnetic doping, or multiferroicity were discovered and investigated in transition-metal oxides, representing a prototype class of strongly correlated electronic systems. This development was accompanied by an enormous progress regarding thin film fabrication. Within the past two decades, epitaxial thin films with crystalline quality approaching semiconductor standards became available using laser molecular beam epitaxy. This evolution is reviewed, particularly with emphasis on transition-metal oxide thin films, their versatile physical properties, and their impact on the field of spintronics. First, the physics of ferromagnetic half-metallic oxides, such as the doped manganites, the double perovskites and magnetite is presented together with possible applications based on magnetic tunnel junctions. Second, the wide bandgap semiconductor zinc oxide is discussed particularly with regard to the controversy of dilute magnetic doping with transition-metal ions and the possibility of realizing $p$-type conductivity. Third, the field of oxide multiferroics is presented with the recent developments in single-phase multiferroic thin film perovskites as well as in composite multiferroic hybrids.
\end{abstract}

\vspace{2pc}
\submitto{\JPD}
\maketitle


\section{Introduction}

Oxygen is by mass the most abundant chemical element on our planet. It changed both the landscape and the biosphere of our planet dramatically a very long time ago \cite{Holland2006}. Free oxygen gas was almost non-existent in the early Earth's atmosphere before photosynthetic archaea and bacteria evolved. It first appeared in significant quantities about 2.45 billion years ago when Earth's ecology changed from anaerobic to aerobic. Oxygen evaporating out of the oceans reached 10\% of its present level around 1.7 billion years ago. At that time, the presence of dissolved and free oxygen in the oceans and the atmosphere may have driven most of the anaerobic organisms to extinction. However, the beginning photosynthesis and cellular respiration of O$_2$ facilitated the evolution of ultimately complex multicellular organisms such as plants and animals. At the current rate of photosynthesis it would take about 2,000 years to regenerate the entire free O$_2$ in the present atmosphere \cite{Holland2006}. It is by mass the major component of the oceans (88.8\%) and constitutes 49.2\% of the Earth's crust by forming various types of oxide minerals -- mainly feldspars ((Ba,Ca,Na,K,NH$_4$)(Al,B,Si)$_4$O$_8$) and quartz (SiO$_2$).

A first example of a \emph{functional} oxide are permanent magnets of ``lodestone'' consisting of the ferrimagnetic mineral magnetite (Fe$_3$O$_4$) which were used by people in Sumer, ancient Greece, and China as compasses for navigating their boats across the sea more than 2,000 years ago \cite{Blackman1983}. In the 20th century, magnetic tapes were developed as recording media for audio information in the still fast growing market of consumer electronics. They first consisted of thin films of Fe$_2$O$_3$ which was replaced by CrO$_2$ from the late 1960s. One decade ago, such \emph{ferromagnetic} oxides have become of particular interest \cite{Bibes2007} in the field of magneto- or spin electronics \cite{Prinz1998,Wolf2001,Zutic2004,Bader2010} since some of them were expected to show an almost complete spin polarization of the charge carriers at the Fermi energy. \emph{Ferroelectric} oxides, however, were found much later in history, and an electric hysteresis was first discovered in Rochelle salt (${\rm NaKC_4H_4O_6 \cdot 4H_2O}$) 90 years ago \cite{Valasek1921}. Today, numerous piezo- and ferroelectric oxides are known, BaTiO$_3$ representing the prototype ferroelectric. The year 1986 marked the great breakthrough for \emph{superconducting} oxides. Although oxide superconductors were known before \cite{Raub1988}, the discovery of superconductivity in the Ba-La-Cu-O system was a milestone in condensed matter physics because of its unprecedented high transition temperature of 35\,K \cite{Bednorz1986}. Oxide \emph{semiconductors} such as ZnO or TiO$_2$ show the advantage of a wide band gap in the UV range making them suitable as transparent conducting oxides (TCOs) \cite{Grundmann2010}. In the recent past, \emph{multiferroic} oxides combining two or more ferroic properties in a single phase have come into the focus of research \cite{Spaldin2005,Fiebig2005}.

This development of the field of oxides was accompanied by an enormous progress in thin film deposition technology over the last two decades \cite{Martin2010}. Today, laser molecular beam epitaxy (laser-MBE) employing \textit{in-situ} reflection high energy electron diffraction (RHEED) allows us to grow oxide thin films with crystalline quality approaching semiconductor standards \cite{Gross2000,Gupta1990,Klein1999,Klein2000}. Moreover, in close analogy to GaAs/AlAs heteroepitaxy it is possible to grow complex heterostructures composed of different oxides on suitable substrates in a layer-by-layer or block-by-block mode \cite{Gross2000,Reisinger2003a}. Most recently, oxide interfaces came into the focus as they allow the realization of new effects due to the creation of low-dimensional interfacial electronic systems \cite{Mannhart2010}.

This topical review will describe the properties of functional oxides and heterostructures. I will restrict this review to magnetic, semiconducting, and multiferroic oxide thin film samples grown by laser-MBE. I will discuss intrinsic physical properties as well as extrinsic phenomena like doping and strain effects. I will show how the integration into oxide heterostructures can lead to novel functionalities and pave the way towards various applications. The topical review is organized as follows. In section \ref{sec:oxides}, I will give a brief overview over the development of the field of oxide spintronics in the past years and, in particular, describe the crystallographic structure and the fabrication of multifunctional oxides using laser-MBE as standard fabrication technique. In section \ref{sec:magnetic-oxides}, I will review the development of half-metallic magnetic oxides and magnetic tunnel junctions. Section \ref{sec:ZnO} will deal with the semiconducting oxide ZnO, particularly with respect to dilute magnetic doping and $p$-type conductivity. In section \ref{sec:multiferroics}, I will give some insight into the rapidly growing field of oxide multiferroics and magnetoelectric heterostructures. Finally in section \ref{sec:summary}, I will conclude by giving a summary and a brief outlook to future developments.

\section{Oxides for Spintronics: Structure and Growth Aspects}\label{sec:oxides}

As a result of decreasing device dimensions, conventional semiconductor technology is approaching fundamental physical limits. Making use of the spin of the electron as a degree of freedom additional to its charge opened the field of magneto- or spinelectronics which is considered a replacement technology \cite{Prinz1998}. Starting from metal-based devices in the late 1980s, the field of spintronics soon expanded to transition-metal oxide \cite{Ogale2005} and recently even to organic materials \cite{Sanvito2011}. Within the past two decades, the growth of high-quality oxide thin films and heterostructures showed important advances concerning sample quality \cite{Bibes2007,Ogale2005,Bibes2011}. Whereas the growth of oxide films was first motivated by the discovery of high-temperature superconductivity in perovskite cuprates \cite{Bednorz1986}, the deposition technology was soon applied to other transition-metal oxides with perovskite structure, namely the mixed-valence manganites. The discovery of colossal magnetoresistance in those thin films \cite{Helmolt1993,Jin1994} triggered intense research activities and led to the use of manganites as electrodes in magnetic tunnel junctions. Since then, research into oxide spintronics has been intense, with the latest developments focused on double perovskites, dilute magnetic oxides and multiferroics \cite{Bibes2007,Ogale2005,Bibes2011}.

\subsection{Structural and Physical Properties}\label{subsec:crystallography}

\subsubsection{Perovskites}\label{subsubsec:perovskite}

Earning their name from the mineral \emph{perovskite} (CaTiO$_3$), named after the Russian mineralogist L.\,A.\,Perovskii (1792--1856), the whole family of the perovskites with chemical composition $ABX^{2-}_3$ contains a large number of compounds \cite{Johnsson2008,Vrejoiu2008}. Among them, however, the ideal cubic structure with space group $Fm3m$ is rare and CaTiO$_3$ itself is actually slightly distorted. The perovskite structure consists of corner sharing octahedra of $BX_3$, building up a unit cell with a 12-fold coordinated cation $A$ at its body centre. The $A$ and $X$ atoms in the ideal cubic $ABX_3$ structure, that is realized for example in SrTiO$_3$, form cubic close-packed layers of $A+3X$. For further information on crystallography and chemistry, I refer the reader to ref.~\cite{Johnsson2008}.

\small\begin{figure*}
    \includegraphics[width=10cm]{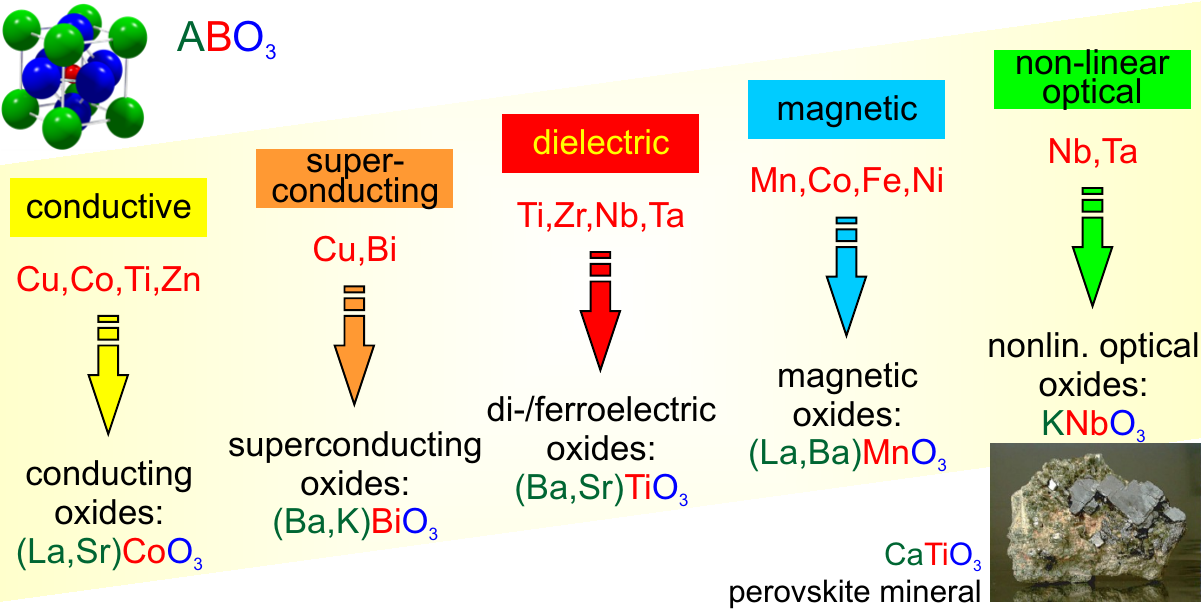}
    \caption{\label{fig:perovskite}
             Functional perovskite oxides. Depending on the combination of $A$ and $B$ ions in the cubic structure of $AB{\rm O}_3$ (top left), different physical properties and ground states are realized. This family of materials earned its name from the mineral \emph{perovskite} (CaTiO$_3$, bottom right).}
\end{figure*}\normalsize

The intense and wide interest in the perovskite family of oxides ($AB$O$^{2-}_3$) is based on the large variety of physical properties and ground states that can be realized (Fig.~\ref{fig:perovskite}). Depending on the combination of cations on the $A$ and $B$ sites, the materials may be conducting (even superconducting) or insulating, may show ferroelectricity or ferromagnetism, or may exhibit nonlinear optical behaviour. It was found that the electrical transport properties and the magnetic or dielectric polarizations in many cases show a large and non-linear response to external magnetic or electric fields. This situation made perovskites prominent key materials for use in numerous applications based on their sensor and switching functionalities. The large number of combinations on the cation site leads to a considerable freedom of ionic radii and ionic charges that can be incorporated into the lattice. The resulting electronic configurations allow bonds with different ionicity or covalency which result in enhanced dielectric or magnetic moments of the polarized electron densities. On the other hand, the O$^{2-}$ anion lattice favours large displacements and dipole moments which may lead to complex oxygen defect structures \cite{Johnsson2008}.

In more detail, substituting \emph{ions with different valences} on the $A$ site will result in different valences for the ion on the $B$ site. If this $B$ site ion carries some functionality, such as a magnetic moment, this may lead to competing interactions which, in turn, alter the physical ground state of the material. Along this line, the antiferromagnetic insulator LaMnO$_3$ will become ferromagnetic conducting and show a colossal magnetoresistive effect if part of La$^{3+}$ is replaced by Sr$^{2+}$ (see section \ref{subsec:manganites}). On the other hand, \emph{ions with different sizes} lead to a distortion of the crystal lattice and a rotation of the oxygen octahedra. The degree of distortion is usually quantified by the Goldschmidt tolerance factor \cite{Goldschmidt1927}
\begin{equation}
    \label{eq:tolerance}
    t = \frac{r_{A} + r_{\rm O}}{\sqrt{2}(r_{B} + r_{\rm O})}
\end{equation}
where $r_{A,B,{\rm O}}$ denote the ionic radii of the $A$, $B$, and O$^{2-}$ ions. A perfect cubic structure as in SrTiO$_3$ with non-tilted oxygen octahedra is represented by $t=1$. Furthermore, the possibility of introducing different ions on the $B$ site extends the range of physical properties and ground states significantly. When placed in an ordered 1:1 fashion, the unit cell will double which results in the realization of double perovskites with the composition $A_2BB^\prime$O$_6$ (see section \ref{subsec:DP}).

In summary, perovskites show a large variety of physical and chemical properties which makes them interesting for basic research and technological applications (Fig.~\ref{fig:perovskite}) \cite{Bhalla2000}. On this basis, the colossal magnetoresistance (CMR) effect \cite{Helmolt1993,Jin1994} as well as high-$T_{\rm c}$ superconductivity \cite{Bednorz1986} were discovered.

\subsubsection{Spinels}\label{subsubsec:spinel}

\small\begin{figure}
    \includegraphics[width=9cm]{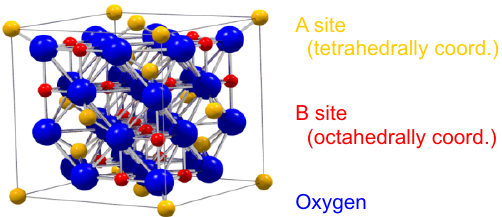}
    \caption{\label{fig:spinel}
             Spinel structure.}
\end{figure}\normalsize

Earning their name from the mineral \emph{spinel} (MgAl$_2$O$_4$), the respective class of materials shows versatile physical properties. Spinels are cubic with space group $Fd3m$ and chemical composition $AB_2{\rm O}_4$ and contain a large number of ions (8 formula units) in their unit cell (Fig.~\ref{fig:spinel}). The $A$ site ion is surrounded by an oxygen tetrahedron whereas the two $B$ site ions are octahedrally coordinated. While in a \emph{normal spinel} the $A$ site ions have a valence of $2+$ and the $B$ site ions have $3+$, in an \emph{inverse spinel} the $A$ site valency is $3+$ which results in a mixed-valence occupancy of $B^{2+}/B^{3+}$ ions on the $B$ site. As discussed above for the perovskites, these mixed valences again may lead to competing interactions and thus change the physical ground state of the material dramatically. A prominent example is the antiferromagnetic spinel zinc ferrite (ZnFe$_2$O$_4$) which transforms into the ferrimagnetic inverse spinel magnetite (Fe$_3$O$_4$) upon replacing Zn$^{2+}$ on the $A$ site by Fe$^{3+}$ (see section \ref{subsec:magnetite}).

\subsubsection{Wurtzite-type compounds}\label{subsubsec:wurtzite}

\small\begin{figure}
    \includegraphics[width=6cm]{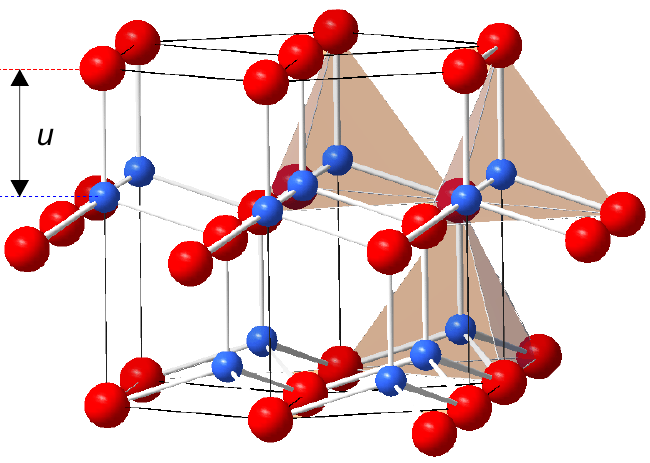}
    \caption{\label{fig:wurtzite}
             Wurtzite structure. Figure provided by W Mader \cite{Mader2009}.}
\end{figure}\normalsize

The wurtzite structure with space group $P6_3mc$ is the hexagonal analog of the zincblende (ZnS) lattice and earns its name from the mineral \emph{wurtzite} ((Zn,Fe)S). It is composed of an hcp cation and an hcp anion sublattice which are shifted by the dimensionless $u$ parameter along the $c$-axis with respect to each other (Fig.~\ref{fig:wurtzite}). $u=1$ indicates a shift by one complete unit cell along $c$. In the ideal coordination, where each atom is tetrahedrally coordinated, this parameter is given by $u = 0.375 = \frac{3}{8}$ \cite{Ney2010b}. The wurtzite structure lacks inversion symmetry. Therefore, wurtzite crystals can (and generally do) display properties such as piezoelectricity and pyroelectricity, which are absent for centrosymmetric crystals. The semiconductors GaN and ZnO crystallize in this structure. Replacing all atoms by C gives the hexagonal diamond structure.

\subsection{Fabrication of Oxide Thin Films using Laser-MBE}

The history of oxide thin films was strongly determined by the development of the appropriate deposition techniques. Due to the high melting points of the starting materials well above $1,000^\circ$C, thermal evaporation of oxides is difficult. Therefore, several alternatives were used to achieve epitaxial growth of superconducting, ferromagnetic, ferroelectric, and multiferroic thin films and nanostructures -- such as sputtering, spin coating, laser-MBE, sol-gel processes, metal-organic chemical vapor deposition, molecular beam epitaxy, and more. However, no other growth method had such a high impact as the development of the novel pulsed laser deposition (PLD) technique in the late 1980s, also referred to as laser-MBE. Detailed information on its history and evolution can be obtained through a large variety of books \cite{PLD1994,PLD2006} and reviews \cite{Martin2010,Christen2008}.

Complex oxide materials moved to the forefront of materials research when the high-$T_{\rm c}$ oxide superconductors were discovered in 1986 \cite{Bednorz1986}. Just one year later, the research of complex oxide materials was revolutionized with the growth of superconducting oxide thin films via laser-MBE \cite{Dijkkamp1987}. The development of laser-MBE as a new deposition technique required the technical development of the appropriate hardware such as ultra-high vacuum systems or short-wavelength lasers and was successfully applied to many new classes of materials. Laser-MBE is based on the stoichiometric transfer of material from a ``target'' to the substrate. The target is usually a pressed and sintered polycrystalline pellet having the desired stoichiometric composition for the thin film. Target and substrate are placed in a high vacuum chamber equipped with several pumps and gas inlets to establish the appropriate process atmosphere and pressure. Film growth is carried out in oxidizing (O$_2$), inert (Ar, N$_2$), or even reducing (H$_2$) atmospheres. Also atomic gas sources for e.g. atomic oxygen or nitrogen are widely used.

\small\begin{figure}
    \includegraphics[width=6cm]{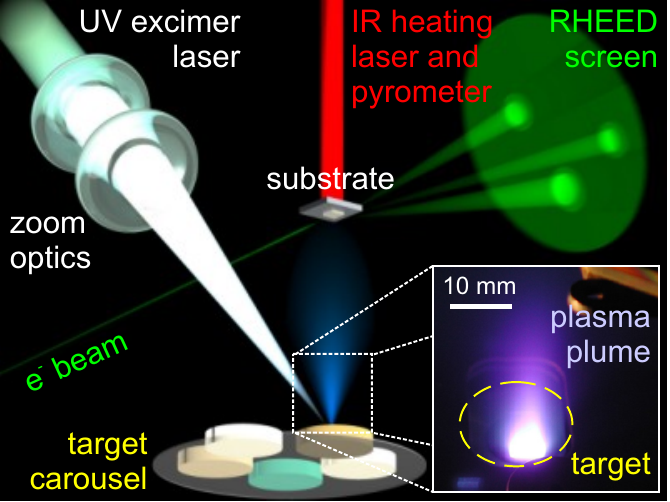}
    \caption{\label{fig:pld}
             Schematic view inside the PLD process chamber, operated at WMI. (Inset) Photograph of the PLD plasma plume.}
\end{figure}\normalsize

The situation inside a typical process chamber is displayed in Figure~\ref{fig:pld}. Several targets are placed on a rotatable plate which is referred to as ``target carousel''. This facilitates a quick change between different materials during the thin film growth processes by just rotating this carousel to its appropriate position without changing the optical path for the PLD laser beam. In the \emph{on-axis} geometry, the substrate is located opposite to the target, both sharing the same out-of-plane axis. Alternatively, the substrate may be placed \emph{off-axis}. Its temperature is controlled via the substrate heater. Recently, contact-less infrared (IR) optical heating systems came into operation reaching temperatures of more than $1,000^\circ$C. From outside the vacuum chamber, the IR light is focused onto the back of the substrate. The temperature of the substrate is determined by a pyrometer, also from outside the chamber. The combination of IR laser heating and pyrometry makes the system very convenient as there are no serviceable parts regarding temperature control present inside the process chamber. The combination of reactive gas pressure and substrate temperature offers access to a wide range of thermodynamic conditions for the thin film growth.

The ablation of the target material is achieved by another external energy source: the PLD laser. Commonly, excimer lasers with a pulse duration of some nanoseconds and a wavelength in the ultraviolet regime are used. Their light is absorbed within the first 100\,nm of the target for most oxide materials leading to an evaporation and ionization of part of the target creating the ``plasma plume''. It consists of different species with different masses like single ions, ionized molecules or even larger particles from the target. At least in the centre of the plume, however, they do not follow different trajectories as their dynamics and kinetics is collision-limited up to a critical distance from the target. This lack of discrimination of masses is crucial for successful deposition in \textit{on-axis} geometry, but sets an upper limit for the distance to the substrate. Laser-MBE systems are usually equipped with a set of optics including apertures, mirrors, and lenses to focus and direct the laser beam into the process chamber with the desired fluence. For a continuous adjustment of the fluence in a wide range of energy densities, some groups use a telescope optics similar to a zoom objective in photography. This offers the possibility to continuously zoom the laser spot size on the target surface without defocusing. The beam shape of the laser is crucial and its energy density has kept to be constant over the illuminated target area following a rectangular intensity profile. An angle of incidence of $45^\circ$ has turned out to be optimum. Moreover, a uniform ablation of the target has to be maintained which can be done best by scanning the beam across its surface while simultaneously rotating the target. The relevant time scales are ns for the absorption of light and the creation of the plasma, $\mathrm{\mu s}$ for the material transfer to the substrate and ms for diffusion processes on the substrate or the thin film surface (see subsection~\ref{subsec:growthmodes}).

A major step forward to higher sample quality was achieved by introducing an \textit{in-situ} characterization of the thin film via reflection high-energy electron diffraction (RHEED) during the deposition process. However, for the relatively high background pressures used for laser-MBE the mean free path for electrons inside the process chamber is short. Therefore, in 1996 the development of a two-stage differential pumping system set another milestone for \textit{in-situ} characterization \cite{Rijnders1997}. \textit{In-situ} RHEED allowed for the controlled growth of oxide thin films layer-by-layer and offered the possibility to monitor the surface morphology in real time or even count the number of monolayers while growing the film. This set an important prerequisite for the controlled deposition of oxide multilayers and heterostructures in the following years.

As a major advantage of laser-MBE, the process is far from thermal equilibrium and, therefore, preserves complex stoichiometries. Further advantages are the simple usage with regard to multiple oxide materials, i.e.~it is easy to replace the targets. Furthermore, it is cost effective for investigating a wide range of materials, and it is excellent for rapid prototyping of different oxides. However, the sample grower has to face some disadvantages as well. Historically, laser-MBE was considered to fabricate thin films with low quality. This is no longer valid as the invention of advanced \textit{in-situ} monitoring processes like RHEED as well as growing experience with the process has led several groups to fabricate very high quality thin films. For sub-optimum growth parameters, laser-MBE can lead to non-uniform target erosion resulting in nonstoichiometric thin films. The biggest drawback from laser-MBE, however, has traditionally been the deposition of macroparticles in the form of explosive ejection of particles, splashing, and fragmentation due to thermal shock. In this context, the formation of macroscopic so-called ``droplets'' with diameters in the micrometer range on top of the growing thin film is a major omnipresent problem. It can be avoided by growing in \textit{off-axis} geometry, however, at the expenses of the deposition rate. By fine-tuning the growth parameters and by carefully optimizing the laser spot on the target surface, droplet formation can be avoided when using \textit{on-axis} geometry as well.

\subsection{Growth Modes} \label{subsec:growthmodes}

\small\begin{figure}
    \includegraphics[width=6cm]{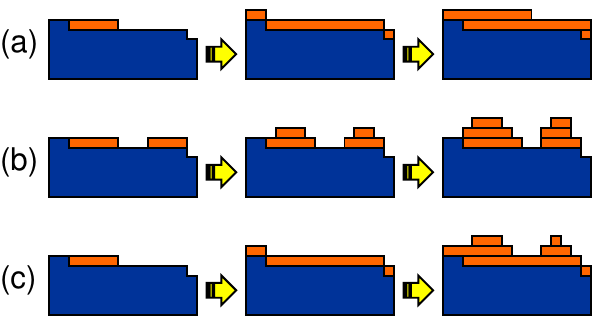}
    \caption{\label{fig:growth}
             Thermodynamic thin film growth modes. (a) ``Frank-van der Merwe'' (layer or 2-dimensional) growth, (b) ``Volmer-Weber'' (island or 3-dimensional) growth, (c) ``Stranski-Krastanov'' (combined layer + island) growth.}
\end{figure}\normalsize

Depending on the process conditions, the thin films grow in one of three major thermodynamic growth modes, see Fig.~\ref{fig:growth}: (a) ``Frank-van der Merwe'' (layer or 2-dimensional) growth, (b) ``Volmer-Weber'' (island or 3-dimensional) growth, or (c) ``Stranski-Krastanov'' (combined layer + island) growth. These different growth modes can be described by simple thermodynamic models for the nucleation and growth of film materials \cite{Martin2010,Chambers2000}. In general, layer growth (Fig.~\ref{fig:growth}(a)) occurs if the depositing atoms or molecules are more strongly bonded to the substrate than to each other, and each layer is progressively less strongly bonded than the previous one. Until the bulk bond strength is reached, the thin film forms planar, 2-dimensional sheets and grows in a layer-by-layer fashion. This growth mode is typically observed for the epitaxial growth of semiconductors or oxide materials. In fact, the entire field of oxide thin film deposition has developed based on the ability to control materials through this and similar growth modes and has finally reached the capability to deposit oxide materials down to the single or sub-unit cell level \cite{Reisinger2003a}. Island growth (Fig.~\ref{fig:growth}(b)), however, occurs when the deposited atoms or molecules are bonded more strongly to each other than to the substrate. Then, the smallest stable clusters nucleate on the substrate and grow in vertical direction forming 3-dimensional islands. This is often the case when film and substrate are dissimilar materials like metals or semiconductors on oxide substrates. Finally, it is fairly common that the growth mode changes during deposition. It may happen that after forming one or more monolayers in a 2-dimensional fashion, continued layer growth becomes energetically unfavourable and islands begin to form (Fig.~\ref{fig:growth}(c)). This ``Stranski-Krastanov'' mode is observed in a number of systems.

\small\begin{figure}
    \includegraphics[width=6cm]{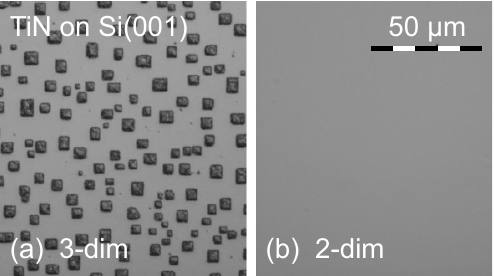}
    \caption{\label{fig:TiN-Si}
             Optical micrographs showing the different morphology of TiN thin films,
             grown on Si(001) at different substrate temperatures.
             (a) Island growth at $650^\circ$C,
             (b) layer growth at $600^\circ$C.
             Reprinted with permission from \cite{Reisinger2003b}. Copyright 2003, American Institute of Physics.}
\end{figure}\normalsize

From a microscopic point of view, thin films grow in a two-step process where particles first impinge onto a surface and then diffuse to sites with low energy. In the above discussed case of thermodynamic equilibrium, this diffusion has to be free and fast with respect to the impinging particle flux. Pulsed laser-deposited oxides, however, usually grow at conditions far away from thermodynamic equilibrium leading to a dominant role of particle kinetics rather than simple thermodynamics. Therefore, kinetically controlled processes, substrate defects, and impurities in the film and substrate material or impinging from the background gas atmosphere have to be taken into account. The kinetics of the arriving particles on the thin film surface is mainly limited by temperature and bonding strength to different adsoprtion sites and can be categorized into intrinsic diffusion and translational mobility processes. The latter is determined by the kinetic energy (temperature) of the particles, the adsorption energy as well as the substrate temperature. After equilibration, only intrinsic diffusion takes place which is separated into surface diffusion (2-dimensional on terraces), step edge diffusion and 3-dimensional volume diffusion - the latter being particularly important for oxides \cite{Henderson1999}. For chemically reactive materials, even interdiffusion between substrate and thin film material can take place. For oxide thin films, the substrate itself represents a significant source of oxygen \cite{Schneider2010}.

Diffusion across step edges was first experimentally observed by Ehrlich and Hudda \cite{Ehrlich1966} and theoretically described by Schwoebel and Sipsey \cite{Schwoebel1966}. At reduced mobility, these ``Ehrlich-Schwoebel'' barriers may limit the free surface diffusion of atoms, thus stimulating three dimensional growth. Commonly, the thin film morphology is explained by ``structure zone models'' introduced first by Movchan and Demchishin for evaporated films \cite{Movchan1969} and extended later on by Thornton \cite{Thornton1986} and many others \cite{Messier1984,Kelly1998,Petrov2003}. They mainly take into account the mobility of the film atoms which depends on the substrate temperature during deposition, the deposition rate as well as the specific growth parameters (background pressure, etc.). An example for this dependence is shown in Fig.~\ref{fig:TiN-Si} for thin films of TiN on (001)-oriented Si substrates. For different temperatures, TiN grows 2- or 3-dimensionally \cite{Reisinger2003b}. According to the structure zone models, morphological changes proceed in steps at which the activation energy for a certain diffusion process is overcome. More details can be found in a recent review \cite{Mahieu2006}. As a result of these considerations, the stress state of thin films deposited far from thermal equilibrium may be different from that of thermally evaporated ones \cite{Koch2010}.

\subsection{Strain Engineering in Epitaxial Thin Films}

\small\begin{figure}
    \includegraphics[width=6cm]{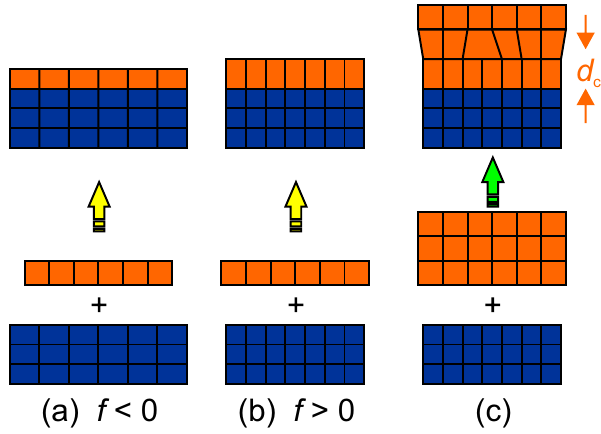}
    \caption{\label{fig:mismatch}
             Heteroepitaxial thin film growth can result in strain as a result of the lattice mismatch $f$.
             (a) In-plane (ip) tensile and out-of-plane (oop) compressive strain,
             (b) ip compressive and oop tensile strain,
             (c) relaxed thin film growth for a thickness $>d_{\rm c}$.}
\end{figure}\normalsize

Thin films with lattice parameters showing a fixed relation to those of the substrate are called \emph{epitaxial}. They form an extended single crystalline film which grows in a pseudomorphic way on top of the respective substrate. In contrast, \emph{textured} thin films are polycrystalline but with a certain lattice plane oriented preferentially parallel to the respective substrate plane. Whether or not a film grows epitaxially over some range of thickness sensitively depends on the lattice mismatch $f$ between the two materials. It is defined as
\begin{equation}
    f = \frac{2(a_{\rm f}-a_{\rm s})}{a_{\rm f}+a_{\rm s}} \simeq \frac{a_{\rm f}-a_{\rm s}}{a_{\rm s}}
\end{equation}
where $a_{\rm f}$, $a_{\rm s}$ denote the in-plane lattice constants $a$ (or multiples of $a$) of the film and the substrate, respectively. Typically $|f| < 10\%$ is required for epitaxy as for larger values so few interfacial bonds are well aligned that there is little reduction in the interfacial energy and the film will not grow epitaxially. As illustrated in Fig.~\ref{fig:mismatch}, $f<0$ results in a tensile (compressive) strain of the film in in-plane (out-of-plane) direction (Fig.~\ref{fig:mismatch}(a)), whereas $f>0$ results in a compressive (tensile) strain in the respective directions (Fig.~\ref{fig:mismatch}(b)). Note that $f$ usually is a function of temperature as the thermal expansion coefficients of the film and the substrate may be different. Therefore, under PLD conditions at high temperatures $f$ is different from its value after growth at room temperature. If the thickness of the film exceeds a critical value $d_{\rm c}$ the lattice of the thin film will relax to its bulk lattice parameters, accompanied with the formation of lattice defects (Fig.~\ref{fig:mismatch}(c)). Frequently, the misfit strain can also be efficiently reduced by forming commensurate superlattices of thin film and substrate during epitaxial growth.

\small\begin{figure*}
    \includegraphics[width=12cm]{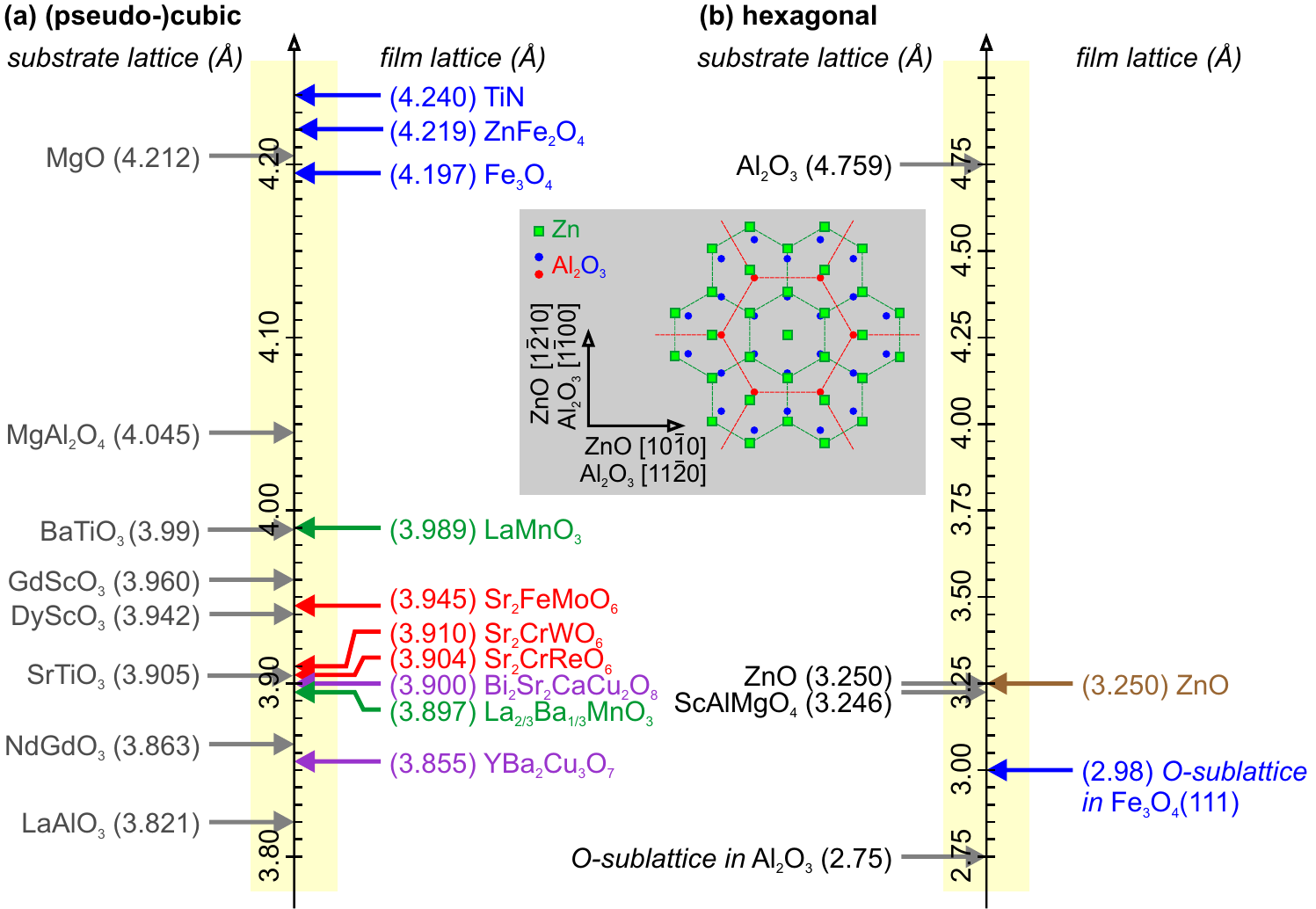}
    \caption{\label{fig:strain}
             Lattice matching between substrates and thin films. The numbers show the in-plane lattice constants $a$ (or integer fractions of $a$) for the (001)/(0001) planes at room temperature for a selection of (a) (pseudo-)cubic/(pseudo-)tetragonal and (b) hexagonal substrates (left) and thin films (right) which are important for this topical review. Perovskites are displayed in green, double perovskites in red, spinels and TiN in blue, the high-$T_{\rm c}$ superconductors in violet, and ZnO in brown. Inspired from \cite{Schlom2007}.
             The inset (shaded gray) displays the epitaxial relation between Al$_2$O$_3$ (red/blue) and the $30^\circ$-rotated ZnO lattice (green) \cite{Chen1998}.}
\end{figure*}\normalsize

This behaviour opens a way to artificially apply stress to the thin film material just by choosing the appropriate substrates (Fig.~\ref{fig:strain}) \cite{Schlom2007}. This enables one to deliberately manipulate the strain state of their functional thin films in order to tailor its physical properties. Along this line, the Curie temperatures of ferroelectric thin films can be shifted by hundreds of degrees by the application of epitaxial strain in the percent regime \cite{Schlom2007}. Moreover, strain was found to have a strong impact on the magnetic properties of ferromagnetic thin films \cite{Martin2010} and is useful to stabilize crystallographic structures of thin films which would be metastable or even unstable when prepared as bulk \cite{Martin2010}.

\subsection{A Brief History of Pulsed Laser-Deposited Functional Oxide Thin Films}

\small\begin{figure}
    \includegraphics[width=6cm]{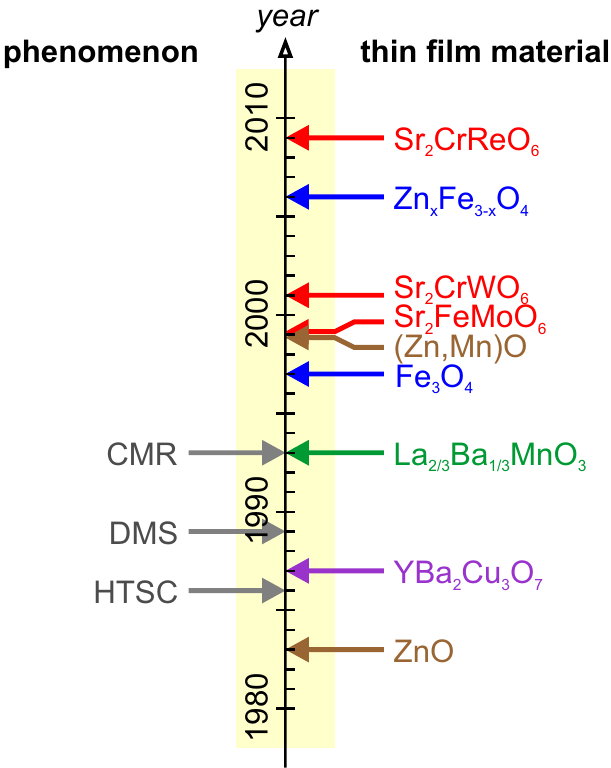}
    \caption{\label{fig:history}
             History of laser-MBE grown oxide thin films (ZnO~\cite{Sankur1983,Nakayama1983}, (Zn,Mn)O~\cite{Fukumura1999}, YBa$_2$Cu$_3$O$_7$~\cite{Dijkkamp1987}, La$_{2/3}$Ba$_{1/3}$MnO$_3$~\cite{Helmolt1993}, Fe$_3$O$_4$~\cite{Gong1997}, Zn$_x$Fe$_{3-x}$O$_4$~\cite{Takaobushi2006}, Sr$_2$FeMoO$_6$~\cite{Manako1999}, Sr$_2$CrWO$_6$~\cite{Philipp2001}, Sr$_2$CrReO$_6$~\cite{Geprags2009})
             and related physical phenomena: high-temperature superconductivity (HTSC), dilute magnetic semiconductors (DMS), and colossal magnetoresistance (CMR).}
\end{figure}\normalsize

Pulsed laser-deposited oxides are known for a long time. Early thin films were reported from the 1960s for semiconductors and dielectrics \cite{Smith1965} as well as titanites \cite{Schwarz1969}. In 1983, the first PLD thin films of the II-VI semiconductor ZnO were prepared \cite{Sankur1983,Nakayama1983}. However, laser-MBE became the method of choice for the growth of oxide thin films not before the discovery of high-$T_{\rm c}$ superconductivity in the cuprates in 1986 \cite{Bednorz1986}. Only one year later (Fig.~\ref{fig:history}), the first superconducting films of the compound YBa$_2$Cu$_3$O$_{7-\delta}$ with transition temperatures of $T_{\rm c} = 85$\,K and 75\,K were fabricated on SrTiO$_3$ and Al$_2$O$_3$ substrates, respectively \cite{Dijkkamp1987}. In the following years, this material evolved to the most widely studied high-$T_{\rm c}$ compound and is today considered the prototype cuprate superconductor. Investigating and clarifying the symmetry of the order parameter in YBa$_2$Cu$_3$O$_{7-\delta}$ represents the first example where thin films turned out to be superior to bulk material. Tsuei, Kirtley and co-workers \cite{Tsuei1994,Kirtley1995} used scanning SQUID (superconducting quantum interference device) microscopy \cite{Vu1993} to demonstrate the half-integer flux quantum effect in tricrystal thin film samples to pin down that cuprate high temperature superconductors have $d$-wave pairing symmetry \cite{Tsuei2000}. Although already discovered in the 1950s \cite{vanSanten1950}, the colossal magnetoresistive (CMR) effect attracted broader attention only after first high quality thin films of the mixed-valence manganites became available in the early 1990s. In 1993, a magnetoresistive effect of $-60\%$ was found at room temperature in thin films of La$_{2/3}$Ba$_{1/3}$MnO$_3$ when applying magnetic fields of up to 5\,T~\cite{Helmolt1993}. With regard to preparation of thin films, this development moved the focus from superconducting cuprates to magnetic manganites. Of particular interest hereby became the fabrication of half-metallic ferromagnetic oxides with Curie temperatures above 300\,K which could be utilized as electrode materials in the rapidly evolving field of spintronics~\cite{Prinz1998}. One of these is CrO$_2$ which represents the only binary ferromagnetic metallic oxide and was predicted as half-metallic in 1986 \cite{Schwarz1986}. However, the fabrication of CrO$_2$ thin films by laser-MBE turned out to be difficult \cite{Gupta2008} and only a few reports exist about pulsed laser deposition of this material \cite{Shima2002}. Ferrimagnetic Fe$_3$O$_4$, however, also predicted to be a half-metal \cite{Yanase1984,Zhang1991}, exhibits a high Curie temperature of 860\,K (as bulk). The first laser-MBE grown thin films were prepared in 1997 showing a magnetic moment at room temperature which was 88\% of the bulk value \cite{Gong1997}. After the discovery of ferromagnetism in the dilute magnetic semiconductors (DMS) (In,Mn)As \cite{Munekata1989} and (Ga,Mn)As \cite{Ohno1996}, Mn-substituted ZnO was also considered a possible candidate for a dilute magnetic system. First thin films were prepared by laser-MBE in 1999 \cite{Fukumura1999}. At that time, room-temperature magnetoresistance was first reported for an ordered double-perovskite material \cite{Kobayashi1998}. This observation stimulated the research into and fabrication of such samples as epitaxial thin films by laser-MBE. The first report was on Sr$_2$FeMoO$_6$ deposited on SrTiO$_3$ which showed metallic conduction as well as room temperature ferromagnetic behaviour with a saturation magnetic moment of 0.8 Bohr magnetons ($\mu_{\rm B}$) per formula unit (f.u.) and a Curie temperature of $T_{\rm C}>400$\,K \cite{Manako1999}. In the following years, Sr$_2$CrWO$_6$ \cite{Philipp2001} and Sr$_2$CrReO$_6$ thin films \cite{Geprags2009} were grown, also on SrTiO$_3$. The first compound displayed a high magnetic moment above $1\,\mu_{\rm B}$/f.u., whereas the second revealed a very large coercive field of 1.1\,T. In the following, I will review these developments in detail, with the emphasis placed on specific improvements in sample quality and physical properties.

\section{Half-metallic Magnetic Oxides and Heterostructures} \label{sec:magnetic-oxides}

Spin electronic (spintronic) devices in which information is encoded in the form of spin additional to charge are considered as a replacement technology for conventional semiconductor-based electronics  \cite{Prinz1998}. Most spintronic device concepts rely on a source of spin-polarized currents, and so there is strong interest in materials that have the largest possible spin polarizations. Most important are materials which display a metallic character for the majority-spin electrons, while the Fermi energy falls in a gap for the minority-spin electrons, i.e. the latter behave semiconducting or even insulating. Such a situation has been found in 1983 for NiMnSb by de Groot \textit{et al}.~\cite{deGroot1983} who introduced the term \emph{half-metallic ferromagnet}. One year later, an analogous behaviour was predicted in oxides where Yanase and Siratori \cite{Yanase1984} calculated a gap for majority-spin electrons, but metallic character for minority-spin ones in Fe$_3$O$_4$. Today, interesting half-metallic material systems are chromium dioxide, the doped manganites, magnetite, and the double perovskites, as they all show Curie temperatures above room temperature. This makes them well suited as electrode materials for spin injection into semiconductors or for the realization of magnetic tunnel junctions with a high tunneling magnetoresistance \cite{Coey2003,Gross2006}. The method of choice to grow such oxide thin films is laser-MBE due to its compatibility with oxygen and other reactive gases. The use of a laser as external energy source results in an extremely clean process without hot filaments or other parts inside the vacuum chamber. Thus, deposition can be carried out in both inert and reactive background atmospheres. As the process is far from thermal equilibrium it preserves complex stoichiometries which is necessary for the successful growth of doped manganites or double perovskites. For the deposition of multilayers like magnetic tunnel junctions, moreover, the flexibility of the laser-MBE process with regard to the possibility of using multiple targets in the same process chamber is a key advantage.

\subsection{Magnetic interactions in transition-metal oxides}

The magnetic ground state in transition-metal oxides is almost always determined by indirect magnetic exchange interactions of transition metal (TM) ions which are separated by an O$^{2-}$ ion in between. In this regard, the orbital structure of the involved ions plays a crucial role as it influences both the sign and the strength of the interaction via the degeneracy and occupancy of the TM-$3d$ orbitals and their overlaps with the O-$2p$ orbitals, respectively. This can make the discussion very complex. Goodenough \cite{Goodenough1955}, Kanamori \cite{Kanamori1959}, and Anderson \cite{Anderson1950}, however, found some simple rules (``GKA'' rules \cite{Goodenough1963}) to predict the magnetic interaction which is in the following illustrated for the antiferromagnetic oxide La$^{3+}$Mn$^{3+}$O$^{2-}_3$ as an example.

\small\begin{figure}
    \includegraphics[width=8cm]{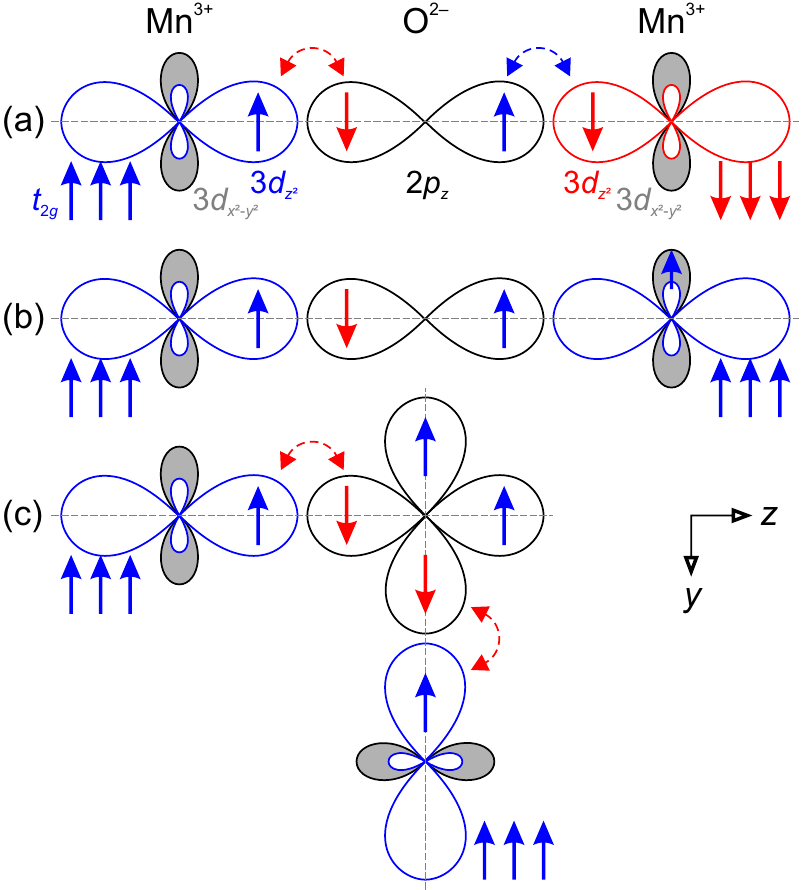}
    \caption{\label{fig:SE}
             Superexchange and Goodenough-Kanamori-Anderson (GKA) rules \cite{Goodenough1963},
             illustrated for the magnetic exchange between the $3d_{z^2}$ orbital of two Mn$^{3+}$ ions
             via the $2p_z$ orbital of an O$^{2-}$ ion, realized e.g.~in LaMnO$_3$.
             The $3d_{x^2-y^2}$ orbital of Mn$^{3+}$ is shaded in grey.
             The core spin $S = 3/2$ of the three (not-shown) $t_{2g}$ orbitals of Mn$^{3+}$ is symbolized
             by triple arrows. Dependent on the occupation with electrons, the resulting interaction can
             be antiferromagnetic (a) or ferromagnetic (b,c).}
\end{figure}\normalsize

The three $2p_x$, $2p_y$, $2p_z$ orbitals of the O$^{2-}$ ion with electronic configuration [He]2s$^2$2p$^6$=[Ne] are fully occupied with six electrons. Mn$^{3+}$ with electronic configuration [Ar]3d$^4$ has four electrons in the five $3d$ orbitals, three of them occupying the three $t_{2g}$ ones ($3d_{xy}$, $3d_{yz}$, $3d_{xz}$) resulting in a \emph{core spin} of $S=3/2$. The fourth is located in one of the two $e_g$ orbitals ($3d_{z^2}$, $3d_{x^2-y^2}$) which point along the crystal axes and therefore overlap with the O-$2p$ orbitals. Virtual hopping of O-$2p$ electrons to these overlapping Mn-$e_g$ orbitals leads to virtual excited states which can result in a reduction of the total energy of the system depending on the spin directions as demonstrated by higher order perturbation theory. We have to keep in mind that (i) the Pauli exclusion principal forces two electrons in the same $e_g$ orbital to possess antiparallel spins whereas (ii) Hund's rule favours electrons with parallel spins in different orbitals ($t_{2g}$ and $e_g$) of the same $3d$ shell. Figure~\ref{fig:SE} illustrates the situation in $z$-direction with regard to the following three GKA rules:

(a) For two Mn$^{3+}$ ions along the $z$ direction with one electron each in their respective Mn-$3d_{z^2}$ orbitals, virtual hopping of the two O-$2p_z$ electrons will reduce the total energy of the system if the Mn core spins are aligned antiparallel. The exchange integral is given by
\begin{equation}
    J^{\rm ex} \approx -\frac{2t^2_{pd}}{\Delta_{pd}}
\end{equation}
where $t_{pd}$ denotes the hopping amplitude and $\Delta_{pd}$ the charge-transfer energy between the O-$2p$ and the Mn-$3d$ orbitals. This $180^\circ$ exchange between two half-filled or two empty orbitals is strong and antiferromagnetic.

(b) For two Mn$^{3+}$ ions along the $z$ direction where the first has one and the second no electron in their respective Mn-$3d_{z^2}$ orbitals, virtual hopping of the two O-$2p_z$ electrons will reduce the total energy of the system if the Mn core spins are aligned parallel. The respective exchange integral is given by
\begin{equation}
    J^{\rm ex} \approx +\frac{2t^2_{pd} J_{\rm H}}{\Delta^2_{pd}}
\end{equation}
where $J_{\rm H}$ is the coupling energy due to Hund's rule at the Mn site. This $180^\circ$ exchange between one half-filled and one empty orbital is weak and ferromagnetic.

(c) For two Mn$^{3+}$ ions with one electron in their respective Mn-$3d_{z^2}$ orbitals corner-sharing the same O$^{2-}$ ion, virtual hopping of one O-$2p_z$ and one O-$2p_y$ electron will reduce the total energy of the system if the Mn core spins are parallel. Here, the reason is Hund's rule coupling of the two remaining O-$2p$ spins at the O site. This $90^\circ$ exchange between two half-filled orbitals is weak and ferromagnetic.

\small\begin{figure*}
    \includegraphics[width=14cm]{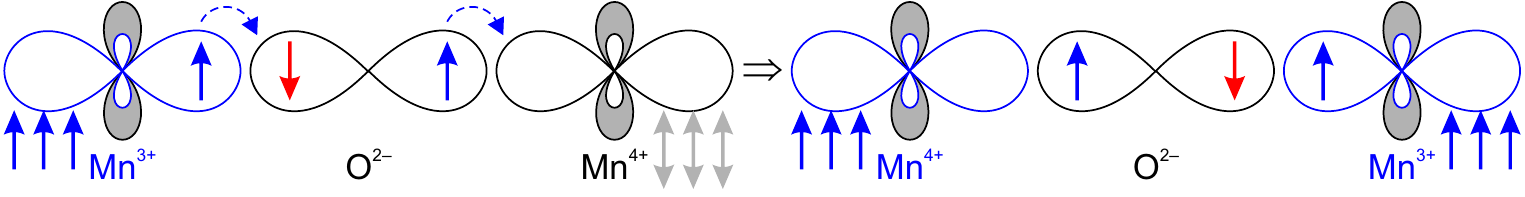}
    \caption{\label{fig:DE}
             Double exchange between the $3d_{z^2}$ orbital of a Mn$^{3+}$ and a Mn$^{4+}$ ion
             via the $2p_z$ orbital of an O$^{2-}$ ion, realized e.g.~in La$_{2/3}$Ba$_{1/3}$MnO$_3$.
             As in Fig.~\ref{fig:SE}, the $3d_{x^2-y^2}$ orbital of Mn is shaded in grey.
             The core spin $S = 3/2$ of the three (not-shown) $t_{2g}$ orbitals of Mn is symbolized
             by triple arrows. The resulting interaction is ferromagnetic, mediated by the itinerant electron of Mn$^{3+}$.}
\end{figure*}\normalsize

The above described oxygen-mediated exchange between TM ions based on \emph{virtual} hopping processes of the O-$2p$ electrons is called ``superexchange'' (SE). In mixed-valence manganites, however, \emph{real} hopping of electrons between Mn$^{3+}$ and Mn$^{4+}$ ions becomes possible and was first theoretically studied by Zener \cite{Zener1951}, Anderson and Hasegawa \cite{Anderson1955}, as well as de Gennes \cite{deGennes1960}. It results in a new magnetic exchange mechanism called ``double exchange'' (DE). Figure~\ref{fig:DE} illustrates the situation for a Mn$^{3+}$ and a Mn$^{4+}$ ion separated by an O$^{2-}$ ion, realized i.e.~in La$_{2/3}$Ba$_{1/3}$MnO$_3$. In contrast to above (Fig.~\ref{fig:SE}), the itinerant electron in the $3d_{z^2}$ orbital of Mn$^{3+}$ can hop via the O-$2p_z$ orbital to the $3d_{z^2}$ orbital of Mn$^{4+}$ without the need to pay the Coulomb energy $U$. If the core spins of the respective Mn ions are aligned parallel the Hund's coupling energy $J_{\rm H}$ will also vanish favouring electronic hopping transport across the oxygen ion. Vice versa, an antiparallel orientation of the core spins would lead to $J_{\rm H} > 0$ and therefore hamper the respective charge transfer. In summary, the itinerant charge carrier mediates a ferromagnetic ground state resulting in a close correlation between electric conductivity and ferromagnetism together with the occurrence of a significant magnetoresistive effect. Note that for $t > U$, a metallic ground state can be realized also in isovalent compounds as the high kinetic energy $t$ overcomes the Coulomb repulsion $U$, resulting in a ferromagnetic state due to Hund's coupling.

The occurrence of these O-mediated exchange interactions is not limited to manganites, but present in numerous other systems like cobaltates \cite{Fuchs2008} or cuprate superconductors \cite{Lyons1988}.

\subsection{Manganites and Carrier Doping}\label{subsec:manganites}

The double exchange model could first and successfully explain the colossal magnetoresistance (CMR) effect observed in mixed-valent manganites near their Curie temperature. The CMR effect was in principle first discovered in the 1950s \cite{vanSanten1950}. At that time, however, the effect was small due to the low sample quality and the small magnetic fields available and was not considered important. In 1954, a reduction of the resistivity by about 7\% in a field of 0.3\,T was reported for La$_{0.8}$Sr$_{0.2}$MnO$_3$ \cite{Volger1954}. In the late 1960s, a large magnetoresistive effect was found in La$_{1-x}$Pb$_x$MnO$_3$ single crystals at temperatures close to the Curie temperature \cite{Searle1969} and discussed in the framework of a double exchange mechanism \cite{Kubo1972}. But it took almost 20 more years when Kusters \textit{et al.}~\cite{Kusters1989}, von Helmolt \textit{et al.}~\cite{Helmolt1993}, and Chainami \textit{et al.}~\cite{Chainami1993} discovered an extremely large negative magnetoresistive effect in epitaxial thin films at room temperature, stimulating many research activities in the field. This was triggered by a wide interest in transition-metal oxides after the discovery of high-temperature superconductivity \cite{Bednorz1986} as well as possible applications of magnetoresistive effects in sensor technology. As the term ``giant'' has already been used for the giant magnetoresistance (GMR) in metallic superlattices \cite{Baibich1988,Binasch1989}, the new phenomenon was called ``colossal''.

To date, a large variety of doped manganites with the composition $(L^{3+}_{1-x}A^{2+}_x)({\rm Mn}^{3+}_{1-x}{\rm Mn}^{4+}_x){\rm O}^{2-}_3$ has been investigated where $L$ is a trivalent lanthanide (or rare earth ion) and $A$ a divalent alkaline earth ion \cite{Coey1999}. In most of these compounds, a CMR effect
\begin{equation}
    {\rm CMR} = -\frac{R(H)-R(0)}{R(H)} = -\frac{\Delta R}{R}
\end{equation}
is observed which at low temperatures and in high magnetic fields $H$ of several tesla can exceed 100,000\%. At room temperature, however, the CMR values are much smaller and reach only few 100\% at fields above 1\,T. These oxides have perovskite-type structure where the Mn ion is surrounded by an oxygen octahedron. Due to the crystal field the five $3d$ states are split into three $t_{2g}$ and two $e_g$ states. In the undoped case ($x=0$), only Mn$^{3+}$ ions are present ($3d^4$). Due to the strong on-site Hund's rule exchange, these four $d$ electrons exhibit parallel spins. The half-filled $e_g$ band is split by the Jahn-Teller effect, making LaMnO$_3$ an antiferromagnetic insulator. Substitution with divalent alkaline earth ions $A_x^{2+}$ introduces holes into the $e_g$ band. If the hole density is sufficiently high the holes will delocalize leading to hopping conductivity \cite{Imada1998}. Due to the strong on-site Hund's coupling, however, this will only become possible if the spins of the itinerant holes (electrons) are aligned antiparallel (parallel) to the local $S = 3/2$ core spin of the $t_{2g}$ electrons, as discussed above. As a result, hopping of the itinerant $e_g$ electron provides a ferromagnetic double exchange. Hence, a ferromagnetic ground state with a maximum Curie temperature between 280\,K and 380\,K is realized for $x = 0.3$ and $A =$ Ca, Ba, Sr.

\small\begin{figure}
    \includegraphics[width=8cm]{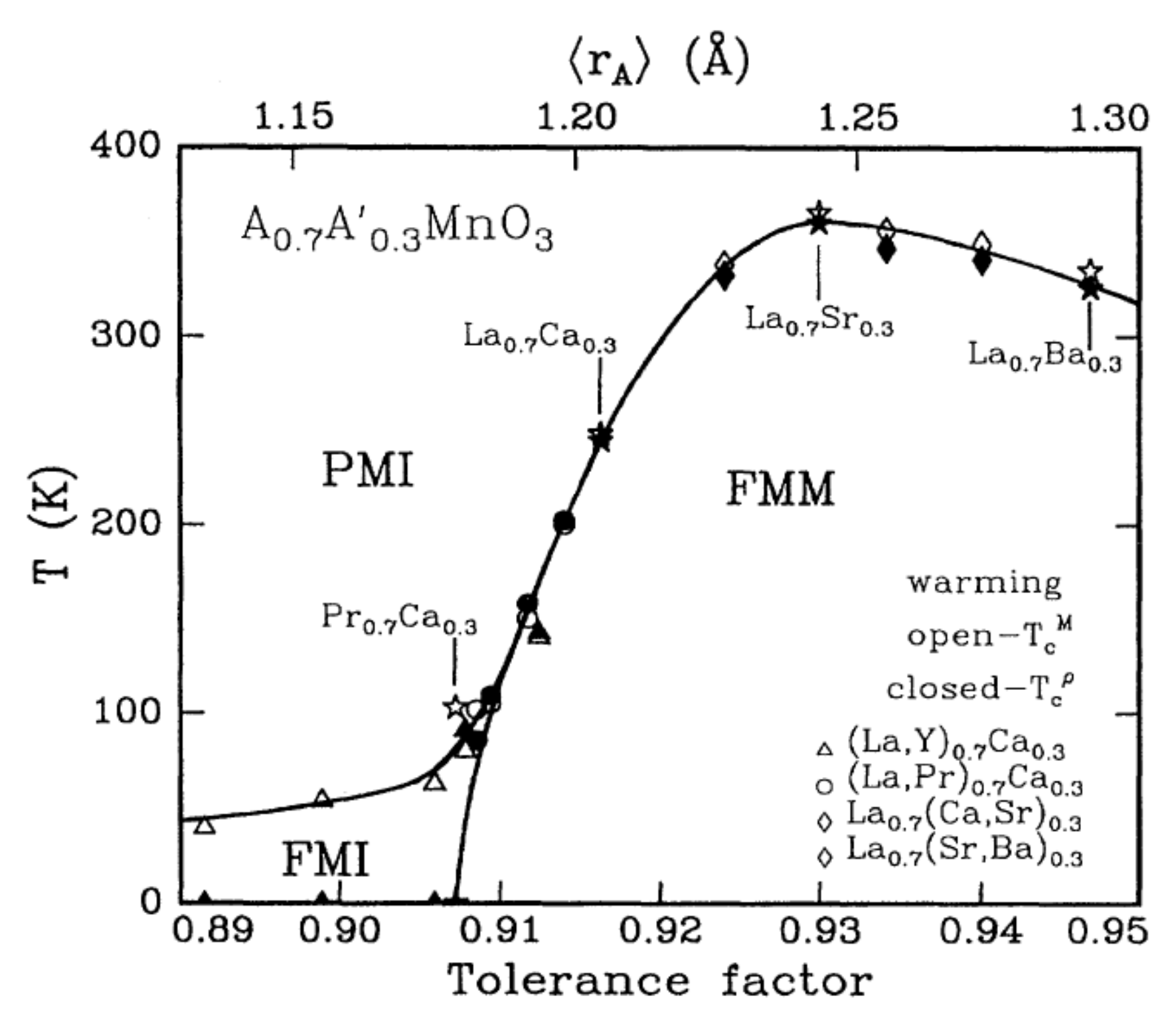}
    \caption{\label{fig:hwang}
             Curie temperature vs.~tolerance factor $t$ from eq.~(\ref{eq:tolerance})
             for the system $L_{0.7}A_{0.3}$MnO$_3$. It displays three distinct regions:
             paramagnetic insulating (PMI), ferromagnetic insulating (FMI), and ferromagnetic
             metallic (FMM). Open symbols ($T_{\rm c}^M$) denote data from magnetization measurements,
             closed symbols ($T_{\rm c}^\rho$) from resistivity measurements.
             Reproduced from \cite{Hwang1995}.
             Copyright 1995 by the American Physical Society.}
\end{figure}\normalsize

In 1995, Hwang \textit{et al.}~\cite{Hwang1995} reported on the phase diagramme of $L_{0.7}A_{0.3}$MnO$_3$. Dependent on the temperature and the tolerance factor $t$, see eq.~(\ref{eq:tolerance}), it displays paramagnetic insulating, ferromagnetic insulating and ferromagnetic metallic phases (Fig.~\ref{fig:hwang}). A deviation from $t=1$ leads to a strained and distorted crystallographic lattice, accompanied by a rotation of the oxygen octahedra. This leads to the occurrence of a characteristic superlattice structure on one hand and to a decrease of the $180^\circ$ Mn--O--Mn bond angles on the other. For La$_{2/3}$Ca$_{1/3}$MnO$_3$, as an example, it is reduced to $160^\circ$. This decrease results in a significant reduction of the overlap between the Mn-$3d$ and the O-$2p$ orbitals and weakens the ferromagnetic exchange interaction.

The literature of doped manganites has been reviewed several times \cite{Coey1999,Tokura1999,Salamon2001,Ziese2002,Haghiri-Gosnet2003,Dorr2006}. Even though their study has deepened our understanding of the physics of magnetic, charge and orbital degrees of freedom and their interrelation, from an application point of view the manganites were disappointing as initial hopes of using them in room-temperature spintronics have not been fulfilled \cite{Bibes2007}. However, it motivated the search for new half-metals with higher Curie temperatures.

\subsection{Double Perovskites and Strain Effects}\label{subsec:DP}

The family of double perovskites with composition $A_2BB'$O$_6$ ($A$: alkaline earth ion; $B$ and $B'$: magnetic and non-magnetic transition metal or lanthanide ion, respectively) first received attention in the 1960s \cite{Longo1961,Sleight1962,Patterson1963,Nakagawa1968}, but it took until 1998 that one of these compounds, Sr$_2$FeMoO$_6$, indicated half-metallic behaviour \cite{Kobayashi1998}. Sarma \textit{et al.}~\cite{Sarma2000a} suggested a model based on a \emph{kinetic energy-driven ferromagnetic exchange} to explain ferromagnetism in the double perovskites, which has been extended to many other systems by Fang, Kanamori and Terakura \cite{Kanamori2001,Fang2001}. In their model, the hybridization of the $t_{2g}$ states of both the $B$ and the $B'$ ion plays the key role in stabilizing ferromagnetism at high Curie temperatures as these hybridized states are located energetically between the exchange-split $d_\uparrow$ and $d_\downarrow$ levels of the magnetic $B$ ion. Hopping of an itinerant electron on the $B,B'$ sublattice results in a finite coupling between states of the same symmetry and spin, causing a finite spin polarization at the Fermi level. This model was confirmed by element-specific magnetization measurements in polycrystalline Sr$_2$CrWO$_6$, Ba$_2$CrWO$_6$, Ca$_2$CrWO$_6$ and Sr$_2$CrReO$_6$ \cite{Philipp2003b,Majewski2005a,Majewski2005c}.

\begin{figure}
    \includegraphics[width=8cm]{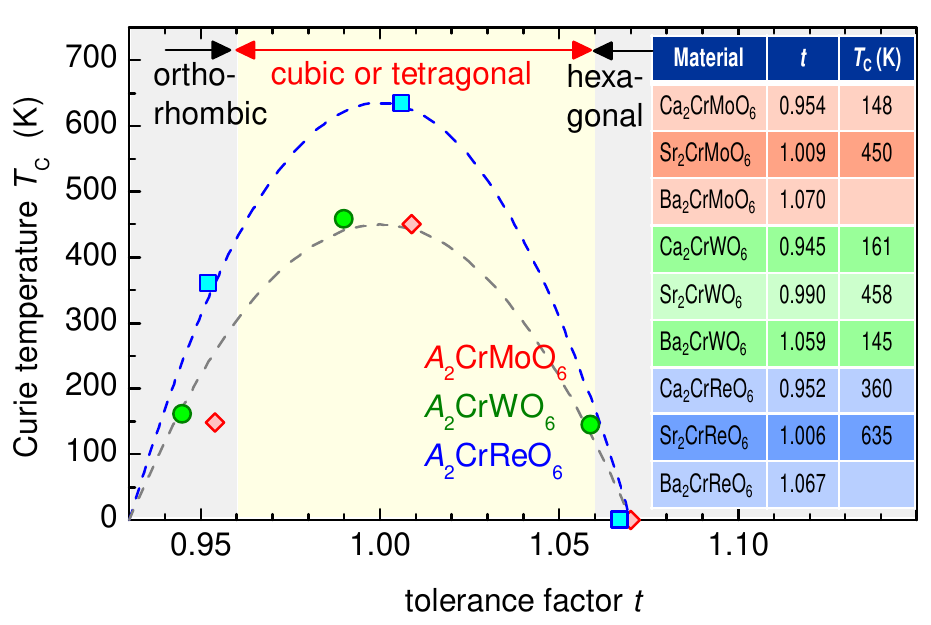}
    \caption{Curie temperatures versus tolerance factor for various Cr-based double perovskites \cite{Philipp2003b}.
             The dashed lines are guides to the eye. Inspired from \cite{Philipp2003b}.}
    \label{fig:philipp}
\end{figure}

As for the doped manganites, the (ferro)magnetic properties of the double perovskites depend sensitively on the orbital overlap and, thus, on the structure of the materials. The maximum Curie temperature for different families of double perovskites is obtained for systems exhibiting a tolerance factor of $t\simeq1$ (Fig.~\ref{fig:philipp}) \cite{Philipp2003b}. These systems display cubic/tetragonal symmetry. For $t<1$, there is a transition to orthorhombic structures below $t\sim0.96$, whereas for $t>1$ a transition to a hexagonal structure is observed above $t\sim1.06$.

The advantages of metallic double perovskites are counteracted by some severe problems. The main is that optimum magnetic properties require a perfect three-dimensional order on the $B$ sublattice which becomes difficult if $B$ and $B'$ ions are of similar size. Cationic disorder, however, systematically decreases the magnetization due to antiferromagnetic coupling between $B$ neighbours at anti-phase boundaries \cite{Balcells2001,Navarro2001a}. Disorder is also detrimental to the spin polarization of the carriers at the Fermi energy \cite{Sarma2000b}.

Despite these difficulties, the growth of double perovskite thin films was carried out in several places. Regarding Sr$_2$FeMoO$_6$, laser-MBE became popular \cite{Manako1999,Venimadhav2004,Sanchez2004,Fix2005,Wang2006} although the fabrication of single phase thin films turned out to be challenging. The growth pressure has to be kept low to avoid the formation of secondary phases \cite{Manako1999,Santiso2002}, but a high growth temperature $>850^\circ$C is required for good Fe/Mo ordering \cite{Santiso2002}. Electron doping via substitution of Sr$^{2+}$ by La$^{3+}$ results in an increase of $T_{\rm C}$ from 370\,K to 410\,K in thin films \cite{Branford2003} (and from 420\,K to 490\,K in polycrystalline samples \cite{Navarro2001b}). For the spin polarization, values of $60\ldots75\%$ have been obtained by Andreev point contact spectroscopy, however, the evaluation of the spectra is not unambiguous \cite{Auth2003}.

\small\normalsize

In 2001, Philipp \textit{et al.}~\cite{Philipp2001,Philipp2003a} reported the successful layer-by-layer growth of Sr$_2$CrWO$_6$ epitaxial thin films on SrTiO$_3$ substrates using laser-MBE. Due to the similar ionic radii of Cr and W, these elements showed no sublattice order. Nevertheless, the Curie temperature was found to be well above 400\,K. As in Sr$_2$FeMoO$_6$, electron doping via La$^{3+}$ substitution leads to an increase in $T_{\rm C}$. However, the material tends to preferentially form a secondary phase of LaCrO$_3$ when prepared as bulk \cite{Geprags2006}. This can be avoided for thin films grown by laser-MBE due to the non-equilibrium growth conditions \cite{Majewski2005b}.

An alternative route to electron doping is represented by $5d$ band filling via substitution on the $B'$ site. Replacing W$^{5+}$ ($5d^1$) by Re$^{5+}$ ($5d^2$) leads to an additional itinerant $t_{2g}$ electron in the minority spin channel. Sr$_2$CrReO$_6$ shows a giant anisotropic magnetostriction \cite{Serrate2007} caused by a large orbital moment on the Re site \cite{Majewski2005c}. In addition, it has a high $T_{\rm C} = 635$\,K \cite{Kato2002} well above room temperature, and a predicted high spin polarization of 86\% \cite{Vaitheeswaran2005}. In 2009, Gepr\"{a}gs \textit{et al.}~\cite{Geprags2009} successfully deposited Sr$_2$CrReO$_6$ thin films on SrTiO$_3$ substrates. Phase-pure films with optimum crystallographic and magnetic properties were obtained by growing in O$_2$ atmosphere ($\sim 5\times10^{-4}$\,mbar) at a substrate temperature of $\sim 700^\circ$C. The films are c-axis oriented and coherently strained with less than 20\% anti-site defects. The magnetization curves reveal a high saturation magnetization of $0.8\,\mu_{\rm B}$/f.u.~and a high coercivity of 1.1\,T, as well as a strong magnetic anisotropy \cite{Geprags2009}. Similar results were obtained by Orna \textit{et al.}~\cite{Orna2010} who, in addition, reported $T_{\rm C} = 481$\,K and a resistivity of 2.8\,m$\Omega$cm at 300\,K. An even higher Curie temperature of 725\,K was found in polycrystalline bulk samples of Sr$_2$CrOsO$_6$ (Os$^{5+}$: $5d^3$) which has a completely filled $5d$ $t_{2g}$ minority spin orbital \cite{Krockenberger2007}. However, no reports on epitaxial thin films exist so far. For none of the Cr-based double perovskites, data concerning their spin polarization are yet available. For further information, I refer the reader to two review articles \cite{Serrate2007,Philipp2004}.

\subsection{Magnetite and Zn Substitution}\label{subsec:magnetite}

\small\begin{figure}
    \includegraphics[width=8cm]{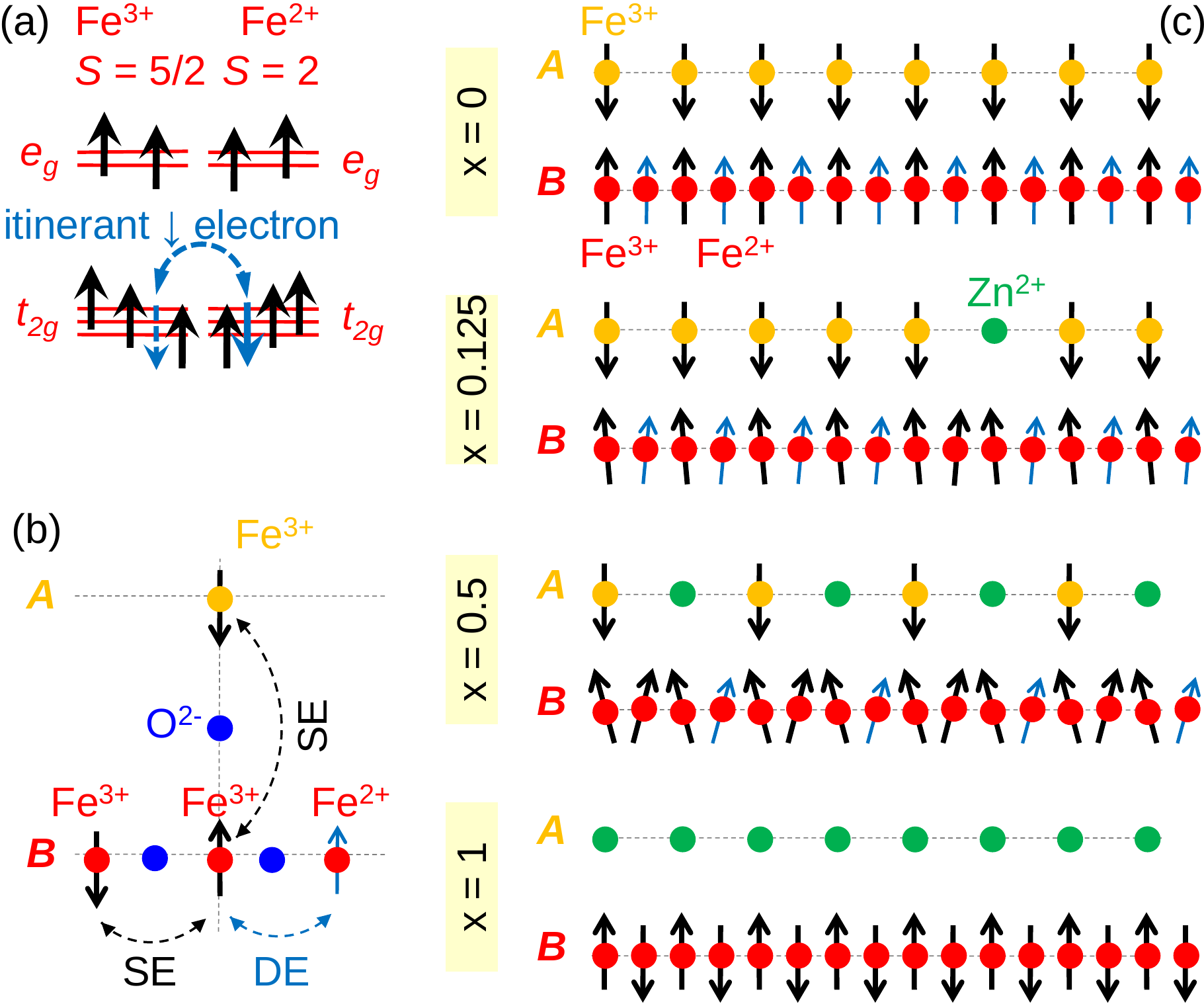}
    \caption{Electronic and magnetic structure of Fe$_{3-x}$Zn$_x$O$_4$.
             (a) The $t_{2g}$ spin-down electron can hop between Fe$_B^{2+}$ and Fe$_B^{3+}$ ions
             with its spin coupled anti-parallel to the local moment formed by the spin-up electrons owing to Hund's rule coupling.
             (b) Superexchange (SE) between Fe$^{3+}$ ions, double exchange (DE) between Fe$^{3+}$ and Fe$^{2+}$ ions.
             (c) Evolution of the sublattice configuration from $x=0$ (ferrimagnetic) to $x=1$ (antiferromagnetic).
             Reproduced from \cite{Venkateshvaran2009}. Copyright 2009 by the American Physical Society.}
    \label{fig:Fe3O4}
\end{figure}\normalsize

The oxide ferrimagnet magnetite (Fe$_3$O$_4$) shows a high $T_{\rm C} \simeq 860$\,K allowing for room temperature spintronic applications \cite{Gorter1955} and crystallizes in the inverse spinel structure. The chemical sum formula can be expressed as [Fe$^{3+}$]$_A$[Fe$^{3+}$Fe$^{2+}$]$_B$O$_4$. The $A$ sites are occupied by Fe$_A^{3+}$ ions ($3d^5, S = 5/2$), whereas on the $B$ sites there is an alternating arrangement of Fe$_B^{2+}$ ($3d^6, S = 2$) and Fe$_B^{3+}$ ($3d^5, S = 5/2$) (Fig.~\ref{fig:Fe3O4}(a)). The density of itinerant charge carriers is determined by the density of the $t_{2g}$ spin-down electrons on the $B$ site, i.e.~by the density of Fe$_B^{2+}$, leading to hopping transport on the $B$ sublattice (Fig.~\ref{fig:Fe3O4}(a)). Below $T_{\rm V} = 123$\,K, bulk Fe$_3$O$_4$ undergoes a structural phase transition from cubic to monoclinic, first reported by Verwey \cite{Verwey1939}. Whether or not this ``Verwey'' transition is accompanied by charge ordering of the Fe$_B^{2+}$ and Fe$_B^{3+}$ on the $B$ sublattice in a process reminiscent to Wigner crystallization \cite{Mott1967} is unclear and still a matter of intense debate \cite{Walz2002,Subias2004,Leonov2004,McQueeney2007,Piekarz2007,Lorenzo2008,Garcia2009,Blasco2011}. The ``Verwey'' transition leads to a discontinuity in both the resistivity and the magnetization of Fe$_{3-\delta}$O$_4$ at $T=T_{\rm V}$, being sensitively dependent on perfect stoichiometry and vanishing for $\delta \gtrsim 0.0117$ \cite{Shepherd1985}. Below 38\,K, the material is reported to become ferroelectric with a switchable polarization of $11\,\mu$C/cm$^2$ below 20\,K \cite{Alexe2009}.

The magnetic exchange is governed by a combination of antiferromagnetic superexchange (SE) between the Fe$^{3+}$ ions on the $A$ and $B$ sites ($J_{AB}$ ($A$--O--$B$)) and ferromagnetic double exchange (DE) between the mixed-valent Fe$^{3+}$ and Fe$^{2+}$ ions on the $B$ sites ($B$--O--$B$), see Fig.~\ref{fig:Fe3O4}(b). Owing to Hund's rule coupling, the spin of the itinerant $t_{2g}$ electron is aligned antiparallel to the localized $3d$ spin-up electrons (Fig.~\ref{fig:Fe3O4}(a)). Since the antiparallel Fe$_A^{3+}$ and Fe$_B^{3+}$ moments compensate each other, a saturation magnetization of $4\mu_{\rm B}$/f.u.~is expected from the remaining Fe$_B^{2+}$ ($S=2$) moments (Fig.~\ref{fig:Fe3O4}(c)). This simple N\'{e}el model \cite{Neel1948} has been extended by Yafet and Kittel \cite{Yafet1952}. They introduced two Fe$_B^{2+}$ and Fe$_B^{3+}$ sub-sublattices and considered a $B$--O--$B$ SE interaction resulting in spin canting expressed by a finite Yafet-Kittel angle and thus a reduction of the saturation magnetization (Fig.~\ref{fig:Fe3O4}(c)). More recent works model the magnetic properties of Fe$_3$O$_4$ in more detail by considering competing exchange interactions \cite{McQueeney2007,Loos2002,Rosencwaig1969}. In 1984, Yanase and Siratori \cite{Yanase1984} reported on augmented plane-wave (APW) calculations \cite{Slater1937}. Regarding the spin density of states, they found a gap for majority-spin electrons, but metallic character for minority-spin carriers, thus at the first time representing half-metallicity in an oxide. Later, Zhang and Satpathy \cite{Zhang1991} confirmed this result by local spin density approximation calculations. They further confirmed that electronic conduction in Fe$_3$O$_4$ is restricted to hopping of itinerant electrons between the Fe-$t_{2g}$ orbitals on the $B$ sublattice. In 2002, Dedkov \textit{et al.}~\cite{Dedkov2002} reported a spin polarization of up to $P = -(80 \pm 5)$\,\% near the Fermi energy at room temperature from spin-resolved photoemission spectroscopy in $(111)$-oriented, thermally deposited and oxidized epitaxial thin films. In following publications (see \cite{Fonin2008} and references therein), the authors confirmed their results and presented additional evidence for the existence of a half-metallic state with a (111) orientation.

In 1997, Gong \textit{et al.}~\cite{Gong1997} reported on the first PLD-grown thin films. They deposited Fe$_3$O$_4$ on (001)-oriented MgO substrates (lattice mismatch $f=-0.36\%$) and showed a saturation magnetic moment at room temperature of $3.3\,\mu_{\rm B}$/f.u. For such films, however, Fe$_3$O$_4$ is known to form structural defects as its lattice constant $a$ is nearly twice as large as the one of MgO. When different Fe$_3$O$_4$ islands meet, they might be shifted by $\frac{1}{2}a$ along $\{100\}$ or $\frac{1}{4}\sqrt{2}a$ along $\{110\}$ or rotated by $90^\circ$ with respect to each other, thus yielding a so-called ``anti-phase boundary (APB)'' \cite{Margulies1996}. Across this APB, the oxygen lattice stays undistorted while the cation lattice is displaced resulting in a perturbation of the magnetic order on the $A$ and $B$ sublattices. Eerenstein \textit{et al.}~\cite{Eerenstein2002} demonstrated that the anti-phase domain size $D$ directly depends on the film thickness $d$ following $D\propto\sqrt{d}$. This results in a high APB density for ultrathin films (i.e.~few nm) which strongly increases its electrical resistivity. This effect also accounts for a reduced saturation magnetization as compared to the expected value of $4\,\mu_{\rm B}$/f.u. In 1998, Ogale \textit{et al.}~\cite{Ogale1998} reported the successful growth on SrTiO$_3$(001) ($f=7.5\%$) and Al$_2$O$_3$(0001) ($1.9\%$) where Fe$_3$O$_4$ was oriented in (001) and (111) direction, respectively. In 2003, Reisinger \textit{et al.}~\cite{Reisinger2003a,Reisinger2003b} reported the epitaxial growth of Fe$_3$O$_4$ thin films on MgO(001) and on Si(001) substrates, the latter utilizing a buffer layer of TiN/MgO. The magnetic and electrical properties of these films were found to be close to those of single crystals \cite{Reisinger2004a}. For $40\ldots50$\,nm thin films, the saturation magnetization and electrical resistivity were $3.6\,\mu_{\mathrm B}$/f.u.~and 4.4\,m$\Omega$cm at room temperature, respectively. From Hall measurements, they deduced a carrier density of 0.22/f.u. Both normal and anomalous Hall effect were found to show negative sign.

\small\begin{figure}
    \includegraphics[width=8cm]{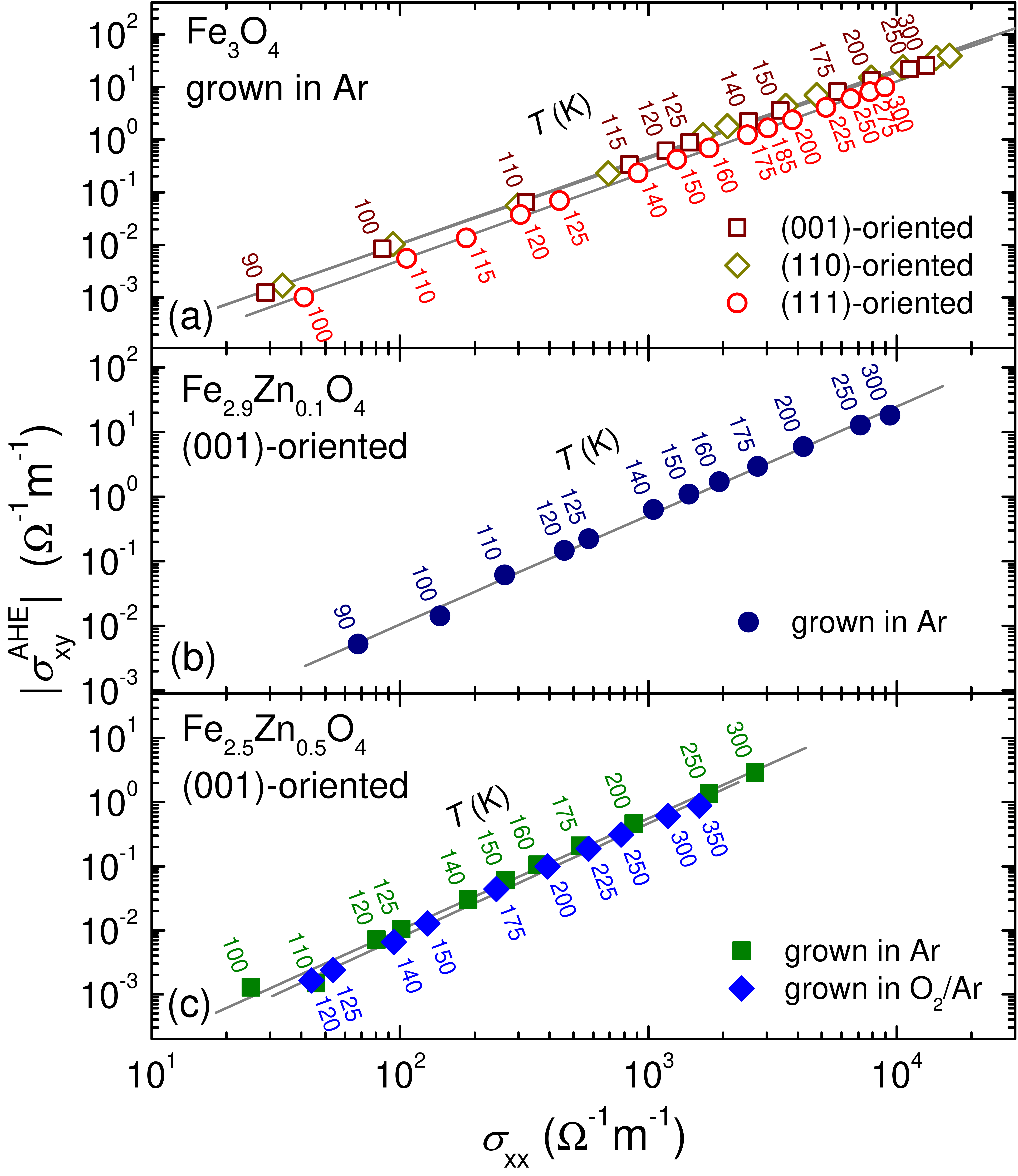}
    \caption{Modulus of the anomalous Hall conductivity, $|\sigma_{xy}^\mathrm{AHE}|$, plotted versus longitudinal
             conductivity $\sigma_{xx}$ in a double logarithmic representation for different epitaxial
             Fe$_{3-x}$Zn$_x$O$_4$ films in the temperature regime between 90 and 350\,K.
             The linear fits to the data (lines) yield a slope of $1.69\pm0.08$.
             Reproduced from \cite{Venkateshvaran2008}. Copyright 2008 by the American Physical Society.}
    \label{fig:venkateshvaran}
\end{figure}\normalsize

In 2008, Fern\'{a}ndez-Pacheco \textit{et al.}~\cite{Fernandez-Pacheco2008a,Fernandez-Pacheco2008b} presented a careful magnetotransport investigation of epitaxial Fe$_3$O$_4$ thin films grown on MgO(001) with particular emphasis on the different Hall effects in ferromagnets \cite{Nagaosa2010}. The authors found a negative ordinary Hall resistivity with a huge enhancement below $T_{\rm V}$, reaching values of $-1$\,m$\Omega$cm. Furthermore, they found a giant planar Hall coefficient with a record value at room temperature of $\approx 60\,\mu\Omega$cm for a 5\,nm thin film, also displaying a huge enhancement when crossing the Verwey transition temperature. The anomalous Hall conductivity $\sigma_{xy}^\mathrm{AHE}$ was found to scale as
\begin{equation}
    \sigma_{xy}^\mathrm{AHE} \propto \sigma_{xx}^{1.6}
\end{equation}
over four decades of the longitudinal conductivity $\sigma_{xx}$ in several samples. This was remarkable as according to the conventional models one expects $\sigma_{xy}^\mathrm{AHE} \propto \sigma_{xx}$ for skew scattering \cite{Luttinger1958} or $\sigma_{xy}^\mathrm{AHE} \sim const.$~for the side jump mechanism \cite{Nozieres1973}. However, the exponent $\alpha = 1.6$ was in agreement with a unified quantum transport theory for the anomalous Hall effect in ferromagnets, predicting three different regimes as a function of the carrier relaxation time: a superclean limit with $\alpha=1$, a moderately dirty limit ($\alpha=0$) and a scaling regime with $\alpha=1.6$ \cite{Onoda2006,Onoda2008}. This behaviour could be confirmed by Venkateshvaran \textit{et al.}~\cite{Venkateshvaran2008} who reported $\alpha = 1.69 \pm 0.08$ for a whole series of epitaxial films of Fe$_{3-x}$Zn$_x$O$_4$ ($x = 0, 0.1, 0.5$) in (100), (110) and (111) orientation (Fig.~\ref{fig:venkateshvaran}). For further information on the anomalous Hall effect and its scaling behaviour, I refer the reader to a recent comprehensive review by Nagaosa \textit{et al.}~\cite{Nagaosa2010}.

\small\begin{figure}
    \includegraphics[width=6cm]{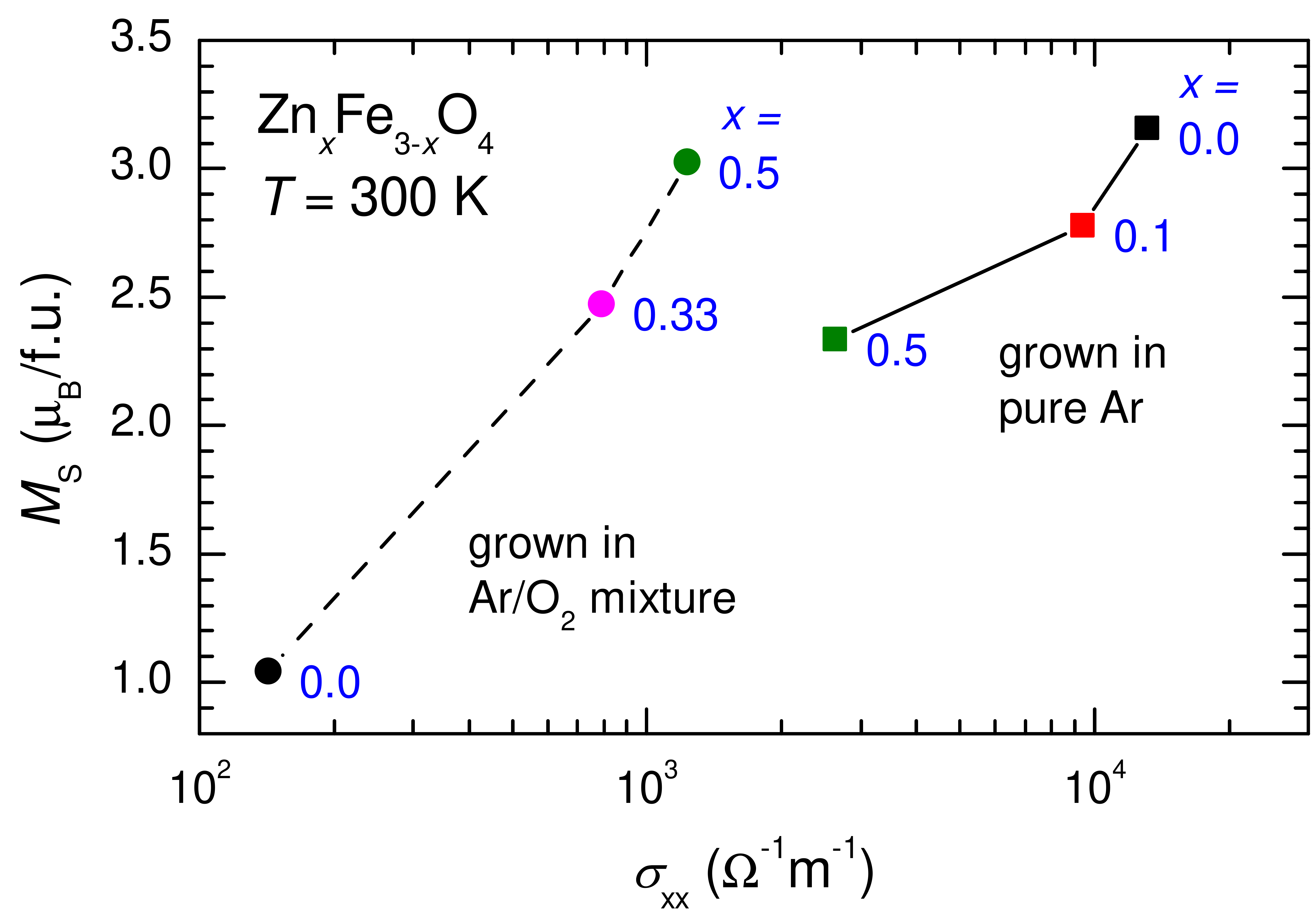}
    \caption{Saturation magnetization $M_{\rm S}$ plotted versus the
             longitudinal conductivity $\sigma_{xx}$ for epitaxial Zn$_x$Fe$_{3-x}$O$_4$ thin films,
             grown by PLD in Ar (squares) or Ar/O$_2$ (99:1) (circles).
             Reproduced from \cite{Venkateshvaran2009}. Copyright 2009 by the American Physical Society.}
    \label{fig:venkateshvaran8}
\end{figure}\normalsize

In a following publication, Venkateshvaran \textit{et al.}~\cite{Venkateshvaran2009} reported in detail on Zn substitution in Zn$_x$Fe$_{3-x}$O$_4$ ($0 \leq x \leq 0.9$) thin films. The nonmagnetic divalent ion Zn$^{2+}$ preferably occupies the tetrahedrally coordinated $A$ site \cite{Harris1996} replacing Fe$_A^{3+}$ and converting Fe$_B^{2+}$ to Fe$_B^{3+}$ on the $B$ sublattice due to charge neutrality (see Fig.~\ref{fig:Fe3O4}(c)) \cite{Venkateshvaran2009}. That is, the amount of itinerant charge carriers mediating the DE on the $B$ sublattice is reduced as shown by photoemission spectroscopy \cite{Takaobushi2007} leading to a reduction of electrical conductivity. The same is true for the fabrication of thin films in excess oxygen where the presence of O$^{2-}$ ions again requires a reduced (increased) amount of Fe$_B^{2+}$ (Fe$_B^{3+}$) ions \cite{Venkateshvaran2009}. Venkateshvaran \textit{et al.}~\cite{Venkateshvaran2009} found that both the electrical conductivity and the saturation magnetization can be tuned over a wide range ($10^2\ldots10^4$\,$\Omega^{-1}$m$^{-1}$ and $1.0\ldots3.2\,\mu_{\rm B}$/f.u.~at room temperature) by Zn substitution and/or finite oxygen partial pressure during growth. Both electrical conduction and magnetism are determined by the density and the hopping amplitude of the itinerant charge carriers on the $B$ sublattice, providing electrical conduction and ferromagnetic exchange. A decrease (increase) in charge carrier density results in a weakening (strengthening) of DE and thereby a decrease (increase) in the conductivity and the saturation magnetization. This scenario was confirmed by the observation that the saturation magnetization scales with the longitudinal conductivity over two orders of magnitude (Fig.~\ref{fig:venkateshvaran8}).

In conclusion, Fe$_3$O$_4$ is the half-metallic oxide with the highest Curie temperature and offers various opportunities to tailor its electrical and magnetic properties. Such tunable ferromagnetic materials are desired for oxide spintronic devices with regard to operation at or above room temperature.

\subsection{Magnetic Tunnel Junctions}

Magnetic tunnel junctions (MTJs) are key elements for the realization of non-volatile magnetic random access memory (MRAM) devices~\cite{Akerman2005}, magnetic sensors, or programmable logic elements~\cite{Ney2003}. MTJs based on simple ferromagnetic metals and alloys have been well known for many years. The tunneling magnetoresistive (TMR) effect was first reported in 1975 in Fe/GeO$_x$/Co MTJs by Julli\'{e}re \cite{Julliere1975}. In the early stage of research, following an approach developed by Tedrow and Meservey \cite{Tedrow1970,Tedrow1971}, the TMR effect has been related to the spin polarizations $P_{1,2}$ of the two electrodes via Julli\'{e}re's equation \cite{Julliere1975}
\begin{equation}
    \frac{R_{\rm ap}-R_{\rm p}}{R_{\rm p}} = \mathrm{TMR} = \frac{2 P_1 P_2}{1 - P_1 P_2}
    \label{eq:julliere}
\end{equation}
with $R_{\rm ap}$ and $R_{\rm p}$ denoting the tunnel resistance for the antiparallel and parallel magnetization directions, respectively. Further experiments by de Teresa \textit{et al.}~\cite{deTeresa1999a,deTeresa1999b} have shown that the spin polarization determined from tunneling depends on the barrier material and can show different signs for the same ferromagnetic electrodes. While in the late 1990s the achieved TMR values at room temperature could be increased up to around 10\% \cite{Moodera1999}, it was predicted in 2000 that a further dramatic increase should be possible for Fe/MgO/Fe MTJs using highly textured MgO(100) tunneling barriers due to the Bloch state filtering of the electronic wave functions \cite{Butler2001,Mathon2001}. Indeed, in 2004 a high TMR effect above 200\% was reported for Fe \cite{Yuasa2004} or CoFe electrodes \cite{Parkin2004}, and today $\mathrm{TMR} > 600$\% is observed in CoFeB/MgO/CoFeB MTJs at room temperature \cite{Ikeda2008}. More detailed information can be found in several reviews \cite{Zutic2004,Bibes2007,Tsymbal2003,Miao2011}.

\small\begin{figure*}
    \includegraphics[height=4.4cm]{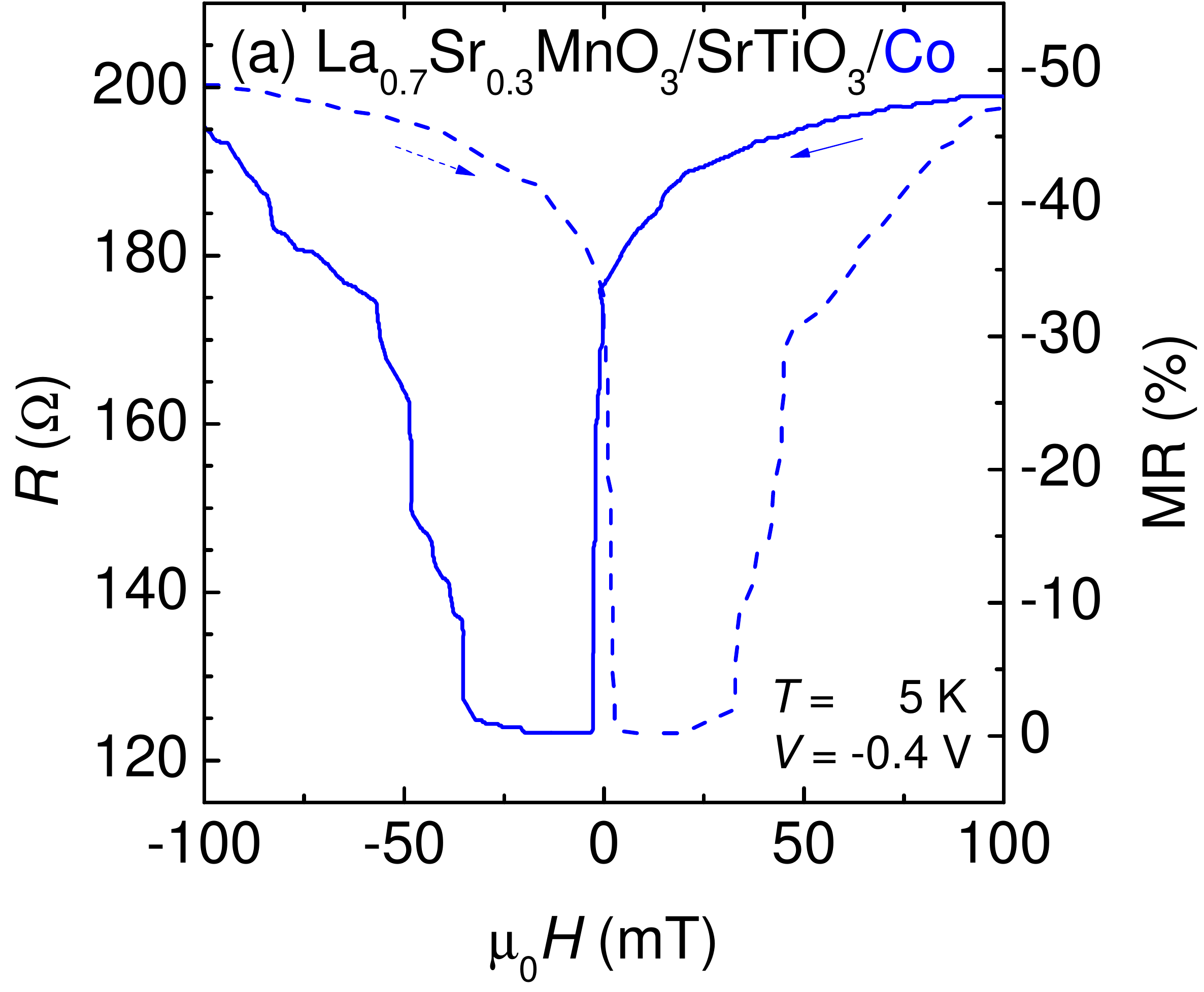}
    \includegraphics[height=4.3cm]{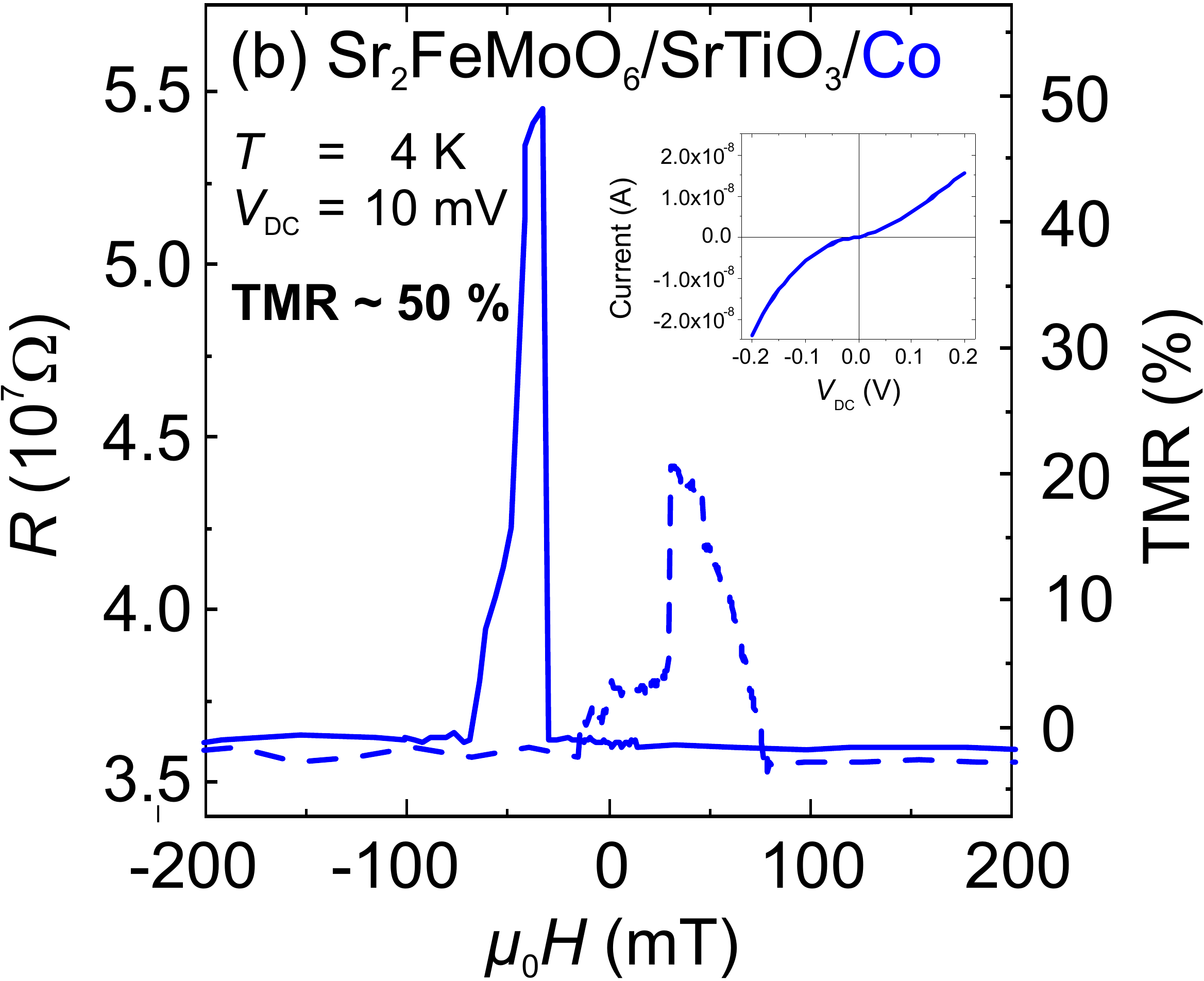}
    \includegraphics[height=4.4cm]{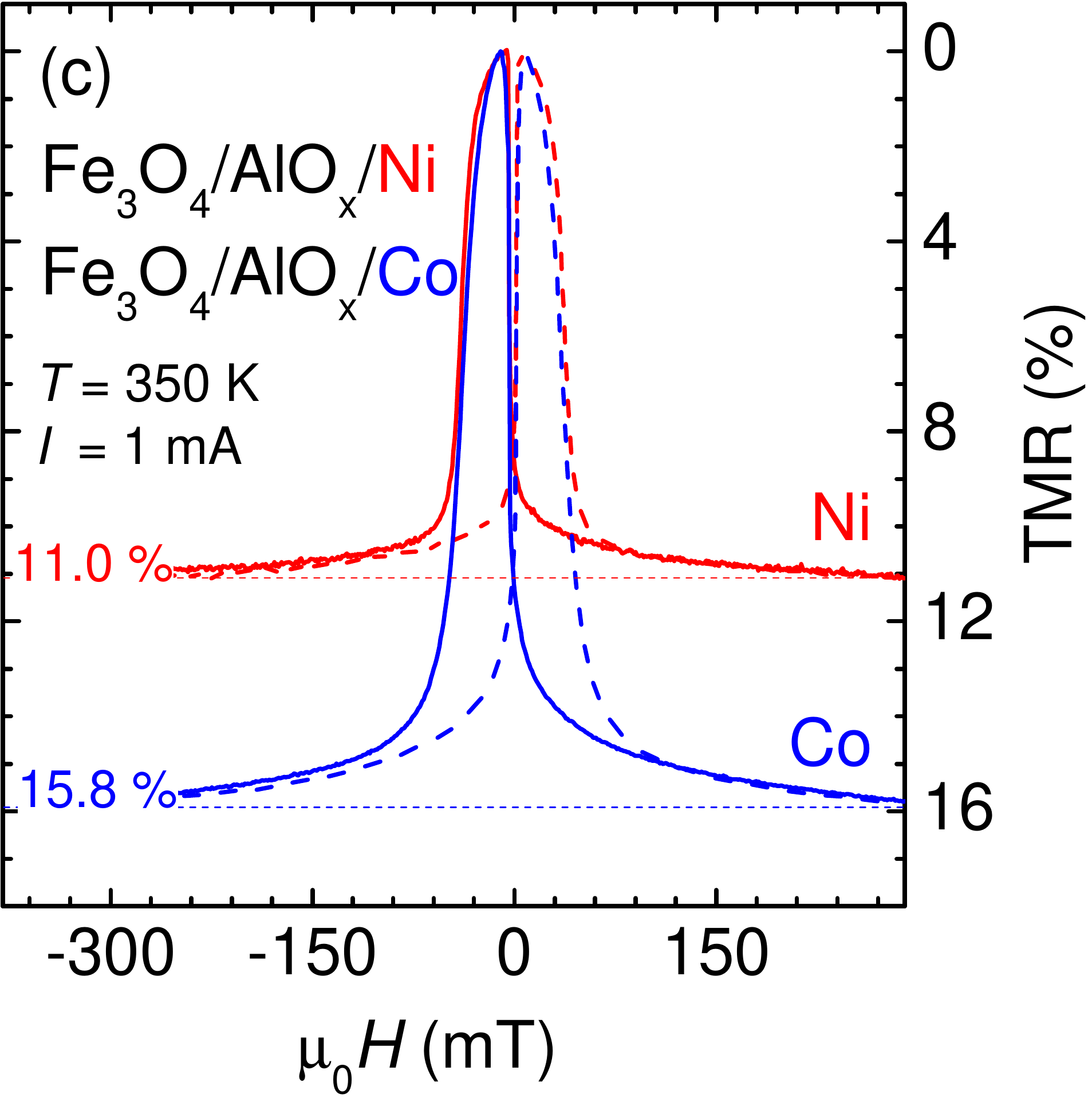}
    \caption{TMR for different oxide-based MTJs.
             (a) La$_{0.7}$Sr$_{0.3}$MnO$_3$/SrTiO$_3$/Co.
             Reproduced from \cite{deTeresa1999b}. Copyright 1999 by the American Physical Society.
             (b) Sr$_2$FeMoO$_6$/SrTiO$_3$/Co \cite{Bibes2010}.
             (c) Fe$_3$O$_4$/AlO$_x$/Ni (red) and Fe$_3$O$_4$/AlO$_x$/Co (blue).
             Reproduced with permission from \cite{Opel2011}. Copyright Wiley-VCH Verlag GmbH \& Co.~KGaA.}
    \label{fig:TMR}
\end{figure*}\normalsize

Extremely high TMR values are expected for MTJs based on half-metallic ferromagnets even without wave function filtering. Along this line, half-metallic ferromagnetic oxides are promising candidates. The first MTJs with manganite electrodes were reported in 1996 by Lu \textit{et al.}~\cite{Lu1996} and Sun \textit{et al.}~\cite{Sun1996} who obtained the best results for optimally doped La$_{2/3}$Sr$_{1/3}$MnO$_3$ electrodes and SrTiO$_3$ barriers with a thickness in the range of $3\ldots6$\,nm. A maximum TMR of 83\% was found at 4.2\,K \cite{Lu1996}, which after eq.~(\ref{eq:julliere}) corresponds to a spin polarization of 54\%. In the following years, the TMR values increased up to 1,850\% as reported by Bowen \textit{et al.} in 2003 \cite{Bowen2003}. This record TMR corresponds to a spin polarization of 95\%, i.e.~a virtually half-metallic character for La$_{2/3}$Sr$_{1/3}$MnO$_3$. De Teresa \textit{et al.}~\cite{deTeresa1999b} investigated MTJs with Co counter electrodes. They found a negative TMR of $-50\%$ at 5\,K in La$_{0.7}$Sr$_{0.3}$MnO$_3$/SrTiO$_3$/Co indicating a negative spin polarization of Co at the SrTiO$_3$ interface (Fig.~\ref{fig:TMR}(a)).

For double perovskites there are only very few reports on MTJs, all of them dealing with Sr$_2$FeMoO$_6$ electrodes. Bibes \textit{et al.}~\cite{Bibes2003} investigated Sr$_2$FeMoO$_6$/SrTiO$_3$/Co tunnel junctions and found a positive TMR of $10\ldots50\%$ at low temperatures (Fig.~\ref{fig:TMR}(b)). In view of the negative TMR in the La$_{0.7}$Sr$_{0.3}$MnO$_3$/SrTiO$_3$/Co system \cite{deTeresa1999b}, this indicates that the spin polarization of Sr$_2$FeMoO$_6$ is comparable, but with opposite sign, i.e.~at least $-80\%$ \cite{Bibes2003}. In 2005, Asano \textit{et al.}~confirmed the positive TMR and reported 10\% at 4\,K in Sr$_2$FeMoO$_6$-based junctions using a native oxide as tunnel barrier and Co as top electrode \cite{Asano2005}.

From the viewpoints of theory (half-metallicity) and application (highest Curie temperature), Fe$_3$O$_4$ should be a superior material for the realization of MTJs. However, so far the maximum achieved TMR values fall far behind the expectations. Different groups investigated Fe$_3$O$_4$-based MTJs with barriers of MgO or Al$_2$O$_3$ and counter electrodes of CoCr$_2$O$_4$ \cite{Li1998}, Co \cite{Seneor1999}, CoO \cite{vanderZaag2000}, CoFe \cite{Matsuda2002,Aoshima2003,Yoon2004}, or NiFe \cite{Aoshima2003,Park2003}. At low temperatures, Seneor \textit{et al.}~\cite{Seneor1999} reported a TMR of $+43\%$ in Fe$_{3-x}$O$_4$/Al$_2$O$_3$/Co junctions. The positive sign was also observed by Aoshima and Wang \cite{Aoshima2003}, Yoon \textit{et al.}~\cite{Yoon2004} and Bataille \textit{et al.}~\cite{Bataille2007}. At room temperature, however, all authors found TMR values below $+15\%$.

Reisinger \textit{et al.}~\cite{Reisinger2004b,Opel2011} investigated Fe$_3$O$_4$-based MTJs with different epitaxial (MgO, NdGaO$_3$) or polycrystalline (SiO$_2$, AlO$_x$) barriers. The highest TMR values were found for a 2.5\,nm thin AlO$_x$ barrier. At 350\,K, they reported positive TMR values of $+11.0\%$ or $+15.8\%$ for Ni or Co counter electrodes, respectively (Fig.~\ref{fig:TMR}(c)). For an improved ring-shaped geometry, maximum TMR values of up to $+20\%$ were observed for Fe$_3$O$_4$/AlO$_x$/Co MTJs at 350\,K. Moreover, a giant geometry-induced TMR effect of several 1,000\% due to the high resistivity of the electrode material could be experimentally demonstrated and theoretically modeled. This calls into question the widely accepted quantitative deduction of the spin polarization $P$ of the electrode material from simple transport experiments in MTJs as the electric current distribution is a crucial parameter. Note that not even the determination of the sign of $P_{\rm Fe_3O_4}$ can be done unambiguously by evaluating the TMR effect in MTJs as it depends on the sign of $P$ for the Ni or Co counter electrode. From theory, $P_{\rm Ni,Co}<0$ is predicted and was experimentally observed for Ni in spin-polarized scanning tunneling microscopy experiments \cite{Alvarado1992}. After eq.~(\ref{eq:julliere}), the positive TMR in Fe$_3$O$_4$/AlO$_x$/(Ni,Co) then leads to $P_{\rm Fe_3O_4}<0$ as reported by spin-resolved photoelectron spectroscopy \cite{Dedkov2002,Fonin2008}, even through an Al$_2$O$_3$ barrier \cite{Bataille2007}. However, $P_{\rm Ni,Co}>0$ was found in electrical transport experiments across AlO$_x$ interfaces \cite{Moodera1999} suggesting $P_{\rm Fe_3O_4}>0$ from Fe$_3$O$_4$/Al$_2$O$_3$/Co transport experiments \cite{Bataille2007} which stands in contrast to theory calculations \cite{Zhang1991}. To solve this problem, one might argue that for Fe$_3$O$_4$-based MTJs the $3d$ electrons play the dominant role which show a negative spin polarization in Ni or Co, but were not probed in early experiments with metallic electrodes \cite{Moodera1999}. Then, the positive TMR in Fe$_3$O$_4$/AlO$_x$/Co is consistent with $P_{\rm Fe_3O_4}<0$ \cite{Opel2011}. A negative $P_{\rm Fe_3O_4}$ was also found by Suzuki \textit{et al.}~who reported ${\rm TMR}<0$ in junctions of Fe$_3$O$_4$/$I$/La$_{0.7}$Sr$_{0.3}$MnO$_3$ with $I =$ CoCr$_2$O$_4$ \cite{Hu2002}, MgTi$_2$O$_4$ or FeGa$_2$O$_4$ \cite{Alldredge2006}.

\section{Zinc Oxide as II-IV Semiconductor}\label{sec:ZnO}

In the past ten years, the research on zinc oxide (ZnO) as a semiconductor underwent a great revival \cite{Janotti2009,Klingshirn2010} together with a rapid expansion of the field towards magnetism on the one hand and device physics and (opto-/spin-)electronic applications on the other. Being initially considered as a substrate material for GaN thin films and related alloys sharing the wurtzite lattice (see section \ref{subsubsec:wurtzite}), the availability of high-quality large bulk single crystals \cite{Park1967}, the strong luminescence demonstrated in optically pumped lasers \cite{Reynolds1996} and the prospects of gaining control over its electrical conductivity have brought ZnO into the focus of current interest. For a comprehensive review on ZnO materials and devices, I refer the reader to \cite{Ozgur2005}.

ZnO is already widely used as a transparent conducting oxide (TCO) and is a promising candidate material for future semiconductor device applications \cite{Ogale2005,Janotti2009,Ozgur2005,Wenckstern2009,Nickel2005,Jagadish2006}. It shows a direct and wide band gap of 3.3\,eV (at 300\,K) in the near-ultraviolet range \cite{Thomas1960,Mang1995,Srikant1998,Reynolds1999} together with an electron mobility of 200\,cm$^2$/Vs and a large free-exciton binding energy \cite{Thomas1960,Mang1995,Reynolds1999} making it interesting for (opto)electronics. Excitonic emission processes can persist at or even above room temperature \cite{Reynolds1996}. Recently, a high electron mobility of 180,000\,cm$^2$/Vs was reported in (Mg,Zn)O/ZnO heterostructures together with the observation of a fractional quantum Hall effect \cite{Tsukazaki2010}. Moreover, ZnO displays a small spin-orbit coupling \cite{Fu2008} resulting in a large spin coherence length which is a prerequisite for the creation, transport, and detection of spin-polarized currents in semiconductor spintronics. The bulk lattice parameters are $a = 3.2459$\,{\AA} and $c = 5.2069$\,{\AA}, and a $u$ parameter of 0.382 is reported for its hexagonal wurtzite lattice \cite{Kisi1989}. Although its properties have been extensively studied for more than 50 years \cite{Brown1957} using ZnO for semiconducting electronics was handicapped because of the difficult control over its electrical conductivity. ZnO crystals are natively always $n$-type which has been a case of extensive debate and research so far \cite{Ogale2005,Janotti2009,Ozgur2005,Nickel2005,Jagadish2006}. One decade ago, ZnO came back into the focus of interest in the framework of establishing (i) room-temperature dilute magnetic semiconductors (DMS) via substitution of Zn$^{2+}$ by $3d$ transition metal ions \cite{Dietl2000} and (ii) stable $p$-type conductivity via nitrogen doping and demonstrating the fabrication of $pn$-junctions for ultraviolet light emitting diodes \cite{Tsukazaki2005}. In the following subsections, these two aspects will be reviewed with respect to recent developments. As for the half-metallic magnetic oxides discussed above, ZnO thin films grown by laser-MBE turned out to be superior because of the enormous intrinsic flexibility of this deposition technique. Transition metal doping in ZnO thin films could be simply realized by growing from different stoichiometric targets. Possible $p$-type conductivity could be investigated in films deposited in different background atmospheres, from different target materials or while adding atomic gaseous species during deposition. An essential advantage hereby is the compatibility of the laser-MBE-process with oxygen and other reactive gases.

\subsection{Dilute Magnetic Doping in ZnO} \label{subsec:DMS}

\small\begin{figure}
    \includegraphics[width=6cm]{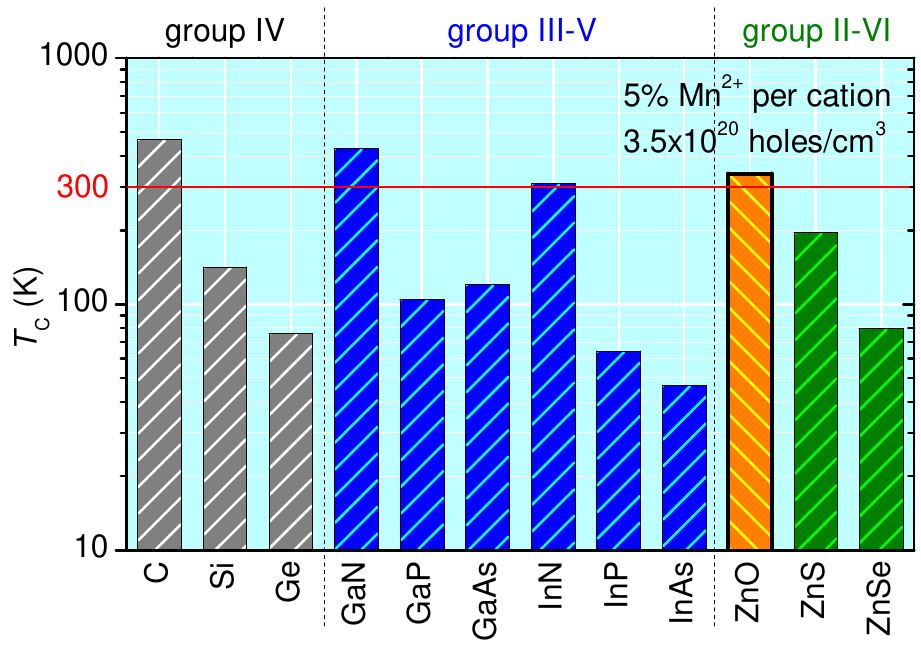}
    \caption{\label{fig:dietl}
             Computed values of the Curie temperature $T_{\rm C}$ for various $p$-type semiconductors
             containing 5\% of Mn$^{2+}$ and $3.5 \times 10^{20}$ holes per cm$^3$ \cite{Dietl2001}.}
\end{figure}\normalsize

Integrating the spin degree of freedom into semiconductor devices and the fabrication of dilute magnetic semiconductors (DMS) exhibiting intrinsic ferromagnetism is a widely discussed issue in the multifunctional materials research community \cite{Dietl2010}. Discovered in the 1960s, EuO must be considered the first intrinsic ferromagnetic semiconductor, however, with a low Curie temperature of only $T_{\rm C} \simeq 77$\,K \cite{Matthias1961}. The alternative approach to dilutely incorporate magnetic ions into the host lattice of a conventional non-magnetic semiconductor in a substitutional way, instead, seemed to be more promising with regard to tunability of the resulting magnetic properties. Hereby, it is of particular importance to avoid the formation of secondary magnetic phases because they significantly influence the magnetic properties of DMS \cite{Sato2007}, are difficult to detect \cite{Opel2008}, and lead to contradictory interpretations -- as we will see below. Munekata \textit{et al.}~\cite{Munekata1989} first reported ferromagnetism in the III-V semiconductor InAs after substituting 18\% of In by the transition metal Mn. Seven years later, Ohno {\it et al.}~\cite{Ohno1996} investigated GaAs where 3.5\% of Ga was replaced by Mn and found ferromagnetism below 60\,K. Up to now, however, the $T_{\rm C}$ values in these dilute magnetic III-V systems do not exceed 170\,K, making room-temperature operation impossible. In 2000, Dietl {\it et al.}~\cite{Dietl2000,Dietl2001} calculated the Curie temperatures for different semiconductors with a Mn substitution level of 5\%. Using an extended Zener model approach \cite{Zener1950} based on the exchange interaction between holes as charge carriers with a concentration of $3.5 \times 10^{20}$\,cm$^{-3}$ and diluted, localized Mn$^{2+}$ spins, they calculated $T_{\rm C}$ for several group IV, III-V, and II-VI semiconductors (Fig.~\ref{fig:dietl}). Interestingly, the wide bandgap materials GaN and ZnO were predicted to exhibit ferromagnetism above room temperature. This marked the starting point of an exciting race for room temperature DMS which continues to date and pushed ZnO into the focus of materials research. For a more detailed understanding, I refer the reader to Refs.~\cite{Ney2010b,Dietl2010,Chambers2010,Ogale2010}.

Ueda {\it et al.}~\cite{Ueda2001} examined for the first time ferromagnetism in $3d$ transition metal (TM)-substituted $n$-type ZnO films grown on Al$_2$O$_3$ substrates (lattice mismatch $f=18.2\%$) using laser-MBE, though Fukumura {\it et al.}~\cite{Fukumura1999} had studied this system earlier with respect to solubility, electrical properties, and magnetoresistance. Ueda {\it et al.}~\cite{Ueda2001} found that a few Zn$_{1-x}$Co$_x$O samples showed ferromagnetic features, whereas most others displayed spin glass-like behaviour. Also, the magnetic properties of their films were seen to depend on the concentration of Co ions and carriers. The reproducibility of the method was stated to be poor ($< 10\%$), and substitution with Cr, Ni, or Mn gave no indication for ferromagnetism.

Ferromagnetism at room temperature for cobalt-substituted ZnO was reported by Schwartz, Kittilstved \textit{et al.}~\cite{Schwartz2004,Kittilstved2005,Kittilstved2006a} who prepared nanocrystalline films by a direct chemical synthesis route in which TM$^{2+}$:ZnO colloids were used as precursors for spin coating \cite{Kittilstved2005}. They reported room temperature ferromagnetism for Co$^{2+}$:ZnO, however, only for $n$-type conducting material. $p$-type Co$^{2+}$:ZnO was found to behave paramagnetic. The situation was vice versa for Mn$^{2+}$:ZnO which was ferromagnetic for $p$-type and paramagnetic for $n$-type. This demonstrated a first close link between the electronic structure and polarity-dependent high-$T_{\rm C}$ ferromagnetism in TM-substituted ZnO and was attributed to the different ligand-to-metal charge transfer electronic structures in the two cases.

\small\begin{figure}
    \includegraphics[width=6cm]{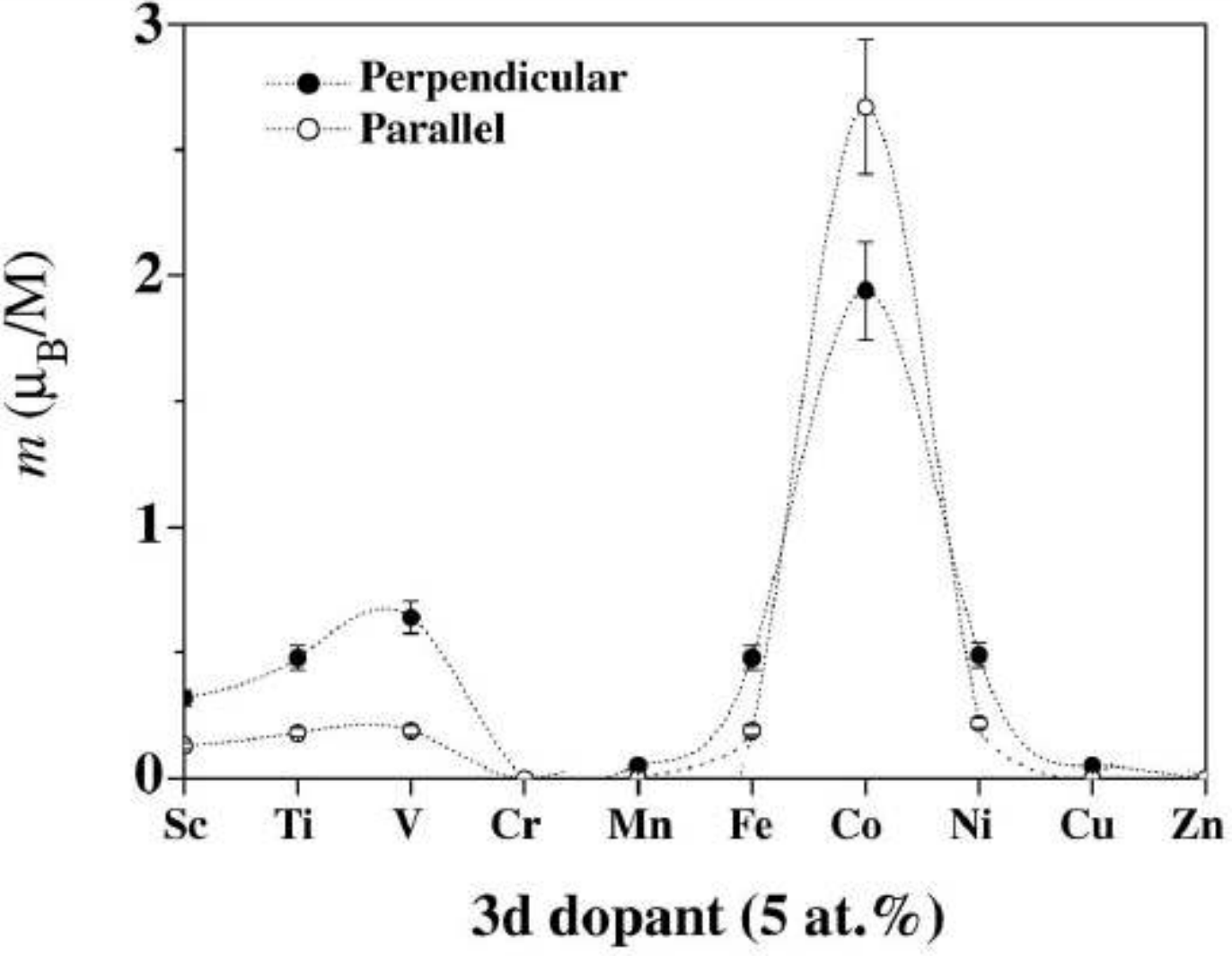}
    \caption{\label{fig:venkatesan}
             Magnetic moment of Zn$_{0.95}M_{0.05}$O films ($M =$ Sc,Ti,\ldots), measured at room temperature
             for the external field applied perpendicular to the film plane (solid symbols) or
             in plane (open symbols). The moment is expressed as $\mu_{\rm B}/M$.
             Reproduced from \cite{Venkatesan2004a}. Copyright 2004 by the American Physical Society.}
\end{figure}\normalsize

In the field of laser deposition, Coey and Venkatesan {\it et al.} \cite{Venkatesan2004a,Coey2005a,Coey2005b} fabricated a whole set of TM:ZnO thin films using the complete $3d$-TM series from Sc$^{3+}$ to Cu$^{2+}$ with a substitution level of 5\%. All films were $n$-type and deposited on $(1\bar{1}02)$-oriented Al$_2$O$_3$ substrates. At 300\,K, they observed maxima in the magnetic saturation moments for Co$^{2+}$:ZnO ($1.9\,\mu_{\rm B}$/Co) and V$^{2+}$:ZnO ($0.5\,\mu_{\rm B}$/V) with significant anisotropy (Fig.~\ref{fig:venkatesan}). Only a very small moment was found for Cr$^{2+}$:ZnO and Mn$^{2+}$:ZnO. The authors proposed that the ferromagnetic exchange was mediated by shallow donor electrons forming bound magnetic polarons which overlap to create a spin-split impurity band \cite{Coey2005a}. Two details of this study are remarkable: (i) Investigating the concentration dependence revealed a giant moment of about $6\,\mu_{\rm B}$/Co in the low concentration limit for the cobalt-substituted compound. (ii) A small moment of $0.3\,\mu_{\rm B}$/Sc was also seen for scandium substitution although Sc$^{3+}$ is a non-magnetic ion. This could indicate an extrinsic origin (substrate, sample holder, structural defects, magnetic precipitates, \textit{etc}.) of the observed magnetic moments, sometimes referred to as ``phantom magnetism''.

Kaspar {\it et al.}~\cite{Kaspar2008a} also explored PLD-grown epitaxial thin films of Co$^{2+}$:ZnO on $(0001)$- and $(1\bar{1}02)$-oriented Al$_2$O$_3$ substrates. The films showed high structural quality, and element-specific X-ray absorption spectroscopy (XANES and EXAFS) confirmed well-ordered Co$^{2+}$ substitution for Zn$^{2+}$ without indicating secondary phases. The authors varied deposition conditions, post-deposition processing, and Al codoping to achieve a wide range of $n$-type conductivities of ${\rm 10^{-5} \ldots 10^4\,\Omega^{-1}cm^{-1}}$. No significant room-temperature ferromagnetism was seen in any of these films. This confirmed earlier observations by Kittilstved {\it et al.} \cite{Kittilstved2006b} that itinerant conduction band electrons alone are not sufficient to induce ferromagnetism in Co$^{2+}$:ZnO.

At present, the situation is puzzling. Although many experimental reports claim room temperature ferromagnetic order \cite{Ueda2001,Venkatesan2004a,Coey2005a,Kittilstved2006a,Tietze2008,Behan2008,Akdogan2008,Liu2009}, more recently an increasing number of publications describe its absence \cite{Opel2008,Kaspar2008a,Kolesnik2004,Lawes2005,Pacuski2006,Yin2006,Chang2007,Sati2007,Ney2010c,Ney2010a}. In particular, several groups made statements on ferromagnetic behaviour only based on the integral magnetic moment determined by SQUID magnetometry \cite{Venkatesan2004a,Tietze2008}, although element-specific techniques failed to establish its presence \cite{Tietze2008,Barla2007}. The situation got even more confused after the discovery of different magnetic regimes depending on the carrier density \cite{Behan2008} and of controlling magnetism via the gate effect \cite{Lee2009}. Furthermore, calculations show a magnetic moment at oxygen-rich surfaces in e.g. ZrO$_2$ or Al$_2$O$_3$ \cite{Gallego2005} which even calls into question the necessity of TM-substitution for ferromagnetism. Finally, unexpected magnetic coupling has also been found in nominally undoped oxides like HfO$_2$ or TiO$_2$ \cite{Venkatesan2004b,Hong2006} which again might point to an extrinsic origin of the observed magnetic moments (``phantom magnetism'').

In this context, phase segregation and the formation of (ferro-)magnetic clusters are important issues concerning intrinsic DMS. Dietl was already pointing out that spinodal decomposition may be a general problem \cite{Dietl2006}. While those clusters have long been regarded as detrimental for spintronic applications, it has been shown that they may be utilized to control high-temperature ferromagnetism in semiconductors and to tailor spintronic functionalities \cite{Kuroda2007a}. TM-substituted ZnO is known to form nanosized (inter)metallic inclusions \cite{Wi2004,Zhou2006,Li2007,Sudakar2007,Wei2009} which may be responsible for the observed RT magnetic response. The same situation was reported for other TM-substituted semiconductors like Mn:Ge \cite{Ahlers2006,Jaeger2006} or Gd:GaN \cite{Ney2008}. Magnetic nanoclusters in TM:ZnO thin films cannot be ruled out, as shown recently by careful x-ray diffraction (XRD) analysis \cite{Opel2008,Venkatesan2007} or depthprofiling x-ray photoelectron spectroscopy (XPS) \cite{Kaspar2008b}. Only in rare cases were such phase-separated clusters directly imaged by cross-sectional transmission electron microscopy (TEM) \cite{Opel2008,Ney2010a,Jedrecy2009}.

\small\begin{figure*}
    \includegraphics[width=15cm]{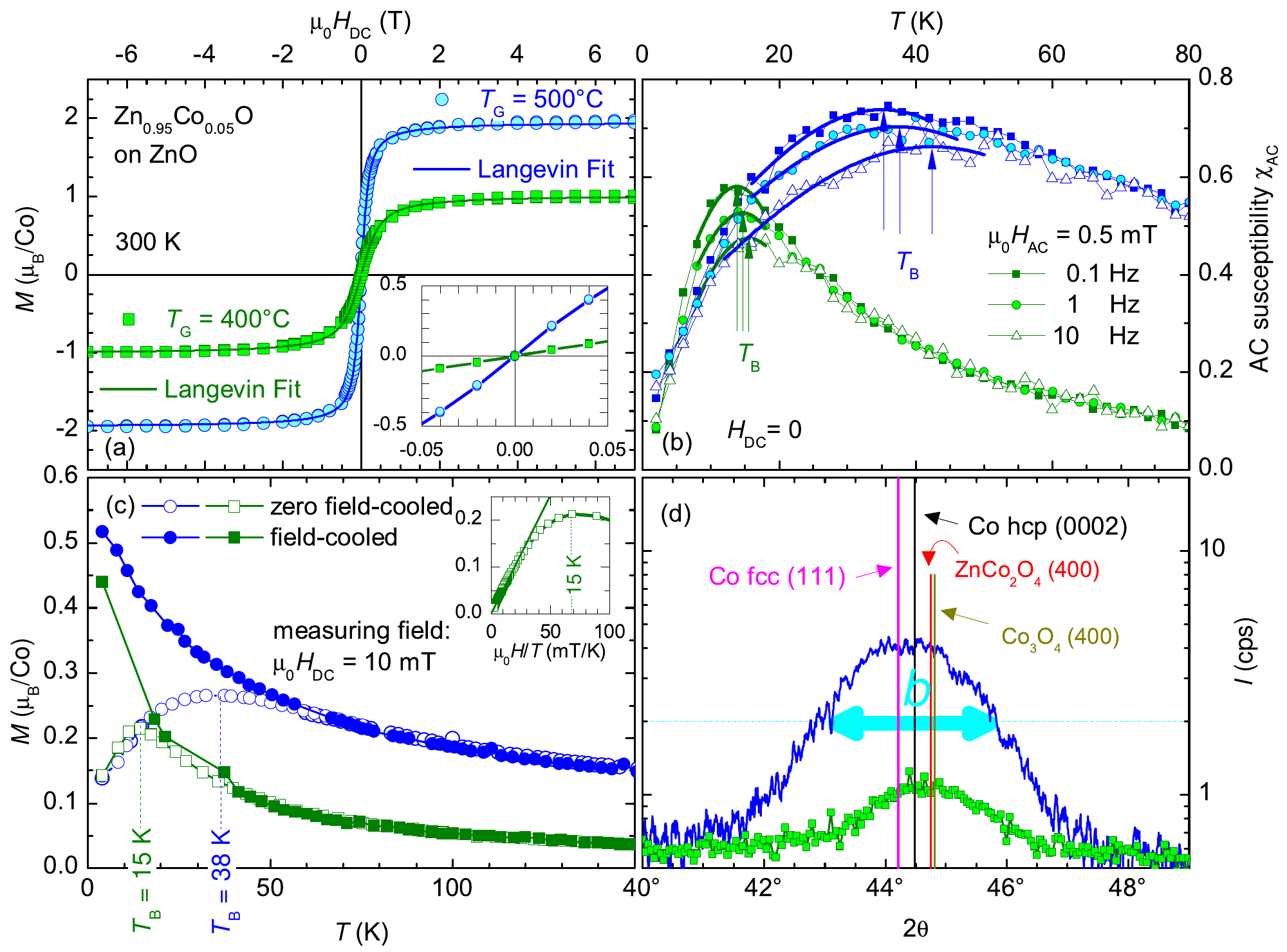}
    \caption{Integral magnetic and x-ray characterization of Zn$_{0.95}$Co$_{0.05}$O thin films, grown by laser-MBE on ZnO substrates at $400^\circ$C (green) and $500^\circ$C (blue).
    (a) Magnetization $M$ versus magnetic field $H$, applied in the film plane at a temperature of $T=300$\,K. The data points are fitted to a Langevin function (lines) after eq.~(\ref{eq:langevin}). The inset shows the absence of any hysteresis.
    (b) AC susceptibility shows maxima around $T_{\rm B}$ (vertical arrows). Their positions move to higher $T$ when increasing the frequency of the AC magnetic field.
    (c) $M(T)$, measured at 10\,mT after cooling in zero field (open symbols) or in 7\,T (closed symbols). The curves indicate blocking behaviour with blocking temperatures of $T_{\rm B}$ = 15\,K and 38\,K.
    (d) XRD diagrammes show a weak reflection in the out-of-plane $\omega$-$2\theta$-scans close to the angles where reflections from secondary Co-rich phases are expected (vertical lines). The cyan arrow denotes the width $b$ of the reflection peak.
    Reproduced from \cite{Opel2008} with kind permission of The European Physical Journal (EPJ).}
    \label{fig:opel}
\end{figure*}\normalsize

Opel {\it et al.}~\cite{Opel2008} performed a careful investigation of the magnetic properties of Co$^{2+}$:ZnO thin films with a Co substitution level of 5\%, grown by laser-MBE on (0001)-oriented ZnO and Al$_2$O$_3$ substrates at different growth temperatures $T_{\rm G}$. For the films grown on ZnO at intermediate temperatures, they found a clear magnetic signal at room temperature of up to $2\,\mu_{\rm B}$/Co (Fig.~\ref{fig:opel}(a)). As a function of the in-plane applied magnetic field, it shows an ``S''-shaped behaviour which cannot be fitted by a Brillouin function for paramagnetic Co$^{2+}$ ions. At first sight, this seemed to indicate ferromagnetism, however, there was no hysteresis detectable in the samples (inset in Fig.~\ref{fig:opel}(a)). On the other hand, they could easily fit the magnetic signal by the Langevin function
\begin{equation}
    M(B) = M_{\rm S} \left(\coth \frac{\mu B}{k_{\rm B} T} - \frac{k_{\rm B} T}{\mu B} \right)
    \label{eq:langevin}
\end{equation}
with the magnetic induction $B$, the Boltzmann constant $k_{\rm B}$, the measuring temperature $T = 300$~K, and the moment $\mu$ of magnetic particles within the thin film. Fitting the data (solid lines in Fig.~\ref{fig:opel}(a)) resulted in $\mu = 2370\,\mu_{\rm B}$ and $5910\,\mu_{\rm B}$ for the films grown at $400^\circ$C and $500^\circ$C, respectively. This called into question the widely accepted interpretation of the room-temperature magnetization of Co$^{2+}$:ZnO thin film samples. With the saturation magnetization of $1.7\,\mu_{\rm B}/{\rm Co}$ for metallic Co and assuming a hexagonal crystallographic structure, the diameter of possible metallic Co clusters in the samples was estimated to about $3 \ldots 4$~nm to yield the moments given above. Comparing the evolution of the magnetization of the samples with temperature after field- (FC) and zero field-cooling (ZFC) displayed maxima for the ZFC curves at low temperatures (Fig.~\ref{fig:opel}(c)) indicating blocking behaviour of superparamagnetic particles with blocking temperatures between $T_{\rm B} = 15 \ldots 38$~K. In the same way, the AC susceptibility displayed pronounced maxima which shifted to higher temperature for increasing AC frequency (Fig.~\ref{fig:opel}(b)). Finally, x-ray diffraction (XRD) revealed a weak signal in the $\omega$-$2\theta$ scans around $44^\circ$ where reflections from different Co-rich compounds are expected (Fig.~\ref{fig:opel}(d)). From the width $b$ of the reflections, using Scherrer's expression \cite{Cullity2001} the authors estimated the size of these inclusions to about 3~nm. These indications for superparamagnetic, nanometer-sized, Co-rich precipitates in the Co$^{2+}$:ZnO thin films were confirmed by microscopic, element-specific techniques. X-ray magnetic circular dichroism (XMCD) \cite{Schutz1987} was performed at the Co $L_{2,3}$ edges both in fluorescence (FY) and in total electron yield (TEY) mode to distinguish between surface (TEY) and bulk magnetic properties (FY). These were the first room-temperature XMCD data for Zn$_{0.95}$Co$_{0.05}$O at the Co $L_{2,3}$ edges in FY mode. Applying the magneto-optical sum rules \cite{Thole1992,Carra1993,Chen1995}, a magnetic moment at room temperature of up to $0.05\,\mu_{\rm B}$/Co was found with a flat, paramagnetic field dependence in TEY mode, whereas in FY the signal went up to $0.45\,\mu_{\rm B}$/Co and displayed the same ``S''-shape as the integral SQUID magnetization data. Moreover, X-ray absorption (XAS) provided clear evidence for the existence of metallic cobalt in the bulk of the samples rather than Co$^{2+}$. Finally, the metallic cobalt was made visible by means of energy-filtering transmission electron microscopy (EF-TEM). The authors illustrated Co-rich inclusions with typical diameters of 5~nm in agreement with the previous results from the Langevin fits and from the XRD analysis. These combined data provided clear evidence that the observed room temperature magnetism is not related to a bulk homogeneous DMS, but must rather be explained by the presence of superparamagnetic metallic cobalt precipitates. Similar TEM results were reported by Jedrecy \textit{et al.}~\cite{Jedrecy2009}.

\small\begin{figure*}
    \includegraphics[width=12cm]{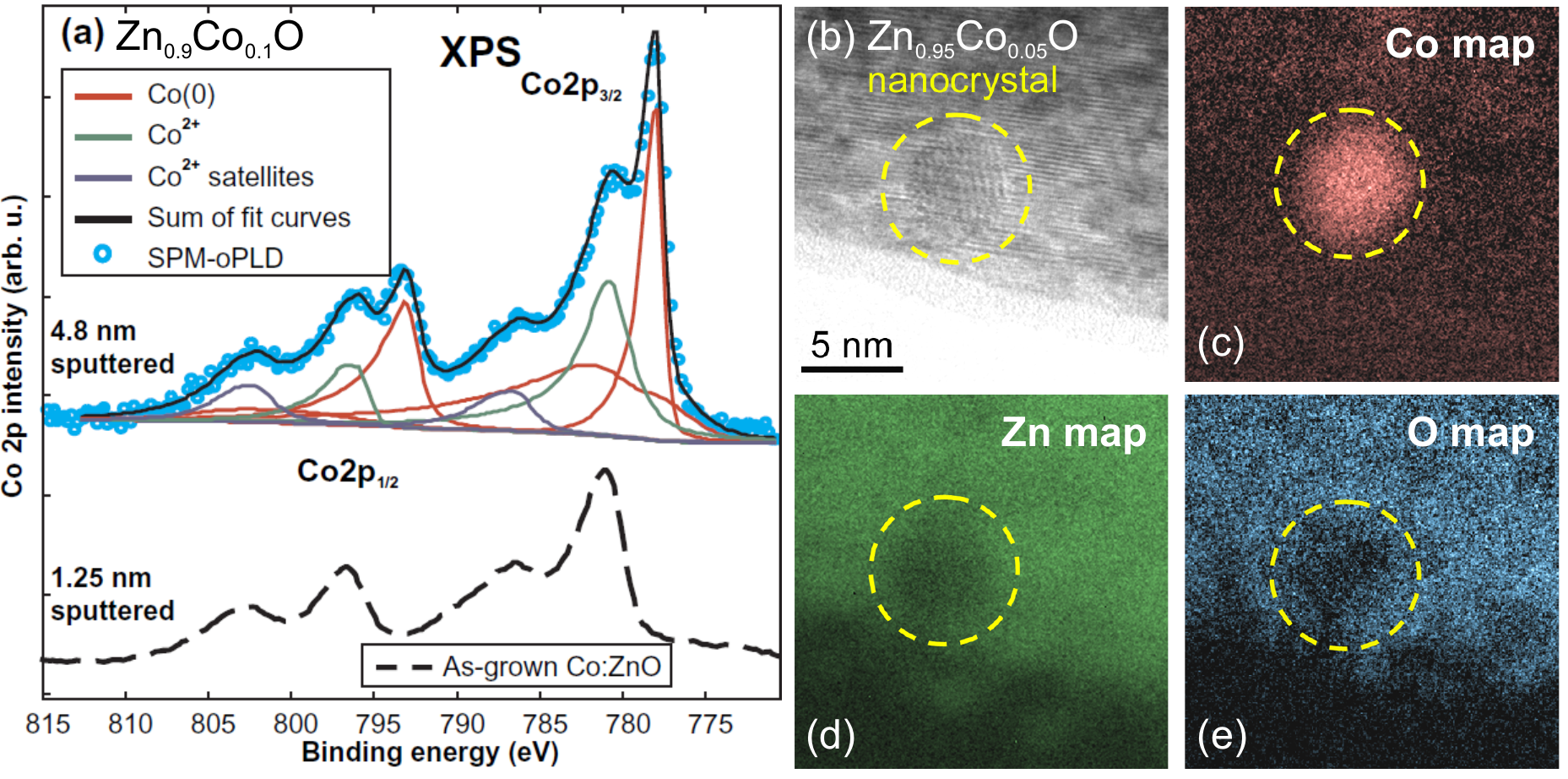}
    \caption{Phase separation in superparamagnetic Co$^{2+}$:ZnO.
             (a) XPS spectrum from Zn$_{0.9}$Co$_{0.1}$O (symbols) after sputter removal of the topmost 4.5~nm of the film. The fits (lines) indicate a superposition of Co$^{2+}$ and Co(0), revealing a fraction of metallic Co. For comparison an XPS spectrum for the as-grown sample after removal of the topmost 1.25~nm is shown below.
             (b) HR-TEM image from a metallic Co nanocrystal in Zn$_{0.95}$Co$_{0.05}$O.
             This region is rich in Co (c), but deficient in Zn (d) and oxygen (e), as demonstrated by EF-TEM.
             Reproduced from \cite{Ney2010a}.}
    \label{fig:mader-ney}
\end{figure*}\normalsize

In a subsequent work, Ney {\it et al.}~\cite{Ney2010a} investigated a comprehensive set of Co$^{2+}$:ZnO epitaxial thin film samples fabricated using three deposition methods in four different laboratories, all of which were subjected to the same SQUID magnetization and synchrotron-based x-ray measurement protocols. They found that advanced x-ray spectroscopic techniques involving both synchrotron and lab-based x-ray sources are invaluable in answering critically important questions. All samples which exhibited a non-paramagnetic magnetization showed indication for secondary magnetic phases from advanced x-ray investigations: (i) X-ray photoelectron spectroscopy (XPS) demonstrated the presence of metallic Co(0) (Fig.~\ref{fig:mader-ney}(a)). (ii) XMCD at the Co $K$ edge indicated the presence of metallic Co in the pre-edge feature. (iii) The X-ray linear dichroism (XLD) exhibited unusual high values. (iv) In X-ray diffraction (XRD), an additional reflection appears (see below in Fig.~\ref{fig:opel-ney}, inset). Again, they were able to prove the existence of nanometer-sized metallic Co inclusions by EF-TEM unambiguously. Figure~\ref{fig:mader-ney}(b) shows a high-resolution TEM micrograph with a region of different contrast indicating a nanocrystal with a diameter of about 4~nm. The Co map (Fig.~\ref{fig:mader-ney}(c)) displays a higher signal in the same region whereas the Zn and O maps (Figs.~\ref{fig:mader-ney}(d,e)) reveal a depletion of these elements. The comprehensive study established the substantial advantage afforded by a combination of integral and microstructural techniques. This experimental approach is of significant potential value to a wide range of researches dealing with dilute systems or other complex materials. The authors have established that, contrary to numerous claims in the literature, phase-pure, crystallographically excellent Co$^{2+}$:ZnO is uniformly paramagnetic, and that ferromagnetism arises only when phase separation or extensive defect formation occur. Interestingly, this effect is related to the choice of the proper substrate. Zn$_{0.95}$Co$_{0.05}$O films are superparamagnetic and contain secondary phases when grown on ZnO(0001) ($f=0$), whereas Zn$_{0.95}$Co$_{0.05}$O films are paramagnetic when deposited on Al$_2$O$_3$(0001) ($f=18.2\%$) from the same PLD target with the same growth parameters (Fig.~\ref{fig:opel-ney}).

\small\begin{figure}
    \includegraphics[width=6cm]{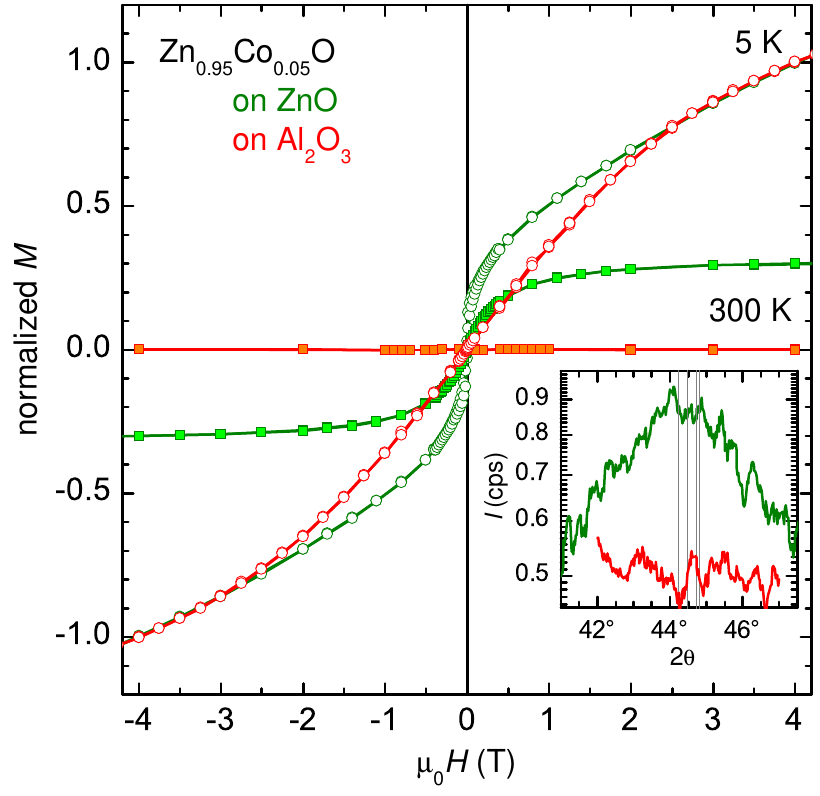}
    \caption{Comparison of the magnetization $M$ between Zn$_{0.95}$Co$_{0.05}$O grown on ZnO (green, superparamagnetic)
             and on Al$_2$O$_3$ (red, paramagnetic) using the same growth parameters.
             For both samples, $M$ was normalized to its respective values at 5~K and 4~T.
             The inset shows XRD diagrammes of the out-of-plane $\omega$-$2\theta$-scans around 44$^\circ$.
             The superparamagnetic sample displays a weak signal (0.9 counts per second) close to the angles
             where reflections from secondary Co-rich phases are expected (vertical grey lines),
             the paramagnetic sample does not.}
    \label{fig:opel-ney}
\end{figure}\normalsize

In summary, very detailed and thorough studies provide increasing evidence that many of the previously published results on ferromagnetic behaviour in Co$^{2+}$:ZnO most likely must be attributed to ferromagnetic or superparamagnetic nanometer-sized precipitates \cite{Opel2008,Kaspar2008a,Ney2010c,Ney2010a,Wi2004,Zhou2006,Li2007,Sudakar2007,Jedrecy2009,Coey2006,Kundaliya2004a,Garcia2005a}. Unfortunately, several groups reported ferromagnetic behaviour in TM:ZnO only based on integral SQUID magnetometry. However, an unambiguous clarification of the origin of magnetism in DMS systems requires a comprehensive study which systematically combines macroscopic and integral experimental techniques with a detailed microscopic and element-specific analysis.

\subsection{p-type ZnO} \label{subsec:p-type}

Although its properties have been extensively studied for more than 50 years \cite{Brown1957}, using ZnO for semiconducting electronics is handicapped because of the difficult control over its electrical conductivity. With regard to implementing ZnO into devices, an important issue is the reproducible fabrication of low-resistivity $p$-type ZnO. Unfortunately, ZnO crystals are natively always $n$-type which has been a case of extensive debate and research so far \cite{Janotti2009,Ozgur2005}. The non-availability of high quality $p$-type ZnO is still hampering the application of ZnO in spintronic and optoelectronic devices \cite{Ogale2005}. Both group-I (Li, Na, K) and group-V elements (N, P, As) are discussed as acceptor atoms, the former replacing Zn, the latter O (Tab.~\ref{tab:p-type}). But, as for other wide band gap semiconductors such as GaN or ZnSe, $p$-doping is difficult \cite{Tsukazaki2005,Kumar2006,Wei2007,Senthil2010a}. On the one hand, dopants may be compensated by low energy native defects such as interstitial zinc (Zn$_{\rm I}$) or oxygen vacancies (V$_{\rm O}$) or background impurities such as hydrogen \cite{Walukiewicz1994}. On the other hand, low solubility of the dopants in the host material is a problem \cite{vandeWalle1993}. Many of these dopants form deep acceptors and do not contribute significantly to $p$-type conduction (Tab.~\ref{tab:p-type}) \cite{Park2002}.

\small\begin{table}
    \caption{\label{tab:p-type}Calculated nearest-neighbour bond lengths $R$ and defect energy levels $\epsilon_{\rm I}$
             for several negatively charged substitutional dopant elements, data taken from \cite{Park2002}. For comparison, the table also lists the
             energies $\Delta E$ required to form positively charged AX centers which convert acceptors into deep donors.
             As $\Delta E <0$ for P and As, they are no suitable $p$-dopants.}
    \begin{indented}
    \item[]\begin{tabular}{@{}c|cccc}
        \br
         &Element & $R$ ({\AA}) & $\epsilon_{\rm I}$ (eV) & $\Delta E$\\
        \mr
        Group I & Li & 2.03 & 0.09 & 0.21\\
         & Na & 2.10 & 0.17 & 1.04\\
         & K & 2.42 & 0.32 & 1.38\\
        \mr
        Group V & N & 1.88 & 0.40 & 0.13\\
         & P & 2.18 & 0.93 & $-0.46$\\
         & As & 2.23 & 1.15 & $-0.18$\\
        \br
    \end{tabular}
    \end{indented}
\end{table}\normalsize

In 2005, Tsukazaki {\it et al.}~\cite{Tsukazaki2005} reported promising progress by using the ``repeated temperature modulation epitaxy'' in conjunction with laser-MBE for $p$-doping of ZnO with nitrogen. They deposited thin films of ZnO on insulating ScAlMgO$_4$ substrates ($f=0.3\%$) \cite{Wessler2002} in an O$_2$ atmosphere of $1.33 \times 10^{-6}$~mbar with simultaneous operation of a nitrogen atom source at 350~W. To achieve $p$-doping, they repeatedly applied the following procedure: (i) At a substrate temperature of $450^\circ$C, a 15~nm thin film of ZnO:N with high nitrogen concentration was deposited. (ii) At $1,050^\circ$C, this film was then annealed and an additional 1~nm thin layer with low N concentration was deposited in order to activate nitrogen as an acceptor and to recover surface smoothness, respectively. This technique overcomes the problem that the nitrogen concentration decreases from a few $10^{22}$\,cm$^{-3}$ at a growth temperature below $500^\circ$C to about $10^{18}$\,cm$^{-3}$ at $700^\circ$C. The authors were able to fabricate a p--i--n junction consisting of ZnO:N/ZnO/ZnO:Ga and demonstrated violet luminescence \cite{Tsukazaki2005}. However, their results could not be confirmed by other groups so far.

Very recently, Kumar \textit{et al.}~\cite{Senthil2010a} reported stable $p$-type conductivity for (Li,Ni)-codoped, laser-MBE grown ZnO thin films on Al$_2$O$_3$ substrates at $400^\circ$C in O$_2$ atmosphere. In a narrow window of O$_2$ pressure ($10^{-3} \ldots 10^{-2}$~mbar), their films exhibited $p$-type conductivity with a maximum hole concentration of $\sim 8.2 \times 10^{17}$~cm$^{-3}$, determined via Hall effect measurements. However, these first results could not be confirmed by subsequent work so far \cite{Senthil2010b}. Note that, in general, the sign of the charge carriers is difficult to deduce from Hall measurements alone if more than one conduction band is involved or if charge transport is via hopping.

In summary, despite the above described success the problem of realizing $p$-type ZnO thin films is still unsolved with particular regard to stability, reproducibility, or efficiency. Neither group-I nor group-V dopants (nor codoping with either two different acceptors or with an acceptor and a donor species) led to a breakthrough toward high, stable, and reproducible $p$-type conductivity for the fabrication of $pn$-junctions \cite{Klingshirn2010}. However, substitutional effects in ZnO were and are exploited for the realization of electronic high-speed Schottky diodes \cite{Wenckstern2009} and metal-semiconductor field effect transistors \cite{Frenzel2010}.

\subsection{ZnO-based oxide heterostructures for spin injection experiments} \label{subsec:ZnO-hetero}

In semiconductor spintronics, spin-polarized currents must be injected, manipulated and detected in the semiconducting material. In this context, the spin dephasing time $T_2^*$ of mobile charge carriers -- and the associated length scale for coherent spin transport -- are fundamental parameters. While other semiconductors like GaAs and related III-V compounds have been studied extensively \cite{Awschalom2002}, only very few reports on spin-coherent properties in ZnO exist \cite{Ghosh2005,Liu2007,Ghosh2008,Bratschitsch2008}. In their pioneering experiments from 2005, Ghosh {\it et al.}~\cite{Ghosh2005} observed electron spin coherence at room temperature in epitaxial ZnO thin films with a spin dephasing time of $T_2^* \simeq 188$~ps being susceptible to electric fields \cite{Ghosh2008}. For bulk single crystals, they found $T_2^* \simeq 20$\,ns at low temperatures. Different authors report similar values for ZnO nanocrystals \cite{Liu2007,Bratschitsch2008}.

In 2011, Schwark \textit{et al.}~\cite{Schwark2011} investigated $T_2^*$ evaluated by time-resolved Faraday rotation (TRFR) experiments in laser-MBE grown ZnO thin film samples on (0001)-oriented Al$_2$O$_3$ substrates. Interestingly, the TRFR signal depended on the excitation wavelength with a direct correlation between the maximum TRFR signal and the prominent $I$-lines in the photoluminescence (PL) spectra \cite{Meyer2004}. $T_2^*$ showed a maximum at a wavelength of 368.9~nm corresponding to an exciton bound to an aluminum donor ($I_6$ line \cite{Meyer2004}). The spin dephasing times were found to $\sim 15$~ns, which is world record for PLD-grown thin films and in the same order of magnitude as reported earlier for bulk crystals. For MBE-grown ZnO thin films, they reported an even higher $T_2^* \simeq 33$\,ns. A second maximum present at 368.3~nm corresponds to the $I_3$ line related to an ionized donor-bound exciton \cite{Meyer2004} yielding $T_2^* \simeq 10$~ns. Donor-bound excitons thus allow the storage of spin information on time scales longer than the lifetime of a free exciton in ZnO. In summary, the experiments consistently showed that the electron spin dephasing time in ZnO is $\gg 1$~ns. However, substantial experimental effort is still required to unambiguously identify the mechanism(s) limiting spin coherence which is mandatory to exploit the unique properties of electron spins in ZnO-based spintronic devices.

Injection of spin-polarized currents into ZnO via ferromagnetic electrodes faces the problem of a large conductivity mismatch to metallic $3d$ ferromagnets preventing an efficient spin injection \cite{Schmidt2000}. This can be circumvented either via the introduction of Schottky or tunnel barriers at the interface \cite{Hanbicki2003} or by the use of non-metallic ferromagnetic materials with low conductivity mismatch. Half-metallic oxide ferromagnets with a spin polarization close to 100\% (see section \ref{sec:magnetic-oxides}) are thus most promising. One candidate is the oxide ferrimagnet Fe$_3$O$_4$ (see section \ref{subsec:magnetite}) with an almost complete spin polarization \cite{Zhang1991,Dedkov2002,Fonin2008}, a high Curie temperature of $T_{\rm C} \simeq 860$~K and the possibility to tune its electrical conductivity $\sigma \approx 200\,\Omega^{-1}{\rm cm}^{-1}$ at room temperature \cite{Reisinger2004a,Venkateshvaran2008} via Zn substitution \cite{Venkateshvaran2009}. However, early Fe$_3$O$_4$/semiconductor heterostructures revealed difficulties in obtaining Fe$_3$O$_4$ thin films with high crystalline quality on both group IV and III-V semiconductors and displayed secondary phases at the interfaces \cite{Reisinger2003b,Kennedy1999,Lu2004,Lu2005,Watts2004,Boothman2007,Paul2009}. Meanwhile, Fe$_3$O$_4$ was successfully deposited in high structural and magnetic quality on GaN \cite{Wong2010} and ZnO \cite{Nielsen2008}.

\small\begin{figure}
    \includegraphics[width=6cm]{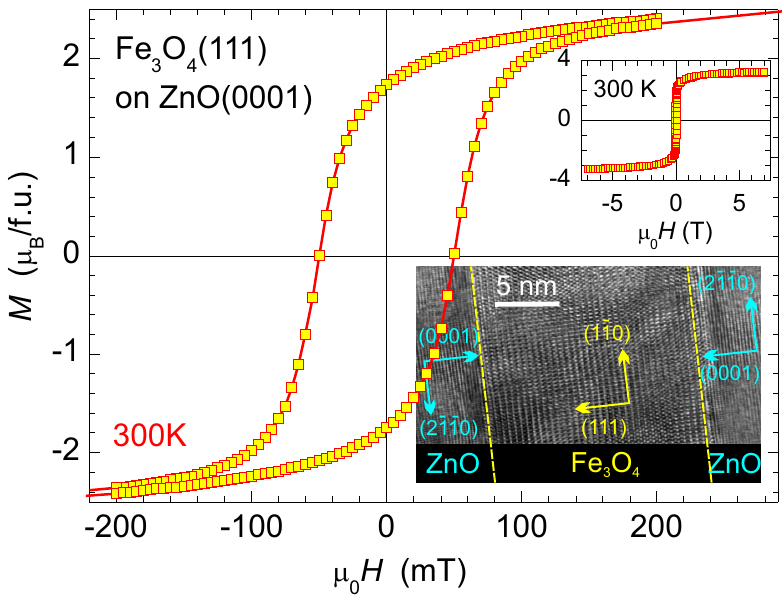}
    \caption{$M(H)$ of a 34\,nm thin, (111)-oriented Fe$_3$O$_4$ film grown on (0001)-oriented ZnO, taken at 300\,K with $H \| [1\overline{1}0]$.
             The upper inset shows the same data on an enlarged field scale,
             the lower illustrates the structural quality by HR-TEM \cite{Nielsen2008}.}
    \label{fig:nielsen}
\end{figure}\normalsize

Nielsen \textit{et al.}~\cite{Nielsen2008} reported that (111)-oriented Fe$_3$O$_4$ can be epitaxially grown on ZnO(0001) substrates ($f=-8.3\%$) using laser-MBE. The authors also demonstrated the epitaxial growth of ZnO thin films on these ferrimagnetic $(111)$-oriented Fe$_3$O$_4$ epilayers representing the first all-oxide ferromagnet/semiconductor epitaxial heterostructures. A detailed characterization of the samples showed that the magnetic and structural properties of the films on ZnO are state of the art, with sharp Fe$_3$O$_4$/ZnO interfaces (Fig.~\ref{fig:nielsen}). These heterostructures are a prerequisite for electrical spin injection from Fe$_3$O$_4$ into ZnO. Preliminary experiments~\cite{Beschoten2011} reported the electrical injection of spin-polarized carriers from metallic Co electrodes, as demonstrated by an optical detection scheme. The observed signals vanished for temperatures above 40~K, indicating a correlation between donor-bound excitons and the electrical spin injection signal.

\section{Multiferroic Oxides and Heterostructures}\label{sec:multiferroics}

\begin{figure}
    \includegraphics[width=6cm]{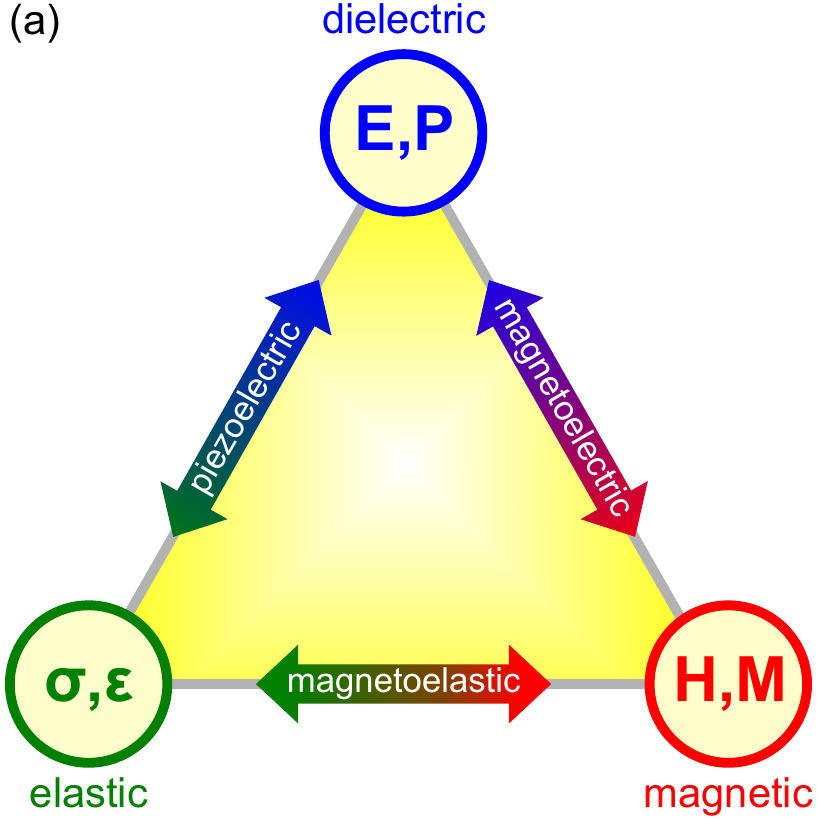}
    \includegraphics[width=8cm]{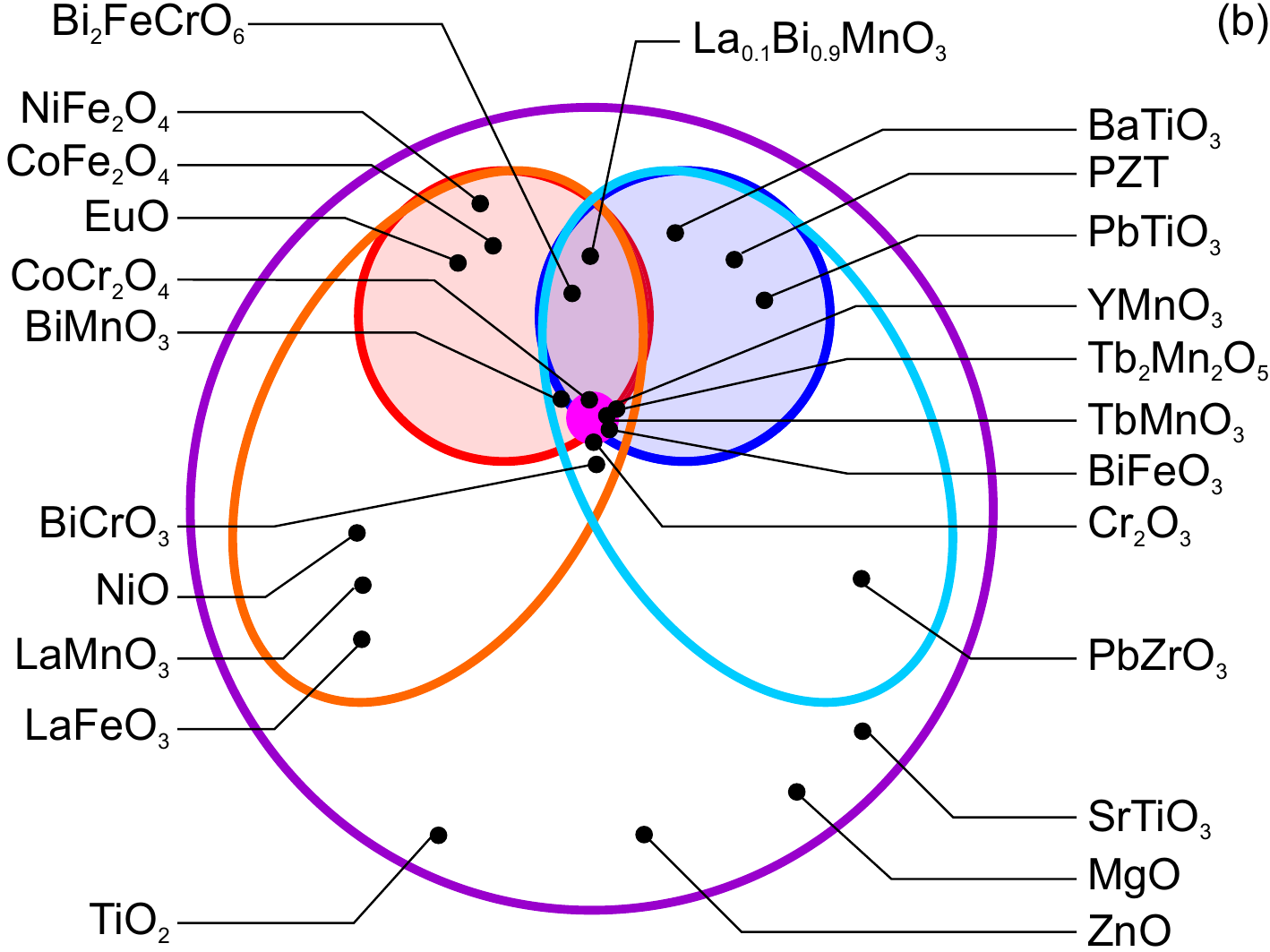}
    \caption{(a) The response of condensed matter to external electric fields ($\vec{E}$),
             magnetic fields ($\vec{H}$), or mechanical stress ($\vec{\sigma}$) are
             polarization ($\vec{P}$), magnetization ($\vec{M}$), or strain
             ($\vec{\epsilon}$), respectively. For useful multiferroic materials exhibiting spontaneous
             $\vec{P}$, $\vec{M}$, or $\vec{\epsilon}$ a strong coupling between the
             dielectric, magnetic, and elastic properties is required~\cite{Spaldin2005}.
             Reproduced with permission from \cite{Opel2011}. Copyright Wiley-VCH Verlag GmbH \& Co.~KGaA.
             (b) Classification of insulating oxides.
             The largest circle represents all insulating oxides among which one finds
             polarizable (cyan ellipse) and magnetizable
             materials (orange ellipse). Within each ellipse, the circle represents
             materials with a spontaneous polarization (ferroelectrics, blue) and/or a spontaneous
             magnetization (ferro- and ferrimagnets, red). Depending on the
             definition, multiferroics correspond to the intersection between the
             ellipses or the circles. The small circle in the centre denotes systems
             exhibiting a magnetoelectric coupling (magenta). Reproduced from \cite{Bea2008b}.}
    \label{fig:MF-Dreieck}
\end{figure}

An extremely active and promising field of research is the integration of different ferroic ordering phenomena such as ferromagnetism, ferroelectricity, or ferroelasticity in one and the same ``multiferroic'' material \cite{Fiebig2005,Schmid1994,Bea2008b,Eerenstein2006,Ramesh2007,Spaldin2010}. The coexistence of ferroelectricity and ferromagnetism in novel multi-functional materials is particularly interesting, since this could allow the realization of new functionalities of electro-magnetic devices, such as the electric field-control of magnetization. Unfortunately, it turned out that there are very few intrinsic magnetoelectric oxide perovskites because the standard microscopic mechanisms driving ferroelectricity or ferromagnetism are incompatible. They usually require either empty or partially filled transition metal orbitals, respectively \cite{Hill2000}. Furthermore, breaking the inversion symmetry is necessary to establish ferroelectricity and, of course, the material must be insulating as otherwise the mobile charge carriers would screen out the electric polarization. However, many ferromagnets tend to be metallic whereas most magnetic insulators turn out to be antiferromagnetic. The need for insulating behaviour can also cause problems if samples show leak currents, as this will suppress ferroelectricity even if the structure is non-centrosymmetric. Unfortunately, magnetic transition metal ions usually accommodate a wider range of valence states than their diamagnetic counterparts resulting in non-stoichiometry and hopping conductivity, preventing a magnetic ferroelectric ground state. Therefore, most of the multiferroics reported so far are antiferromagnets which are not expected to respond noticeably when applying magnetic fields. This has initiated the search for mechanisms favoring the coexistence of ferroelectricity and magnetic order in single-phase materials as, for example, in BiMnO$_3$ \cite{Kimura2003a} or in extrinsic, two-component multiferroic thin film heterostructures \cite{Spaldin2005}. For the realization of useful multiferroic materials, however, the different ferroic order parameters do not only have to coexist, but must be coupled to each other (Fig.~\ref{fig:MF-Dreieck}) to enable, e.g., an electric field-control of the magnetization in magnetoelectric multiferroics \cite{Lawes2011}. Laser-MBE is the growth technique of choice as it again offers many advantages in this field of research. Most important is its intrinsic flexibility which results from the use of non-volatile, solid starting materials which are held at room temperature. The ability to easily change the deposited thin film composition \textit{in situ} without breaking the vacuum is another unique advantage enabling the development of novel materials, including metastable phases and artificial, multiferroic superlattices and heterostructures.

\small\normalsize

\subsection{Single-Phase Multiferroic Oxides}

As pointed out by Ramesh and Spaldin \cite{Ramesh2007}, a single-phase magnetoelectric oxide perovskite can be configured in four general ways: (i)~A large $A$-site cation can provide ferroelectricity via a lone pair of electrons due to stereo-chemical activity while a small $B$-site cation rules ferromagnetism. This mechanism is realized in the Bi-based compounds BiFeO$_3$, BiCrO$_3$, and BiMnO$_3$, see below. (ii)~Ferroelectricity can be driven geometrically as, e.g., in hexagonal YMnO$_3$ \cite{vanAken2004,Fennie2005}. (iii)~Ferroelectricity can be induced by a structural phase transition to a magnetic ground state that lacks inversion symmetry as, e.g., in orthorhombic $R$MnO$_3\,(R={\rm Y}; {\rm Ho},{\rm Er},\ldots{\rm Lu})$ \cite{Kimura2003b}. Below the phase transition temperature, those materials enlarge their unit cell and develop an electric dipole moment induced by a non-linear coupling to non-polar lattice distortions (buckling of the $R$-O planes or shifts of the Mn-O bipyramids). The resulting polarization is small but strong as it is caused directly by the magnetic ordering. (iv)~Non-centrosymmetric charge-ordering arrangements can cause ferroelectricity in magnetic materials as in LuFe$_2$O$_4$ \cite{Ikeda2005,Subramanian2006}. I finally note that most of the work reported on multiferroics has been performed in bulk materials, but increasing effort is made to obtain high-quality thin films \cite{Bibes2007}.

\small\begin{figure}
    \includegraphics[width=6cm]{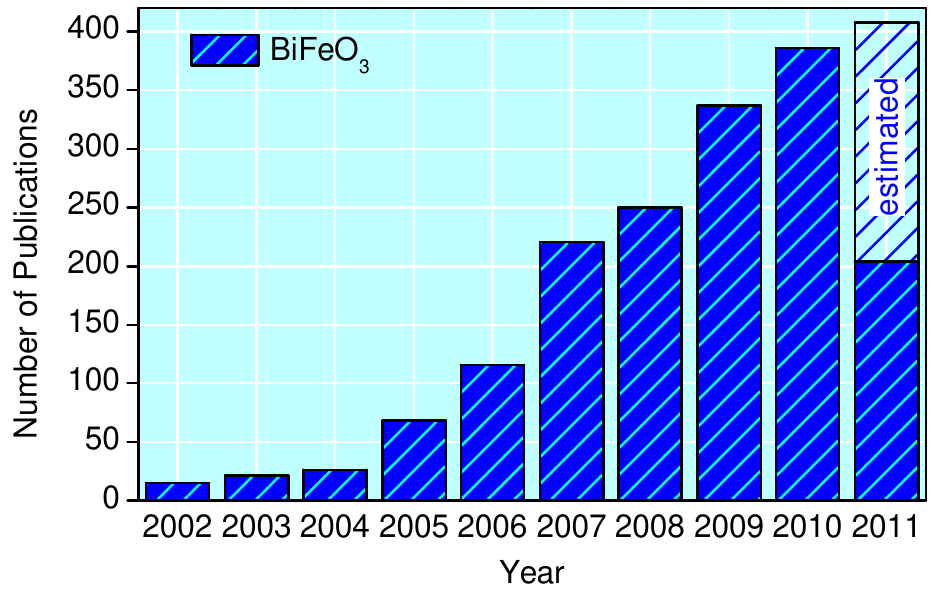}
    \caption{\label{fig:BiFeO3}
             Publications per year with topic ``BiFeO$_3$'' until 30 June 2011 (from Web of Science).}
\end{figure}\normalsize

The perovskite BiFeO$_3$, being one of the few robust materials with ferroelectric and antiferromagnetic order well above room temperature, is in the focus of more than 300 publications per year (Fig.~\ref{fig:BiFeO3}). Bulk BiFeO$_3$ is antiferromagnetic below the N\'{e}el temperature of $T_{\rm N} = 643$\,K and ferroelectric below $T_{\rm C} = 1103$\,K \cite{Kiselev1963,Smolenskii1961}. Strained thin films of BiFeO$_3$ attracted renewed interest after in 2003 Wang \textit{et al.}~\cite{Wang2003} had reported a high ferroelectric polarization of $60\,{\rm \mu C/cm^2}$ together with a high residual magnetic moment of $M_{\rm S} = 1\,{\rm \mu_B}$/f.u. However, there is an ongoing controversial discussion about the origin of this saturation magnetization \cite{Eerenstein2005,Wang2005} as the initially reported values for relaxed BiFeO$_3$ thin films could not be reproduced by subsequent work \cite{Eerenstein2005,Bea2005,Bea2006b,Bea2009a,Geprags2007}. In 2007, Gepr\"{a}gs \textit{et al.}~\cite{Geprags2007} found $M_{\rm S} = 0.02 \,{\rm \mu_B}$/f.u.~in strained BiFeO$_3$ thin films which is in full agreement with other recent publications \cite{Eerenstein2005,Bea2005,Bea2006b,Bea2009a} and older density functional calculations \cite{Ederer2005a}. Meanwhile, it was shown that strain-free bulk crystals of BiFeO$_3$ do not show any parasitic ferromagnetism down to 2\,K at all \cite{Lu2010}. Whether or not the small magnetic moments reported for BiFeO$_3$ thin films originate from nanoscale Fe-rich precipitates \cite{Bea2006b} or Fe$^{3+}$ spin canting still remains unresolved \cite{Catalan2009}.

In 2006, Zhao \textit{et al.}~\cite{Zhao2006} reported on changing the antiferromagnetic domain structure in BiFeO$_3$ thin films by external electric fields. The authors simultaneously applied X-ray photoemission electron microscopy (PEEM) and piezoresponse force microscopy (PFM) to show that electric field-induced ferroelastic switching events resulted in a corresponding rotation of the magnetization plane. B\'{e}a \textit{et al.}~\cite{Bea2006a,Bea2008a,Lebeugle2010} discussed and extensively investigated the use of such thin films for exchange-biasing other ferromagnetic oxide layers with electrical control in spintronic multilayer structures, as did some other groups \cite{Bibes2008,Chu2008}. The ferroelectric polarization in BiFeO$_3$ is oriented along the diagonals of the monoclinically distorted tetragonal unit cell, giving rise to eight different possible polarization states and three types of domain walls, separating regions with polarization orientations differing by $180^\circ$, $109^\circ$, and $71^\circ$ \cite{Bea2009b}. In 2009, Seidel \textit{et al.}~\cite{Seidel2009} found indication for electric conductance of the $180^\circ$ and $109^\circ$ domain walls via local probe measurements, consistent with first-principles density functional calculations \cite{Lubk2009}. From WO$_3$, is it well known that selective doping of ferroelastic twin boundaries with oxygen vacancies can change their conductance, even leading to superconductivity~\cite{Aird1998}. Domain boundary engineering~\cite{Salje2010} in BiFeO$_3$ could open the door for a number of possible applications, such as sensors, memory devices, and switches \cite{Bea2009b}. For further information on BiFeO$_3$, I refer the reader to \cite{Martin2010,Catalan2009}.

\begin{figure}
    \includegraphics[width=6cm]{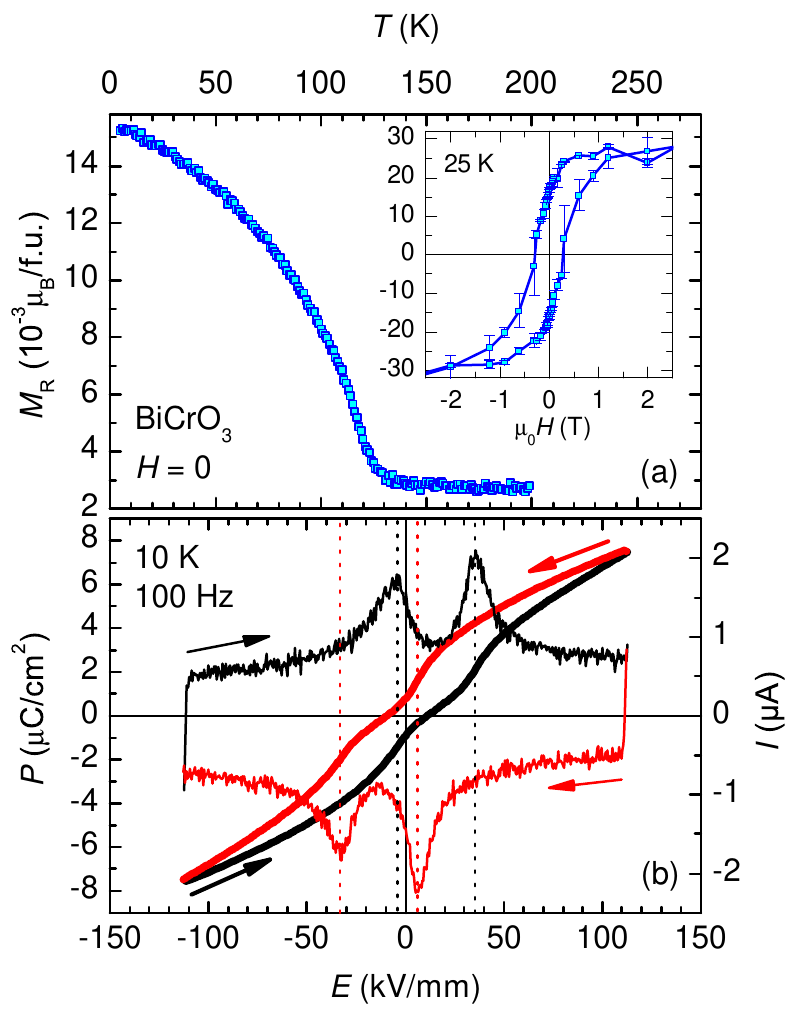}
    \caption{\label{fig:geprags}
             Magnetic and electric properties of BiCrO$_3$ thin films.
             (a) Remnant magnetization, measured after field cooling at 7\,T in zero field.
             The inset shows a $M(H)$ loop recorded at 25\,K.
             (b) Dielectric polarization (thin solid lines, left scale) and dielectric current (thick solid
             lines, right scale). The field-up sweeps are shown in black, the field-down sweeps in red. The
             dashed vertical lines mark the field positions where the polarizations
             of the dielectric sub-lattices change their sign.
             Reproduced with permission from \cite{Opel2011}. Copyright Wiley-VCH Verlag GmbH \& Co.~KGaA.}
\end{figure}

Another interesting candidate material is BiCrO$_3$. Bulk crystals are antiferromagnetic below $T_{\rm N} = 123$\,K~\cite{Sugawara1968} and show a weak ferromagnetic moment, which is attributed to Cr$^{3+}$ spin canting \cite{Geprags2007}. The situation is similar in thin films, where a N\'{e}el temperature between 120\,K \cite{Murakami2006} and 140\,K \cite{Kim2006} is reported. Regarding dielectric properties, the picture is unclear as both ferroelectricity at room temperature \cite{Murakami2006} and antiferroelectricity at 5\,K \cite{Kim2006} have been observed. In tensile strained thin films, Gepr\"{a}gs \textit{et al.}~\cite{Opel2011,Geprags2007} found parasitic ferromagnetism below a critical temperature of $T_{\rm C} = 128$\,K with a weak residual remnant magnetization of $0.015\,\mu_{\rm B}$/f.u.~at low temperatures (Fig.~\ref{fig:geprags}(a)). This observation is consistent with the picture that the Cr$^{3+}$ spins are coupled antiferromagnetically and that a slight canting of the spins results in a weak ferromagnetic signal \cite{Ederer2005a,Ederer2005b}. At 10\,K, the electric polarization $P$ showed an antiferroelectric hysteresis with $P \simeq 8\,{\rm \mu C/cm^2}$ for high electric fields (Fig.~\ref{fig:geprags}(b)) \cite{Opel2011}.

In summary, one has to say that neither BiFeO$_3$ nor BiCrO$_3$ thin films exhibited the expected coexistence of (weak) ferromagnetism and ferroelectricity. (i)~BiFeO$_3$ is antiferromagnetic. The existence of the reported parasitic ferromagnetic phase could not be reproduced and most likely is an artefact. (ii)~BiCrO$_3$ is antiferroelectric instead of showing ferroelectric behaviour. (iii)~There is no clear evidence for intrinsic magnetoelectric coupling for BiFeO$_3$ nor BiCrO$_3$ \cite{Catalan2009}.

\small\normalsize

(La,Bi)MnO$_3$ is one of the very few single-phase ferroelectric and ferromagnetic materials. In bulk, it shows a distorted perovskite structure (monoclinic space group $C2$, $a = 9.5323$\,{\AA}, $b = 5.6064$\,{\AA}, $c = 9.8535$\,{\AA}, $\beta = 110.667^\circ$) \cite{Atou1999,Moreira2002a}. In 2002, Moreira dos Santos \textit{et al.}~\cite{Moreira2002a,Moreira2002b} reported a small polarization of $\sim\,100$\,nC/cm$^2$ with a ferroelectric Curie temperature around 450\,K. The compound displays ferromagnetism below 105\,K \cite{Sugawara1968} as a result of an unusual orbital ordering triggered by the Bi $6s^2$ electron lone pairs \cite{Seshadri2001} leading to ferromagnetic superexchange and a magnetic moment of $3.6\,\mu_{\rm B}$/f.u.~\cite{Chiba1997}. Magneto-capacitance experiments performed by Kimura \textit{et al.}~\cite{Kimura2003a} and theoretical models by Zhong \textit{et al.}~\cite{Zhong2004} suggested the presence of magnetoelectric coupling in this multiferroic compound. They found an anomaly in the dielectric constant at the Curie temperature which is not present when sweeping the temperature in a large magnetic field.

Few double perovskite materials were also expected to simultaneously exhibit ferromagnetism and ferroelectricity \cite{Baettig2005}. Recently, thin films of Bi$_2$FeCrO$_6$ were deposited on SrTiO$_3$ substrates \cite{Nechache2009}. They exhibit a remanent electric polarization of $\sim 55 \mu$C/cm$^2$ and a magnetic moment of $2\,\mu_{\rm B}$/f.u.~at 10\,K. Also for Bi$_2$NiMnO$_6$, deposited on SrTiO$_3$, both ferromagnetic ($T_{\rm C} \sim 100$\,K) and ferroelectric order with a spontaneous polarization of $\sim 5 \mu$C/cm$^2$ was found \cite{Sakai2007}. To stabilize its single-phase composition, the partial replacement of Bi by La was suggested \cite{Langenberg2009}. However, these Bi-based double perovskites are difficult to fabricate due to the high volatility of bismuth and the crucial problem of $B$-site ordering \cite{Opel2011}.

\subsection{Composite Multiferroic Hybrids}

An alternative route to realize multiferroic behaviour at room temperature are composite hybrids, consisting of spatially separated materials with different ferroic properties. Magnetoelectric systems operate by coupling the magnetic and electric degrees of freedom between two materials, generally a ferroelectric and a ferromagnet, via the elastic channel (Fig.~\ref{fig:MF-Dreieck}(a)). An applied electric field creates a piezoelectric strain in the ferroelectric producing a corresponding stress on the ferromagnetic material and a subsequent piezomagnetic change in magnetization or the magnetic anisotropy \cite{Gonnenwein2010}. For more detailed information, I refer the reader to a recent review on composite multiferroic magnetoelectrics \cite{Ma2011}.

Work started several decades ago with bulk composites, but experimental magnetoelectric voltage coefficients were far below those calculated theoretically \cite{Boomgaard1974}. In the 1990s, calculations showed possible strong magnetoelectric coupling in multilayer configurations, an ideal structure to be examined by the growing field of complex oxide thin film growth \cite{Avellaneda1994}. A large number of materials was investigated in the following years, including ferroelectrics such as PbZr$_x$Ti$_{1-x}$O$_3$ (PZT) \cite{Ryu2001a,Ryu2007,Ryu2001b,Srinivasan2001,Srinivasan2002,Srinivasan2004,Brandlmaier2008}, PbMg$_{0.33}$Nb$_{0.67}$O$_3$-PbTiO$_3$ (PMN-PT) \cite{Ryu2002}, BaTiO$_3$ \cite{Lee2000,Dale2003,Murugavel2005,Chopdekar2006,Eerenstein2007,Tian2008,Vaz2009,Czeschka2009,Geprags2010}, and ferromagnets such as NiFe$_2$O$_4$ \cite{Ryu2007,Srinivasan2001}, CoFe$_2$O$_4$ \cite{Srinivasan2004,Chopdekar2006}, Ni$_{0.8}$Zn$_{0.2}$-Fe$_2$O$_4$ \cite{Ryu2001b}, (La,Sr)MnO$_3$ \cite{Lee2000,Srinivasan2002,Dale2003,Eerenstein2007}, La$_{0.7}$Ca$_{0.3}$MnO$_3$ \cite{Srinivasan2002}, Pr$_{0.85}$Ca$_{0.15}$MnO$_3$ \cite{Murugavel2005}, Fe$_3$O$_4$ \cite{Tian2008,Vaz2009}, Sr$_2$CrReO$_6$ \cite{Czeschka2009}, and others \cite{Martin2008}. Some of these experiments showed great promise and magnetoelectric voltage coefficients up to $\Delta E/\Delta H = 4680$\,mV/(cm\,Oe). A larger problem with regard to thin film structures, however, is the clamping effect of the substrate on the ferroelectric phase \cite{Bichurin2003} which can reduce the magnetoelectric voltage coefficient by a factor of up to five.

\small\begin{figure}
    \includegraphics[width=6cm]{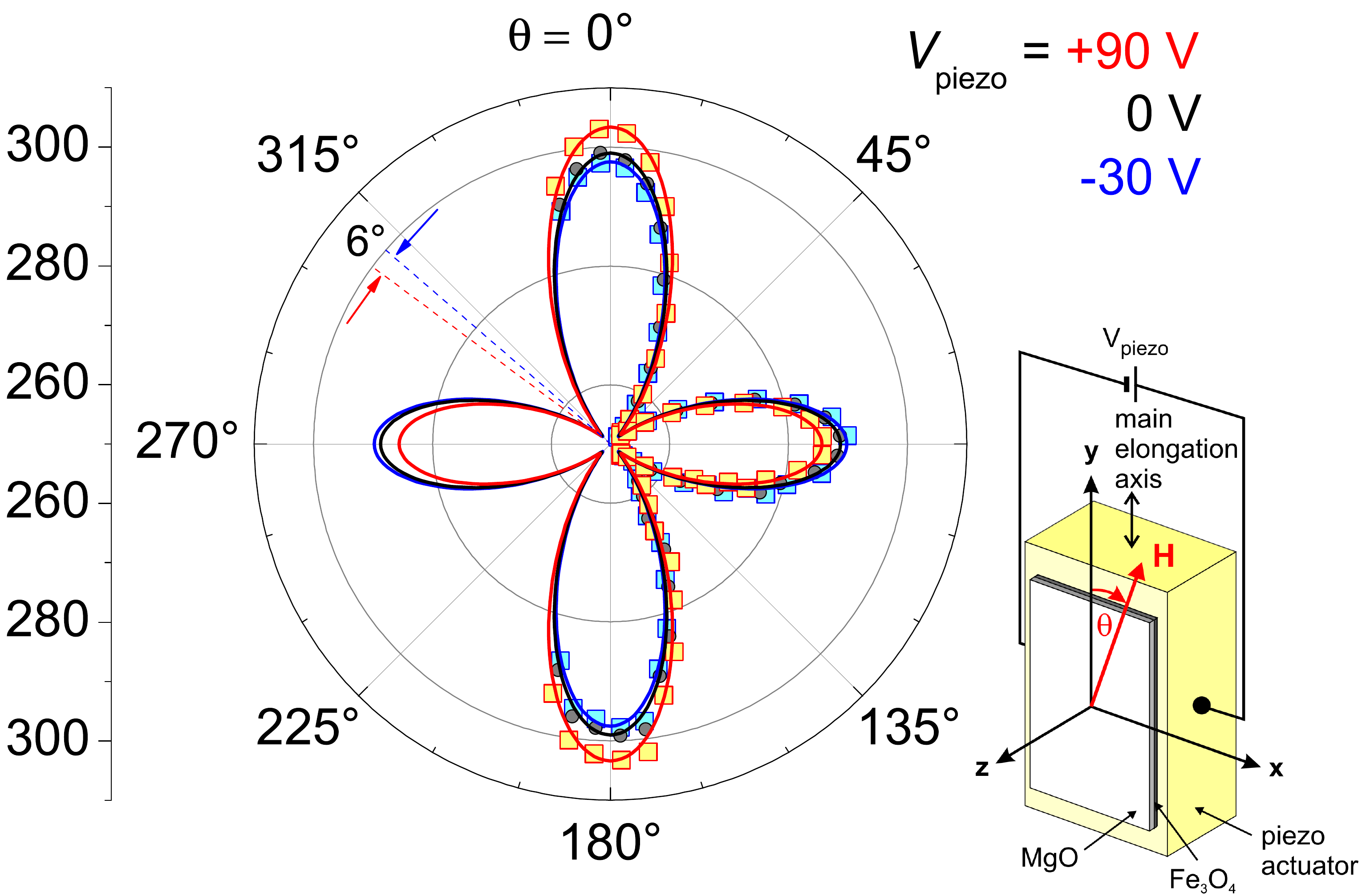}
    \caption{Ferromagnetic resonance (FMR) fields of the Fe$_3$O$_4$/piezo system
             at room temperature as a function of the angle $\theta$ between magnetic field $H$
             and the main elongation axis of the piezo actuator. The symbols represent data points
             obtained at different applied piezo voltages. The lines are simulations based on
             magnetoelastic theory.
             Reproduced with permission from \cite{Opel2011}. Copyright Wiley-VCH Verlag GmbH \& Co.~KGaA.}
    \label{fig:brandlmaier}
\end{figure}\normalsize

In 2008, Brandlmaier \textit{et al.}~\cite{Brandlmaier2008} demonstrated that the magnetic anisotropy and in turn the magnetization direction of Fe$_3$O$_4$ can be varied by an electric field. After depositing epitaxial thin films on (001)-oriented MgO substrates, the latter was polished down to a thickness of about $50\,{\rm \mu m}$. To introduce an \textit{in-situ} tunable strain, a commercial Pb(Zr,Ti)O$_3$ piezoelectric actuator was cemented on top of the Fe$_3$O$_4$ thin film \cite{Shayegan2003,Botters2006}. As demonstrated earlier for ferromagnetic (Ga,Mn)As \cite{Gonnenwein2008}, the in-plane expansion (or contraction) of the piezoelectric actuator as a function of the applied electric voltage $V_{\rm piezo}$ is directly transferred into the ferromagnetic film, yielding a voltage-controllable strain contribution \cite{Brandlmaier2008}. Applying ferromagnetic resonance (FMR) and magnetoelastic theory, the authors found a strain transmission efficiency factor of 70\% from the piezoactuator to the Fe$_3$O$_4$ thin film. The changes of magnetic anisotropy induced by this piezo-strain result in a reversible shift of the free energy minimum and thus the magnetization orientation by about $6^\circ$ at room temperature (Fig.~\ref{fig:brandlmaier}). Even larger angles of $70^\circ$ were obtained for polycrystalline Ni films which do not show a magnetocrystalline anisotropy \cite{Weiler2009}.

\begin{figure*}
    \includegraphics[width=12cm]{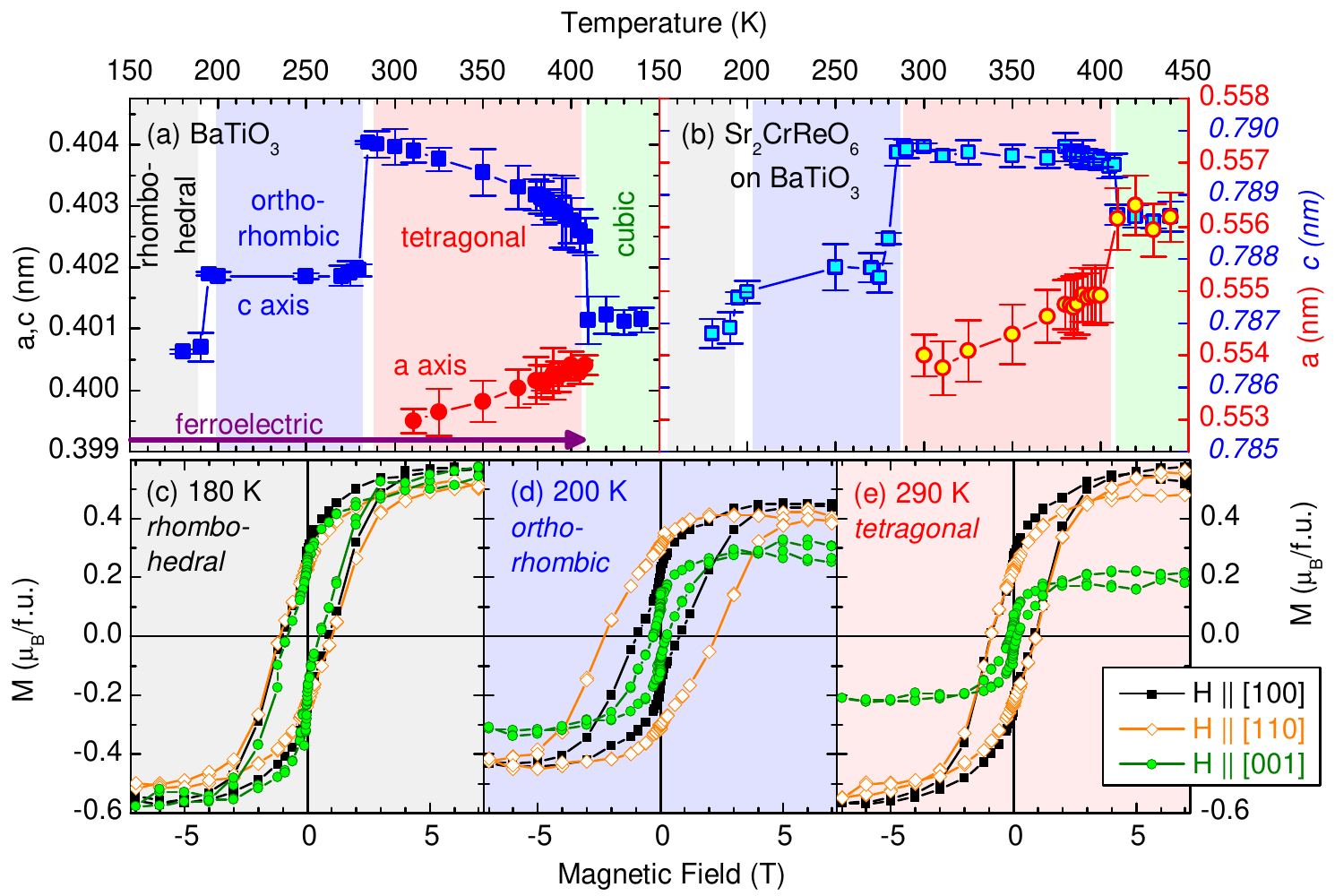}
    \caption{\label{fig:czeschka}
             Structural and magnetic properties of (001)-oriented Sr$_2$CrReO$_6$ on BaTiO$_3$.
             (a,b)~Lattice parameters of the BaTiO$_3$ substrate and the Sr$_2$CrReO$_6$ thin film \cite{Opel2011}.
             (c,d,e)~Ferromagnetic hysteresis loops of Sr$_2$CrReO$_6$ for different crystallographic phases
             of the BaTiO$_3$ substrate \cite{Czeschka2009}.}
\end{figure*}

Higher strains should be possible when replacing the piezoelectric actuator by a ferroelectric substrate. In 2009, Czeschka \textit{et al.}~\cite{Czeschka2009} reported on ferromagnetic Sr$_2$CrReO$_6$, known for a giant anisotropic magnetostriction \cite{Serrate2007}, deposited on BaTiO$_3$(001) substrates, representing the prototype ferroelectric. Its crystallographic structure shows a variety of phase transitions, dependent on the temperature (Fig.~\ref{fig:czeschka}) \cite{Shebanov1981}. Above 393\,K, bulk BaTiO$_3$ is cubic and paraelectric. Below 393\,K, it becomes ferroelectric and its lattice structure changes to tetragonal. Within the ferroelectric state, the lattice symmetry is further reduced to orthorhombic (below 278\,K), and finally to rhombohedral (below 183\,K). The dielectric constant, the spontaneous polarization, as well as the lattice constants change abruptly at these phase transition temperatures, accompanied by a thermal hysteresis~\cite{Kay1949}. The authors found that the lattice parameters of Sr$_2$CrReO$_6$ nicely follow those of the BaTiO$_3$ substrate \cite{Opel2011}. They further reported qualitative changes in the magnetic anisotropy at the BaTiO$_3$ phase transition temperatures \cite{Czeschka2009}. Abrupt changes in the coercive field of up to 1.2\,T along with resistance changes of up to 6.5\% have been observed and were attributed to the high sensitivity of the double perovskites to mechanical deformation \cite{Czeschka2009}.

\small\normalsize

Further experiments have been performed in Ni/BaTiO$_3$ hybrid systems. Gepr\"{a}gs \textit{et al.}~\cite{Geprags2010} demonstrated that the magnetization $M$ can be irreversibly switched as its coercivity was found to depend on the electric field applied to the BaTiO$_3$ substrate. Alternatively, $M$ could be changed reversibly by more than 20\% due to the combined action of electroelastic strain and inverse magnetostriction. Two different remnant magnetization states could be electrically realized. These electroremanent magnetization states were traced back to irreversible domain wall effects in BaTiO$_3$.

\section{Summary and Recent Developments}\label{sec:summary}

Initiated by the progress in thin film technology since the discovery of high-$T_{\rm c}$ superconductors, the field of oxide electronics took off at the end of the 1990s and has been growing exponentially. In this topical review, we have seen that oxides offer versatile physical properties. They are one representative for strongly correlated electronic systems where different degrees of freedom are present and different ordering phenomena occur within the same materials. The competition between structural, charge, spin, and orbital order in transition metal oxides results in new physical ground states such as superconductivity, (anti)ferromagnetism, or ferroelectricity. In addition, the simultaneous development of laser-molecular beam epitaxy allows us to deposit the materials in a controlled layer-by-layer growth mode as thin films or hybrid heterostructures, making them suitable for various device applications.

Half-metallic \emph{ferromagnetic oxides} are superior to conventional $3d$ ferromagnets as they are expected to display a nearly complete spin polarization at the Fermi level. Competing magnetic interactions in the doped manganites, double perovskites, and magnetite stabilize ferromagnetic ground states up to Curie temperatures of 860\,K and display large magnetic anisotropies. These materials were implemented in magnetic tunnel junctions exhibiting a high tunneling magnetoresistance of more than 1,000\% in a few systems. However, in most experiments the maximally achieved TMR values fell far behind the expectations. More effort is needed to obtain a better control of the interface roughnesses or the microstructure of the tunneling barriers. Recently, the use of ferroelectric insulating barriers was suggested to realize four different resistive states depending on the relative orientations of the two magnetizations of the ferromagnetic electrodes and the two possible electric polarizations of the ferroelectric barrier \cite{Velev2009}. Regarding experimental realization, however, it turned out to be problematic that thin barriers are no longer ferroelectric and thick barriers prevent the carriers from tunneling.

\emph{Semiconducting oxides} are in the focus of current research for two reasons. First, zinc oxide was discussed as an ideal starting material for dilute magnetic doping with transition metal ions because Curie temperatures above 300\,K were predicted. In an ongoing controversy, however, it became clear that phase segregation is a major problem. As a result of the formation of secondary magnetic phases at the nanometer scale, such materials may masquerade as room temperature ferromagnets when studied by integral methods although an element-specific investigation fails to support this behaviour. Second, the  control over its electrical conductivity turned out to be difficult. At present, this problem is preventing major (spin)electronic applications of ZnO.

Transition metal oxides are also discussed as possible candidates for \emph{multiferroic oxides} where two or more ferroic order parameters coexist in one single phase. However, it is not yet clear how many oxides belong to the technologically relevant class of ferroelectric ferromagnets as the majority of the discussed materials exhibit an antiferromagnetic ground state. Nevertheless, first applications of single-phase multiferroics are reported, e.g.~as ultrathin barrier material in spin-filter-type tunnel junctions \cite{Gajek2007} or as electrodes in semiconductor devices \cite{Thomas2010}. As an alternative, spatially separated, multiferroic, oxide hybrid structures could be successfully fabricated and offer an electric field-control of the magnetization.

In the wide field of oxide-based thin films, of course, there is much more work to report on than that compiled in this topical review. Having marked the starting point of oxide electronics, \emph{superconducting oxides} are still under investigation with regard to the basic mechanisms leading to Cooper pair formation and phase coherence at temperatures around 100\,K. For thin films, the fabrication of Josephson devices was demonstrated \cite{Rogers1989} as well as the influence of epitaxial strain on their transition temperatures \cite{Locquet1998}. Nevertheless, their technological usage is still restricted to niche applications because of, amongst others, their mechanical properties which are difficult to control. In \emph{ferroelectric} oxides, strains on the percent level were found to have a tremendous effect on the properties of thin films and superlattices \cite{Schlom2007}. They can turn non-ferroelectric materials ferroelectric at any temperature as well as enhance $T_{\rm C}$ by hundreds of degrees and simultaneously enhance their remnant polarization. This is important with regard to technological applications such as ferroelectric memory devices. Recently, \emph{oxide interfaces} moved into the focus of current research after reporting the formation of a two-dimensional electron gas (2DEG) at the interface between the two band insulators LaAlO$_3$ and SrTiO$_3$ \cite{Ohtomo2004}. This high-mobility 2DEG was even found to become superconducting \cite{Reyren2007}. Moreover, controlling superconductivity and the carrier density was demonstrated via the gate effect \cite{Caviglia2008}. This discovery resulted in an intense research into interfaces between polar and non-polar insulating oxides in general.

In summary, artificial oxide heterostructures are particularly promising for the realization of materials with improved or new functionalities (e.g. a magnetoelectric coupling) and novel device concepts (e.g. an electric field-control of the magnetization). Hereby, interface and surface effects play a crucial role. Due to the complexity of the involved oxide materials, the rich variety of physics resulting from band bending effects, magnetic exchange, or elastic coupling at interfaces in heterostructures is far from being understood and needs further detailed studies. Moreover, mixing of the different atomic species deposited in multilayer structures plays an important role and might influence or even dominate the overall physical properties. Therefore, a careful structural and element-specific investigation of the artificial material systems or device structures on an atomic scale is highly desirable for the study of complex functional oxides. The successful story of oxide thin films or heterostructures will continue, whether as new electronic compounds, as sensors, as memory devices, as solar cells or simply for their exciting science \cite{Heber2009}.

\ack
I thank the current and former members of the ``magnetism \& spintronics'' group at the Walther-Mei{\ss}ner-Institut (WMI) for their close cooperation and many stimulating discussions (in alphabetical order): M Althammer, A Brandlmaier, F D Czeschka, S Gepr\"{a}gs, S T B Goennenwein, R Gross, J Lotze, P Majewski, E P Menzel, A Nielsen, K-W Nielsen, J B Philipp, D Reisinger, D Venkateshvaran and M Weiler.
I further thank T Brenninger for continuous technical support, A Erb for preparing the polycrystalline target materials for the PLD process and B S Chandrasekhar for numerous stimulating discussions.
I am grateful to R Gross and the WMI for the support of my work.
I thank the collaborators P Algarabel, B Beschoten, M S Brandt, J C Cezar, L Eng, W Mader, D Mannix, L Morellon, A Ney, M S R Rao, D D Sarma, D Schmei{\ss}er, J M de Teresa, I Vrejoiu and F Wilhelm.
I am grateful to B S Chandrasekhar, R Hackl and A Lerf for carefully reading this manuscript.
Financial support by the Deutsche Forschungsgemeinschaft (DFG) via the priority programmes SPP 1157 (project no.~GR 1132/13) and SPP 1285 (project no.~GR 1132/14) and via the ``Nanosystems Initiative Munich (NIM)'' is gratefully acknowledged.


\section*{References}
\begin{thebibliography}{999}


\bibitem{Holland2006}
    Holland H D 2006 {\it Phil. Trans. R. Soc.} B {\bf 361} 903
\bibitem{Blackman1983}
    Blackman M 1983 {\it Contemp. Phys.} {\bf 24} 319
\bibitem{Bibes2007}
    Bibes M and Barth\'{e}l\'{e}my A 2007 {\it IEEE Transactions on Electron Devices} {\bf 54} 1003
\bibitem{Prinz1998}
    Prinz G A 1998 {\it Science} {\bf 282} 1660
\bibitem{Wolf2001}
    Wolf S A, Awschalom D D, Buhrman R A, Daughton J M, von Molnar S, Roukes M L, Chtchelkanova A Y and Treger D M
    2001 {\it Science} {\bf 294} 1488
\bibitem{Zutic2004}
    \v{Z}uti\'{c} I, Fabian J and Das Sarma S 2004 {\it Rev. Mod. Phys.} {\bf 76} 323
\bibitem{Bader2010} 
    Bader S D and Parkin S S P
    2010 {\it Annu. Rev. Condens. Matter Phys.} {\bf 1} 71
\bibitem{Valasek1921}
    Valasek J 1921 {\it Phys. Rev.} {\bf 17} 475
\bibitem{Raub1988}
    Raub C J 1988 {\it Journal of the Less Common Metals} {\bf 137} 287
\bibitem{Bednorz1986}
    Bednorz J G and M\"{u}ller K A 1986 {\it Z. Phys.} B {\bf 64} 189
\bibitem{Grundmann2010} 
    Grundmann M, Frenzel H, Lajn A, Lorenz M, Schein F and von Wenckstern H
    2010 {\it Phys. Status Solidi} A {\bf 207} 1437
\bibitem{Spaldin2005}
    Spaldin N A and Fiebig M 2005 {\it Science} {\bf 309} 391
\bibitem{Fiebig2005}
    Fiebig M 2005 {\it J. Phys.} D {\bf 38} R123
\bibitem{Martin2010}
    Martin L W, Chu Y-H and Ramesh R 2010 {\it Mater. Sci. Eng.} R {\bf 68} 89
\bibitem{Gross2000}
    Gross R, Klein J, Wiedenhorst B, H\"{o}fener C, Schoop U,
    Philipp J B, Schonecke M, Herbstritt F, Alff L, Lu Y, Marx A,
    Schymon S, Thienhaus S and Mader W 2000 {\it Proc. SPIE} {\bf 4058} 278
\bibitem{Gupta1990}
    Gupta A, Gross R, Olsson E, Segm\"{u}ller A, Koren G and Tsuei C C
    1990 {\it Phys. Rev. Lett.} {\bf 64} 3191
\bibitem{Klein1999}
    Klein J, H\"{o}fener C, Alff L and Gross R 1999 {\it Supercond. Sci. Technol.} {\bf 12} 1023
\bibitem{Klein2000}
    Klein J, H\"{o}fener C, Alff L and Gross R 2000 {\it J. Magn. Magn. Mater.} {\bf 211} 9
\bibitem{Reisinger2003a} 
    Reisinger D, Blass B, Klein J, Philipp J B, Schonecke M, Erb A, Alff L and Gross R 2003
    {\it Appl. Phys.} A {\bf 77} 619
\bibitem{Mannhart2010}
    Mannhart J and Schlom D G 2010 {\it Science} {\bf 327} 1607



\bibitem{Ogale2005}
    Ogale S B 2005 {\it Thin Films and Heterostructures for Oxide Electronics} (New York: Springer)
\bibitem{Sanvito2011} 
    Sanvito S 2011 {\it Chem. Soc. Rev.} {\bf 40} 3336
\bibitem{Bibes2011} 
    Bibes M, Villegas J E and Barth\'{e}l\'{e}my A
    2011 {\it Adv. Phys.} {\bf 60} 5
\bibitem{Helmolt1993}
    von Helmolt R, Wecker J, Holzapfel B, Schultz L and Samwer K
    1993 {\it Phys. Rev. Lett.} {\bf 71} 2331
\bibitem{Jin1994}
    Jin S, Tiefel T H, McCormack M, Fastnacht R A, Ramesh R and Chen L H
    1994 {\it Science} {\bf 264} 413
\bibitem{Johnsson2008}
    Johnsson M and Lemmens P 2008 {\it J. Phys. Condens. Matter} {\bf 20} 264001
\bibitem{Vrejoiu2008}
    Vrejoiu I, Alexe M, Hesse D and G\"{o}sele U 2008 {\it Adv. Funct. Mater.} {\bf 18} 3892
\bibitem{Goldschmidt1927}
    Goldschmidt V M 1926 {\it Skrifter Nordske Videnskaps-Akad. Oslo I, Mat-Naturvidensk Kl.} {\bf 8} 2;
    Goldschmidt V M 1927 {\it Geochemische Verteilungsgesetze der Elemente} (Oslo: Norske Videnskap)
\bibitem{Bhalla2000}
    Bhalla A S, Guo R and Roy R 2000 {\it Mater. Res. Innovat.} {\bf 4} 3
\bibitem{Mader2009}
    Mader W 2009 {\it priv.~communication}
\bibitem{Ney2010b}
    Ney A 2010 {\it Materials} {\bf 3} 3565
\bibitem{PLD1994}
    Chrisey D B and Hubler G K 1994 {\it Pulsed Laser Deposition of Thin Films} (New York: John Wiley \& Sons)
\bibitem{PLD2006}
    Eason R 2006 {\it Pulsed Laser Deposition of Thin Films: Applications-Led Growth of Functional Materials} (Hoboken/NJ: John Wiley \& Sons)
\bibitem{Christen2008}
    Christen H M and Eres G 2008 {\it J. Phys. Condens. Matter} {\bf 20} 264005
\bibitem{Dijkkamp1987}
    Dijkkamp D, Venkatesan T, Wu X D, Shaheen S A, Jisrawi N, Min-Lee Y H, McLean M L and Croft M
    1987 {\it Appl. Phys. Lett.} {\bf 51} 619
\bibitem{Rijnders1997}
    Rijnders G J H M, Koster G, Blank, D H A and Rogalla H
    1997 {\it Appl. Phys. Lett.} {\bf 70} 1888
\bibitem{Chambers2000}
    Chambers S A 2000 {\it Surface Science Reports} {\bf 39} 105
\bibitem{Henderson1999} 
    Henderson M A 1999 {\it Surface Science} {\bf 419} 174
\bibitem{Schneider2010} 
    Schneider C W, Esposito M, Marozau I, Conder K, Doebeli M, Hu Y, Mallepell M, Wokaun A and Lippert T
    2010 {\it Appl. Phys. Lett.} {\bf 97} 192107
\bibitem{Ehrlich1966}
    Ehrlich G and Hudda F G
    1966 {\it J. Chem. Phys.} {\bf 44} 1039
\bibitem{Schwoebel1966}
    Schwoebel D and Sipsey E J
    1966 {\it J. Appl. Phys.} {\bf 37} 3682
\bibitem{Movchan1969}
    Movchan B A and Demchishin A V
    1969 {\it Phys. Met. Metallogr.} {\bf 28} 83
\bibitem{Thornton1986}
    Thornton J 1986 {\it J. Vac. Sci. Technol.} A {\bf 4} 3059
\bibitem{Messier1984}
    Messier R, Giri A P and Roy R A
    1984 {\it J. Vac. Sci. Technol.} A {\bf 2} 500
\bibitem{Kelly1998}
    Kelly P J and Arnell R D
    1998 {\it J. Vac. Sci. Technol.} A {\bf 16} 2858
\bibitem{Petrov2003}
    Petrov I, Barna P B, Hultman L and Greene J E
    2003 {\it J. Vac. Sci. Technol.} A {\bf 21} S117
\bibitem{Mahieu2006}
    Mahieu S, Ghekiere P, Depla D and de Gryse R
    2006 {\it Thin Solid Films} {\bf 515} 1229
\bibitem{Koch2010} 
    Koch R 2010 {\it Surf. Coat. Technol.} {\bf 204} 1973
\bibitem{Reisinger2003b} 
    Reisinger D, Schonecke M, Brenninger T, Opel M, Erb A, Alff L and Gross R
    2003 {\it J. Appl. Phys.} {\bf 94} 1857
\bibitem{Schlom2007}
    Schlom D G, Chen L-Q, Eom C-B, Rabe K M, Streiffer S K and Triscone J-M
    2007 {\it Annu. Rev. Mater. Res.} {\bf 37} 589
\bibitem{Chen1998} 
    Chen Y, Bagnall D M, Koh H-J, Park K-T, Hiraga K, Zhu Z and Yao T
    1998 {\it J. Appl. Phys.} {\bf 84} 3912
\bibitem{Smith1965}
    Smith H M and Turner A F 1965 {\it Appl. Opt.} {\bf 4} 147
\bibitem{Schwarz1969}
    Schwarz H and Tourtellotte H A 1969 {\it J. Vac. Sci. Technol.} {\bf 6} 373
\bibitem{Sankur1983} 
    Sankur H and Cheung J T
    1983 {\it J. Vac. Sci. Technol.} A {\bf 1} 1806
\bibitem{Nakayama1983} 
    Nakayama T 1983 {\it Surface Science} {\bf 133} 101
\bibitem{Tsuei1994}
    Tsuei C C, Kirtley J R, Chi C C, Yu-Jahnes L S, Gupta A, Shaw T, Sun J Z and Ketchen M B
    1994 {\it Phys. Rev. Lett.} {\bf 73} 593
\bibitem{Kirtley1995}
    Kirtley J R, Tsuei C C, Sun J Z, Chi C C, Yu-Jahnes L S, Gupta A, Rupp M and Ketchen M B
    1995 {\it Nature} {\bf 373} 225
\bibitem{Vu1993} 
    Vu L N, Wistrom M S and van Harlingen D J
    1993 {\it Appl. Phys. Lett.} {\bf 63} 1693
\bibitem{Tsuei2000}
    Tsuei C C and Kirtley J R 2000 {\it Rev. Mod. Phys.} {\bf 72} 969
\bibitem{vanSanten1950}
    van Santen H and Jonker G H 1950 {\it Physica} {\bf XVI} 599
\bibitem{Schwarz1986}
    Schwarz K 1986 {\it J. Phys. F: Met. Phys.} {\bf 16} L211
\bibitem{Gupta2008}
    Gupta A 2008 {\it private communication}
\bibitem{Shima2002} 
    Shima M, Tepper T and Ross C A
    2002 {\it J. Appl. Phys.} {\bf 91} 7920
\bibitem{Yanase1984}
    Yanase A and Siratori K 1984 {\it J. Phys. Soc. Japan} {\bf 53} 312
\bibitem{Zhang1991}
    Zhang Z and Satpathy S 1991 {\it Phys. Rev.} B {\bf 44} 13319
\bibitem{Gong1997}
    Gong G Q, Gupta A, Xiao G, Qian W and Dravid V P
    1997 {\it Phys. Rev.} B {\bf 56} 5096
\bibitem{Takaobushi2006}
    Takaobushi J, Tanaka H, Kawai T, Ueda S, Kim J-J, Kobata M, Ikenaga E, Yabashi M, Kobayashi K, Nishino Y, Miwa D, Tamasaku K and Ishikawa T,
    2006 {\it Appl. Phys. Lett.} {\bf 89} 242507
\bibitem{Munekata1989} 
    Munekata H, Ohno H, von Molnar S, Segm\"{u}ller A, Chang L L and Esaki L
    1989 {\it Phys. Rev. Lett.} {\bf 63} 1849
\bibitem{Ohno1996} 
    Ohno H, Shen A, Matsukura F, Oiwa A, Endo A, Katsumoto S and Iye Y
    1996 {\it Appl. Phys. Lett.} {\bf 69} 363
\bibitem{Fukumura1999} 
    Fukumura T, Jin Z, Ohtomo A, Koinuma H and Kawasaki M
    1999 {\it Appl. Phys. Lett.} {\bf 75} 3366
\bibitem{Kobayashi1998}
    Kobayashi K-I, Kimura T, Sawada H, Terakura K and Tokura Y
    1998 {\it Nature} {\bf 395} 677
\bibitem{Manako1999} 
    Manako T, Izumi M, Konishi Y, Kobayashi K I, Kawasaki M and Tokura Y
    1999 {\it Appl. Phys. Lett.} {\bf 74} 2215
\bibitem{Philipp2001} 
    Philipp J B, Reisinger D, Schonecke M, Marx A, Erb A, Alff L and Gross R 2001
    {\it Appl. Phys. Lett.} {\bf 79} 3654
\bibitem{Geprags2009}
    Gepr\"{a}gs S, Czeschka F D, Opel M, Goennenwein S T B, Yu W, Mader W and Gross R
    2009 {\it J. Magn. Magn. Mater.} {\bf 321} 2001



\bibitem{deGroot1983}
    de Groot R A, Mueller F M, van Engen P G and Buschow K H J 1983
    {\it Phys. Rev. Lett.} {\bf 50} 2024
\bibitem{Coey2003}
    Coey J M D and Chien C L 2003 {\it MRS Bull.} {\bf 28} 720
\bibitem{Gross2006}
    Gross R 2006 {\it Nanoscale Devices -- Fundamentals and Applications} eds R Gross, A Sidorenko and L Tagirov (Berlin: Springer) p 49
\bibitem{Goodenough1955}
    Goodenough J B 1955 {\it Phys. Rev.} {\bf 100} 564
\bibitem{Kanamori1959}
    Kanamori J 1959 {\it J. Phys. Chem. Solids} {\bf 10} 87
\bibitem{Anderson1950}
    Anderson P W 1950 {\it Phys. Rev.} {\bf 79} 350
\bibitem{Goodenough1963}
    Goodenough J B 1963 {\it Magnetism and chemical bond} (New York: Interscience Publishers)
\bibitem{Zener1951}
    Zener C 1951 {\it Phys. Rev.} {\bf 82} 403
\bibitem{Anderson1955}
    Anderson P W and Hasegawa H 1955 {\it Phys. Rev.} {\bf 100} 675
\bibitem{deGennes1960}
    de Gennes P G 1960 {\it Phys. Rev.} {\bf 118} 141
\bibitem{Fuchs2008} 
    Fuchs D, Arac E, Pinta C, Schuppler S, Schneider R and von L\"{o}hneysen H
    2008 {\it Phys. Rev.} B {\bf 77} 014434
\bibitem{Lyons1988} 
    K. B. Lyons, P. A. Fleury, L. F. Schneemeyer, and J. V.Waszczak
    1988 {\it Phys. Rev. Lett.} {\bf 60} 732


\bibitem{Volger1954}
    Volger J 1954 {\it Physica} {\bf 20} 49
\bibitem{Searle1969}
    Searle C W and Wang S T 1969 {\it Can. J. Phys.} {\bf 47} 2023
\bibitem{Kubo1972}
    Kubo K and Ohata N 1972 {\it J. Phys. Soc. Jpn.} {\bf 33} 21
\bibitem{Kusters1989}
    Kusters R M, Singleton J, Keen D A, McGreevy R and Hayes W
    1989 {\it Physica} B {\bf 155} 362
\bibitem{Chainami1993}
    Chainami A, Mathew M and Sarma D D
    1993 {\it Phys. Rev.} B {\bf 47} 15397
\bibitem{Baibich1988} 
    Baibich M N, Broto J M, Fert A, van Dau F N, Petroff F, Eitenne P, Creuzet G, Friederich A and Chazelas J
    1988 {\it Phys. Rev. Lett.} {\bf 61} 2472
\bibitem{Binasch1989} 
    Binasch G, Gr\"{u}nberg P, Saurenbach F and Zinn W
    1989 {\it Phys. Rev.} B {\bf 39} 4828
\bibitem{Hwang1995} 
    Hwang H Y, Cheong S-W, Radaelli P O, Marezio M and Batlogg B
    1995 {\it Phys. Rev. Lett.} {\bf 75} 914
\bibitem{Coey1999}
    Coey J M D, Viret M and von Molnar S
    1999 {\it Adv. Phys.} {\bf 48} 167
\bibitem{Imada1998} 
    Imada M, Fujimori Y and Tokura Y,
    1998 {\it Rev. Mod. Phys.} {\bf 70} 1039
\bibitem{Tokura1999}
    Tokura Y ed 1999 {\it Colossal Magnetoresistive Oxides} (London: Gordon and Breach Science Publishers)
\bibitem{Salamon2001} 
    Salamon M B and Jaime M 2001 {\it Rev. Mod. Phys.} {\bf 73} 583
\bibitem{Ziese2002} 
    Ziese M 2002 {\it Rep. Prog. Phys.} {\bf 65} 143
\bibitem{Haghiri-Gosnet2003} 
    Haghiri-Gosnet A-M and Renard J-P 2003 {\it J. Phys. D: Appl. Phys.} {\bf 36} R127
\bibitem{Dorr2006} 
    D\"{o}rr K 2006 {\it J. Phys. D: Appl. Phys.} {\bf 39} R125


\bibitem{Longo1961} 
    Longo J and Ward R 1961 {\it J. Am. Chem. Soc.} {\bf 83} 2816
\bibitem{Sleight1962} 
    Sleight A W, Longo J and Ward R
    1962 {\it Inorg. Chem.} {\bf 1} 245
\bibitem{Patterson1963} 
    Patterson F, Moeller C and Ward R
    1963 {\it Inorg. Chem.} {\bf 2} 196
\bibitem{Nakagawa1968} 
    Nakagawa T 1968 {\it J. Phys. Soc. Jpn.} {\bf 24} 806
\bibitem{Sarma2000a}
    Sarma D D, Mahadevan P, Saha-Dasgupta T, Ray S and Kumar A
    2000 {\it Phys. Rev. Lett.} {\bf 85} 2549;
    2001 {\it Curr. Op. Solid State Mater.} {\bf 5} 261
\bibitem{Kanamori2001}
    Kanamori J and Terakura K 2001 {\it J. Phys. Soc. Jpn.} {\bf 70} 1433
\bibitem{Fang2001}
    Fang Z, Terakura K and Kanamori J
    2001 {\it Phys. Rev.} B {\bf 63} 180407
\bibitem{Philipp2003b} 
    Philipp J B, Klein J, Afilal S, Recher C, Walther T, Mader W,
    Schmid M, Suryanarayanan R, Alff L, Gross R, Topwal D and Sarma D D
    2003 {\it Phys. Rev.} B {\bf 68} 144431
\bibitem{Majewski2005a} 
    Majewski P, Gepr\"{a}gs S, Boger A, Opel M, Erb A, Alff L, Gross R, Vaitheeswaran G S, Kanchana V, Delin A, Wilhelm F and Rogalev A
    2005 {\it Phys. Rev.} B {\bf 72} 132402
\bibitem{Majewski2005c} 
    Majewski P, Gepr\"{a}gs S, Sanganas O, Opel M, Gross R, Wilhelm F, Rogalev A and Alff L
    2005 {\it Appl. Phys. Lett.} {\bf 87} 202503
\bibitem{Balcells2001} 
    Balcells L, Navarro J, Bibes M, Roig A, Martinez B and Fontcuberta J
    2001 {\it Appl. Phys. Lett.} {\bf 78} 781
\bibitem{Navarro2001a} 
    Navarro J, Balcells L, Sandiumenge F, Bibes M, Roig A, Martinez B and Fontcuberta J
    2001 {\it J. Phys. Condens. Matter} {\bf 13} 8481
\bibitem{Sarma2000b} 
    Sarma D D, Sampathkumarana E V, Ray S, Nagarajan R, Majumdar S, Kumar A, Nalini G and Guru Row T N
    2000 {\it Solid State Commun.} {\bf 114} 465
\bibitem{Venimadhav2004} 
    Venimadhav A, Sher F, Attfield J P and Blamire M G
    2004 {\it J. Magn. Magn. Mater.} {\bf 269} 101
\bibitem{Sanchez2004} 
    D. S\'{a}nchez D, Garcia-Hern\'{a}ndez M, Auth N and Jakob G
    2004 {\it J. Appl. Phys.} {\bf 96} 2736
\bibitem{Fix2005} 
    Fix T, Versini G, Loison J L, Colis S, Schmerber G, Pourroy G and Dinia A
    2005 {\it J. Appl. Phys.} {\bf 97} 024907
\bibitem{Wang2006} 
    Wang S, Pan H, Zhang X, Lian G and Xion G
    2006 {\it Appl. Phys. Lett.} {\bf 88} 121912
\bibitem{Santiso2002} 
    Santiso J, Figueras A and Fraxedas J
    2002 {\it Surf. Interface Anal.} {\bf 33} 676
\bibitem{Branford2003} 
    Branford W R, Clowes S K, Bugoslavsky Y V, Miyoshi Y, Cohena L F, Berenov A V, MacManus-Driscoll J L, Rager J and Roy S B
    2003 {\it J. Appl. Phys.} {\bf 94} 4714
\bibitem{Navarro2001b} 
    Navarro J, Frontera C, Balcells L, Mart\'{\i}nez B and Fontcuberta J
    2001 {\it Phys. Rev.} B {\bf 64} 092411
\bibitem{Auth2003} 
    Auth N, Jakob G, Block T and Felser C
    2003 {\it Phys. Rev.} B {\bf 68} 024403
\bibitem{Philipp2003a} 
    Philipp J B, Reisinger D, Schonecke M, Opel M, Marx A, Erb A, Alff L and Gross R
    2003 {\it J. Appl. Phys.} {\bf 93} 6853
\bibitem{Geprags2006} 
    Gepr\"{a}gs S, Majewski P, Ritter C, Alff L and Gross R 2006 {\it J. Appl. Phys.} {\bf 99}  08J102
\bibitem{Majewski2005b} 
    Majewski P, Gepr\"{a}gs S, Boger A, Opel M, Alff L and Gross R 2005
    {\it J. Magn. Magn. Mater.} {\bf 290-291} 1154
\bibitem{Serrate2007} 
    Serrate D, de Teresa J M and Ibarra M R
    2007 {\it J. Phys. Condens. Matter} {\bf 19} 023201
\bibitem{Kato2002}
    Kato H, Okuda T, Okimoto Y, Tomioka Y, Takenoya Y, Ohkubo A, Kawasaki M and Tokura Y
    2002 {\it Appl. Phys. Lett.} {\bf 81} 328
\bibitem{Vaitheeswaran2005}
    Vaitheeswaran G, Kanchana V and Delin A
    2005 {\it Appl. Phys. Lett.} {\bf 86} 032513
\bibitem{Orna2010} 
    Orna J, Morellon L, Algarabel P A, Pardo  J A, Magen C, Varela M, Pennycook S J, de Teresa J M and Ibarra M R
    2010 {\it J. Magn. Magn. Mater.} {\bf 322} 1217
\bibitem{Krockenberger2007} 
    Krockenberger Y, Mogare K, Reehuis M, Tovar M, Jansen M, Vaitheeswaran G, Kanchana V
    Bultmark F, Delin A, Wilhelm F, Rogalev A, Winkler A and Alff L
    2007 {\it Phys. Rev.} B {\bf 75} 020404
\bibitem{Philipp2004}
    Philipp J B, Majewski P, Reisinger D, Gepr\"{a}gs S, Opel M, Erb A, Alff L and Gross R
    2004 {\it Acta Phys. Pol.} A {\bf 105} 7




\bibitem{Gorter1955}
    Gorter E W 1955 {\it Proc. IRE} {\bf 43} 1945
\bibitem{Verwey1939}
    Verwey E J W 1939 {\it Nature} {\bf 144} 327
\bibitem{Mott1967}
    Mott N F 1967 {\it Adv. Phys.} {\bf 16} 49
\bibitem{Walz2002} 
    Walz F 2002 {\it J. Phys.: Condens. Matter} {\bf 14} R285
\bibitem{Subias2004} 
    Sub\'{\i}as G, Garc\'{\i}a J, Blasco J, Grazia Proietti M, Hubert Renevier H and Concepci\'{o}n S\'{a}nchez M
    2004 {\it Phys. Rev. Lett.} {\bf 93} 156408
\bibitem{Leonov2004} 
    Leonov I, Yaresko A N, Antonov V N, Korotin M A and Anisimov V I
    2004 {\it Phys. Rev. Lett.} {\bf 93} 146404
\bibitem{McQueeney2007} 
    McQueeney R J, Yethiraj M, Chang S, Montfrooij W, Perring T G, Honig J M and Metcalf P
    2007 {\it Phys. Rev. Lett.} {\bf 99} 246401
\bibitem{Piekarz2007} 
    Piekarz P, Parlinski K and Ole\'{s} A M
    2007 {\it Phys. Rev.} B {\bf 76} 165124
\bibitem{Lorenzo2008} 
    Lorenzo J E, Mazzoli C, Jaouen N, Detlefs C, Mannix D, Grenier S, Joly Y and Marin C
    2008 {\it Phys. Rev. Lett.} {\bf 101} 226401
\bibitem{Garcia2009} 
    Garc\'{\i}a J, Sub\'{\i}as G, Herrero-Mart\'{\i}n J, Blasco J, Cuartero V, Concepci\'{o}n S\'{a}nchez M, Mazzoli C and Yakhou F
    2009 {\it Phys. Rev. Lett.} {\bf 102} 176405
\bibitem{Blasco2011} 
    Blasco J, Garc\'{\i}a J and Sub\'{\i}as G
    2011 {\it Phys. Rev.} B {\bf 83} 104105
\bibitem{Shepherd1985} 
    Shepherd J P, Arag\'{o}n R, Koenitzer J W  and Honig J M
    1985 {\it Phys. Rev.} B {\bf 32}, 1818
\bibitem{Alexe2009} 
    Alexe M, Ziese M, Hesse D, Esquinazi P, Yamauchi K, Fukushima T, Picozzi S and G\"{o}sele U
    2009 {\it Adv. Mater.} {\bf 21} 4452
\bibitem{Venkateshvaran2009}
    Venkateshvaran D, Althammer M, Nielsen A, Gepr\"{a}gs S, Rao M S R, Goennenwein S T B, Opel M and Gross R
    2009 {\it Phys. Rev.} B {\bf 79} 134405
\bibitem{Neel1948}
    N\'{e}el L 1948 {\it Ann. Phys.} {\bf 3} 137
\bibitem{Yafet1952}
    Yafet Y and Kittel C 1952 {\it Phys. Rev.} {\bf 87} 290
\bibitem{Loos2002}
    Loos J and Novak P 2002 {\it Phys. Rev.} B {\bf 66} 132403
\bibitem{Rosencwaig1969}
    Rosencwaig A 1969 {\it Phys. Rev.} {\bf 181} 946
\bibitem{Slater1937}
    Slater J C 1937 {\it Phys. Rev.} {\bf 51} 846
\bibitem{Dedkov2002} 
    Dedkov Y S, R\"{u}diger U and G\"{u}ntherodt G 2002 {\it Phys. Rev.} B {\bf 65} 064417
\bibitem{Fonin2008} 
    Fonin M, Dedkov Y S, Pentcheva R, R\"{u}diger U and G\"{u}ntherodt G
    2008 {\it J. Phys.: Condens. Matter} {\bf 20} 142201
\bibitem{Margulies1996} 
    Margulies D T, Parker F T, Spada F E, Goldman R S, Li J, Sinclair R and Berkowitz A E
    1996 {\it Phys. Rev.} B {\bf 53} 9175
\bibitem{Eerenstein2002} 
    Eerenstein W, Palstra T T M, Hibma T and Celotto S
    2002 {\it Phys. Rev.} B {\bf 66} 201101
\bibitem{Ogale1998} 
    Ogale S B, Ghosh K, Sharma R P, Greene R L, Ramesh R and Venkatesan T
    1998 {\it Phys. Rev.} B {\bf 57}, 7823
\bibitem{Reisinger2004a}
    Reisinger D, Majewski P, Opel M, Alff L and Gross R
    2004 {\it Appl. Phys. Lett.} {\bf 85} 4980
\bibitem{Fernandez-Pacheco2008a} 
    Fern\'{a}ndez-Pacheco A, de Teresa J M, Orna J, Morellon L, Algarabel P A, Pardo J A and Ibarra M R
    2008 {\it Phys. Rev.} B {\bf 77}, 100403
\bibitem{Fernandez-Pacheco2008b} 
    Fern\'{a}ndez-Pacheco A, de Teresa J M, Orna J, Morellon L, Algarabel P A, Pardo J A, Ibarra M R, Magen C and Snoeck E
    2008 {\it Phys. Rev.} B {\bf 78}, 212402
\bibitem{Nagaosa2010} 
    Nagaosa N, Sinova J, Onoda S, MacDonald A H and Ong N P
    2010 {\it Rev. Mod. Phys.} {\bf 82} 1539
\bibitem{Luttinger1958} 
    Luttinger J M 1958 {\it Phys. Rev.} {\bf 112} 739;
    Berger L 1970 {\it Phys. Rev.} B {\bf 2} 4559
\bibitem{Nozieres1973} 
    Nozi\`{e}res P and Lewiner C 1973 {\it J. Phys. (Paris)} {\bf 34} 901;
    Berry M V 1984 {\it Proc. R. Soc. London} Ser.~A {\bf 392} 45
\bibitem{Onoda2006} 
    Onoda S, Sugimoto N and Nagaosa N
    2006 {\it Phys. Rev. Lett.} {\bf 97}, 126602
\bibitem{Onoda2008} 
    Onoda S, Sugimoto N and Nagaosa N
    2008 {\it Phys. Rev.} B {\bf 77}, 165103
\bibitem{Venkateshvaran2008}
    Venkateshvaran D, Kaiser W, Boger A, Althammer M, Rao M S R, Goennenwein S T B, Opel M and Gross R
    2008 {\it Phys. Rev.} B {\bf 78} 092405
\bibitem{Harris1996} 
    Harris V G, Koon N C, Williams C M, Zhang Q, Abe M and Kirkland J P
    1996 {\it Appl. Phys. Lett.} {\bf 68} 2082
\bibitem{Takaobushi2007} 
    Takaobushi J, Ishikawa M, Ueda S, Ikenaga E, Kim J-J, Kobata M, Takeda Y, Saitoh Y, Yabashi M,
    Nishino Y, Miwa D, Tamasaku K, Ishikawa T, Satoh I, Tanaka H, Kobayashi K and Kawai T
    2007 {\it Phys. Rev.} B {\bf 76} 205108



\bibitem{Akerman2005} 
    {\AA}kerman J 2005 {\it Science} {\bf 308} 508
\bibitem{Ney2003} 
    Ney A, Pampuch C, Koch R and Ploog K H
    2003 {\it Nature} {\bf 425} 485
\bibitem{Julliere1975}  
    Julli\`{e}re M 1975 {\it Phys. Lett.} A {\bf 54} 225
\bibitem{Tedrow1970} 
    Tedrow P M, Meservey R and Fulde P
    1970 {\it Phys. Rev. Lett.} {\bf 25} 1270
\bibitem{Tedrow1971} 
    Tedrow P M and Meservey R 1971 {\it Phys. Rev. Lett.} {\bf 26} 192
\bibitem{deTeresa1999a} 
    de Teresa J M, Barth\'{e}l\'{e}my A, Fert A, Contour J-P, Montaigne F and Seneor P
    1999 {\it Science} {\bf 286} 507
\bibitem{deTeresa1999b} 
    de Teresa J M, Barth\'{e}l\'{e}my A, Fert A, Contour J-P, Lyonnet R, Montaigne F, Seneor P and Vaur\`{e}s A
    1999 {\it Phys. Rev. Lett.} {\bf 82} 4288
\bibitem{Moodera1999} 
    Moodera J S, Nassar J and Mathon G
    1999 {\it Annu. Rev. Mater. Sci.} {\bf 29} 381
\bibitem{Butler2001} 
    Butler W H, Zhang X-G, Schulthess T C and MacLaren J M
    2001 {\it Phys. Rev.} B {\bf 63} 054416
\bibitem{Mathon2001} 
    Mathon J and Umerski A 2001 {\it Phys. Rev.} B {\bf 63} 220403
\bibitem{Yuasa2004} 
    Yuasa S, Nagahama T, Fukushima A, Suzuki Y and Ando K J
    2004 {\it Nature Mater.} {\bf 3} 868
\bibitem{Parkin2004} 
    Parkin S S P, Kaiser C, Panchula A, Rice P M, Hughes B, Samant M and Yang S-H
    2004 {\it Nature Mater.} {\bf 3} 862
\bibitem{Ikeda2008} 
    Ikeda S, Hayakawa J, Ashizawa Y, Lee Y M, Miura K, Hasegawa H, Tsunoda M, Matsukura F and Ohno H
    2008 {\it Appl. Phys. Lett.} {\bf 93} 082508
\bibitem{Tsymbal2003} 
    Tsymbal E Y, Mryasiv O N and LeClair P R
    2003 {\it J. Phys. Condens. Matter} {\bf 15} R109
\bibitem{Miao2011} 
    Miao G-X, M\"{u}nzenberg M and Moodera J S
    2011 {\it Rep. Prog. Phys.} {\bf 74} 036501
\bibitem{Lu1996} 
    Lu Y, Li W, Gong G, Xiao G, Gupta A, Lecoeur P, Sun J, Wang Y and Dravid V
    1996 {\it Phys. Rev.} B {\bf 54} 8357
\bibitem{Sun1996} 
    Sun J Z, Gallagher W J, Ducombe P R, Krusin-Elbaum L, Altman R A, Gupta A, Lu Y, Gong G Q and Xiao G
    1996 {\it Appl. Phys. Lett.} {\bf 69} 3266
\bibitem{Bowen2003}
    Bowen M, Bibes M, Bath\'{e}l\'{e}my A, Contour J-P, Anane A, Lema\^{\i}tre Y and Fert A
    2003 {\it Appl. Phys. Lett.} {\bf 82} 233
\bibitem{Bibes2003} 
    Bibes M, Bouzehouane K, Besse M, Barth\'{e}l\'{e}my A, Fusil S,
    Bowen M, Seneor P, Contour J-P and Fert A
    2003 {\it Appl. Phys. Lett.} {\bf 83} 2629
\bibitem{Bibes2010}
    Bibes M 2010 {\it priv.~communication}
\bibitem{Asano2005} 
    Asano H, Koduka N, Imaeda K, Sugiyama M and Matsui M
    2005 {\it IEEE Trans. Magn.} {\bf 41} 2811
\bibitem{Li1998} 
    Li X W, Gupta A, Xiao G, Qian W and Dravid V P
    1998 {\it Appl. Phys. Lett.} {\bf 73} 3282
\bibitem{Seneor1999} 
    Seneor P, Fert A, Maurice J-L, Montaigne F, Petroff F and Vaures A
    1999 {\it Appl. Phys. Lett.} {\bf 74} 4017
\bibitem{vanderZaag2000} 
    van der Zaag P J, Bloemen P J H, Gaines J M, Wolf R M, van der Heijden P A A, van de Veerdonk R J M and de Jonge W J M
    2000 {\it J. Magn. Magn. Mater.} {\bf 211} 301
\bibitem{Matsuda2002} 
    Matsuda H, Takeuchi M, Adachi H, Hiramoto M, Matsukawa N, Odagawa A, Setsune K and Sakakima H
    2002 {\it Jpn. J. Appl. Phys.} {\bf 41} L387
\bibitem{Aoshima2003} 
    Aoshima K and Wang S X 2003 {\it J. Appl. Phys.} {\bf 93} 7954
\bibitem{Yoon2004} 
    Yoon K S, Koo J H, Do Y H, Kim K W, Kim C O and Hong J P
    2004 {\it J. Magn. Magn. Mater.} {\bf 285} 125
\bibitem{Park2003} 
    Park C, Shi Y, Peng Y, Barmak K, Zhu J-G, Laughlin D E and White R M
    2003 {\it IEEE Transactions on Magnetics} {\bf 39} 2806
\bibitem{Bataille2007} 
    Bataille A M, Mattana R, Seneor P, Tagliaferri A, Gota S, Bouzehouane K,
    Deranlot C, Guittet M-J, Moussy J-B, de Nada C, Brookes N B, Petroff F and Gautier-Soyer M
    2007 {\it J. Magn. Magn. Mater.} {\bf 316} e963
\bibitem{Reisinger2004b} 
    Reisinger D, Majewski P, Opel M, Alff L and Gross R
    2004 {\it unpublished}, arXiv:0407725
\bibitem{Opel2011}
    Opel M, Gepr\"{a}gs S, Menzel E P, Nielsen A, Reisinger D, Nielsen K-W, Brandlmaier A, Czeschka F D,
    Althammer M, Weiler M, Goennenwein S T B, Simon J, Svete M, Yu W, H\"{u}hne S-M, Mader W and Gross R
    2011 {\it Phys. Status Solidi} A {\bf 208} 232
\bibitem{Alvarado1992}
    Alvarado S F and Renaud P 1992 {\it Phys. Rev. Lett.} {\bf 68} 1387
\bibitem{Hu2002} 
    Hu G and Suzuki Y 2002 {\it Phys. Reev. Lett.} {\bf 89} 276601
\bibitem{Alldredge2006} 
    Alldredge L M B, Chopdekar R V, Nelson-Cheeseman B B and Suzuki Y
    2006 {\it Appl. Phys. Lett.} {\bf 89} 182504




\bibitem{Janotti2009}
    Janotti A and van de Walle C G 2009 {\it Rep. Prog. Phys.} {\bf 72} 126501
\bibitem{Klingshirn2010} 
    Klingshirn C, Fallert J, Zhou H, Sartor J, Thiele C, Mair-Flaig F, Schneider D and Kalt H
    2010 {\it Phys. Status Solidi} B {\bf 247} 1424
\bibitem{Park1967}
    Park Y S and Reynolds D C 1967 {\it J. Appl. Phys.} {\bf 38} 756
\bibitem{Reynolds1996}
    Reynolds D C, Look D C and Jogai B 1996 {\it Solid State Commun.} {\bf 99} 873
\bibitem{Ozgur2005}
    \"{O}zg\"{u}r \"{U}, Alivov Y I, Liu C, Teke A, Reshchikov M A, Do\v{g}an S, Avrutin V, Cho S-J and Morko\c{c} H
    2005 {\it J. Appl. Phys.} {\bf 98} 041301
\bibitem{Wenckstern2009} 
    von Wenckstern H, Schmidt H, Brandt M, Lajn A, Pickenhain R, Lorenz M, Grundmann M,
    Hofmann D M, Polity A, Meyer B K, Saal H, Binnewies M, B\"{o}rger A, Becker K-D, Tikhomirov V A and Jug K
    2009 {\it Prog. Sol. Stat. Chem.} {\bf 37} 153
\bibitem{Nickel2005}
    Nickel N H and Terukov E (ed) 2005 {\it Zinc Oxide - A Material for Micro- and Optoelectronic Applications} (Netherlands: Springer)
\bibitem{Jagadish2006}
    Jagadish C and Pearton S J (ed) 2006 {\it Zinc Oxide Bulk, Thin Films, and Nanostructures} (New York: Elsevier)
\bibitem{Thomas1960}
    Thomas D G 1960 {\it J. Phys. Chem. Solids} {\bf 15} 86
\bibitem{Mang1995}
    Mang A, Reimann K and R\"{u}benacke S 1995 {\it Solid State Commun.} {\bf 94} 251
\bibitem{Srikant1998} 
    Srikant V and Clarke D R 1998 {\it J. Appl. Phys.} {\bf 83} 5447
\bibitem{Reynolds1999}
    Reynolds D C, Look D C, Jogai B, Litton C W, Cantwell G and Harsch W C
    1999 {\it Phys. Rev.} B {\bf 60} 2340
\bibitem{Tsukazaki2010} 
    Tsukazaki A, Akasaka S, Nakahara K, Ohno Y, Ohno H, Maryenko D, Ohtomo A and Kawasaki M
    2010 {\it Nature Mater.} {\bf 9} 889
\bibitem{Fu2008} 
    Fu J Y and Wu M W 2008 {\it J. Appl. Phys.} {\bf 104} 093712
\bibitem{Kisi1989} 
    Kisi E H and Elcombe M M 1989 {\it Acta Cryst.} C {\bf 45} 1867
\bibitem{Brown1957}
    Brown M E (ed) 1957 {\it ZnO - Rediscovered} (New York: The New Jersey Zinc Company)
\bibitem{Dietl2000} 
    Dietl T, Ohno H, Matsukara F, Cibert J and Ferrand D,
    2000 {\it Science} {\bf 287} 1019
\bibitem{Dietl2001} 
    Dietl T, Ohno H and Matsukara F
    2001 {\it Phys. Rev.} B {\bf 63} 195205
\bibitem{Tsukazaki2005} 
    Tsukazaki A, Ohtomo A, Onuma T, Ohtani M, Makino T, Sumiya M, Ohtani K, Chichibu S F, Fuke S, Segawa Y, Ohno H, Koinuma H and Kawasaki M
    2005 {\it Nature Materials} {\bf 4} 42


\bibitem{Dietl2010} 
    Dietl T 2010 {\it Nature Mater.} {\bf 9}, 965
\bibitem{Matthias1961} 
    Matthias B T, Bozorth R M and van Vleck J H
    1961 {\it Phys. Rev. Lett.} {\bf 7} 160
\bibitem{Sato2007}
    Sato K, Fukushima T and Katayama-Yoshida H
    2007 {\it Jpn. J. Appl. Phys.} {\bf 46} L682, and refs. therein
\bibitem{Opel2008}
    Opel M, Nielsen K-W, Bauer S, Goennenwein S T B, Cezar J C, Schmeisser D, Simon J, Mader W and Gross R
    2008 {\it Eur. Phys. J.} B {\bf 63} 437
\bibitem{Zener1950}
    Zener C 1950 {\it Phys. Rev.} {\bf 81} 440; 1950 {\it Phys. Rev.} {\bf 83} 299.
    A similar model for the nuclear ferromagnetism was developed by Fr\"{o}hlich and Nabarro
    [1940 {\it Proc. R. Soc. London} Ser. A {\bf 175} 382].
\bibitem{Chambers2010} 
    Chambers S A 2010 {\it Adv. Mater.} {\bf 22} 219
\bibitem{Ogale2010} 
    Ogale S B 2010 {\it Adv. Mater.} {\bf 22} 3125
\bibitem{Ueda2001} 
    Ueda K, Tabata H and Kawai T
    2001 {\it Appl. Phys. Lett.} {\bf 79} 988
\bibitem{Venkatesan2004a} 
    Venkatesan M, Fitzgerald C B, Lunney J G and Coey J M D
    2004 {\it Phys. Rev. Lett.} {\bf 93} 177206
\bibitem{Schwartz2004} 
    Schwartz D A and Gamelin D R
    2004 {\it Adv. Mater.} {\bf 16} 2115
\bibitem{Kittilstved2005} 
    Kittilstved K R, Norberg N S and Gamelin D R
    2005 {\it Phys. Rev. Lett.} {\bf 94} 147209
\bibitem{Kittilstved2006a} 
    Kittilstved K R, Liu W K and Gamelin D R
    2006 {\it Nature Mater.} {\bf 5} 291 (2006)
\bibitem{Coey2005a} 
    Coey J M D, Venkatesan M and Fitzgerald C B
    2005 {\it Nature Mater.} {\bf 4} 173
\bibitem{Coey2005b} 
    Coey J M D 2005 {\it J. Appl. Phys.} {\bf 97} 10D313
\bibitem{Kaspar2008a} 
    Kaspar T C, Droubay T, Heald S M, Nachimuthu P, Wang C M, Shutthanandan V, Johnson C A, Gamelin D R and Chambers S A
    2008 {\it New J. Phys.} {\bf 10}, 055010
\bibitem{Kittilstved2006b} 
    Kittilstved K R, Schwartz D A, Tuan A C, Heald S M, Chambers S A and Gamelin D R
    2006 {\it Phys. Rev. Lett.} {\bf 97} 037203
\bibitem{Tietze2008}
    Tietze T, Gacic M, Sch\"{u}tz G, Jakob G, Br\"{u}ck S and Goering E
    2008 {\it New J. Phys.} {\bf 10}, 055009
\bibitem{Behan2008} 
    Behan A J, Mokhtari A, Blythe H J, Score D, Xu X-H, Neal J R, Fox A M and Gehring G A
    2008 {\it Phys. Rev. Lett.} {\bf 100} 047206
\bibitem{Akdogan2008}
    Akdogan N, Nefedov A, Westerholt K, Zabel H, Becker H-W, Somsen C, Khaibullin R, Tagirov L
    2008 {\it J. Phys. D} {\bf 41}, 165001
\bibitem{Liu2009}
    Liu Y and MacManus-Driscoli J L 2009 \textit{Appl. Phys. Lett.} {\bf 94}, 022503
\bibitem{Kolesnik2004} 
    Kolesnik S, Dabrowski B and Mais J
    2004 {\it J. Appl. Phys.} {\bf 95} 2582
\bibitem{Lawes2005} 
    Lawes G, Risbud A S, Ramirez A P and Seshadri R
    2005 {\it Phys. Rev. B} {\bf 71} 045201
\bibitem{Pacuski2006} 
    Pacuski W, Ferrand D, Cibert J, Deparis C, Gaj J A, Kossacki P and Morhain C
    2006 {\it Phys. Rev.} B {\bf 73} 035214
\bibitem{Yin2006} 
    Yin S, Xu M X, Yang L, Liu J F, R\"{o}sner H, Hahn H, Gleiter H, Schild D, Doyle S, Liu T, Hu T D, Takayama-Muromachi E and Jiang J Z
    {\it Phys. Rev.} B {\bf 73} 224408 (2006)
\bibitem{Chang2007} 
    Chang G S, Kurmaev E Z, Boukhvalov D W, Finkelstein L D, Colis S, Pedersen T M, Moewes A and Dinia A
    2007 {\it Phys. Rev.} B {\bf 75} 195215
\bibitem{Sati2007} 
    Sati P, Deparis C, Morhain C, Sch\"{a}fer S and Stepanov A
    2007 {\it Phys. Rev. Lett.} {\bf 98}, 137204
\bibitem{Ney2010c} 
    Ney A, Ney V, Ye S, Ollefs K, Kammermeier T, Kaspar T C, Chambers S A, Wilhelm F and Rogalev A
    2010 {\it Phys. Rev.} B {\bf 82} 041202
\bibitem{Ney2010a}
    Ney A, Opel M, Kaspar T C, Ney V, Ye S, Ollefs K, Kammermeier T, Bauer S, Nielsen K-W, Goennenwein S T B,
    Engelhard M H, Zhou S, Potzger K, Simon J, Mader W, Heald S M, Cezar J C, Wilhelm F, Rogalev A, Gross R and Chambers S A
    2010 {\it New J. Phys.} {\bf 12} 013020
\bibitem{Barla2007}
    Barla A, Schmerber G, Beaurepaire E, Dinia A, Bieber H, Colis S, Scheurer F,
    Kappler J-P, Imperia P, Nolting F, Wilhelm F, Rogalev A, M\"{u}ller D and Grob J J
    2007 {\it Phys. Rev.} B {\bf 76}, 125201
\bibitem{Lee2009} 
    Lee H-J, Helgren E and Hellman F
    2009 {\it Appl. Phys. Lett.} {\bf 94} 212106
\bibitem{Gallego2005} 
    Gallego S, Bertr\'{a}n J I, Cerd\'{a} J and Mu\~{n}oz M C
    2005 {\it J. Phys.: Condens. Matter} {\bf 17} L451
\bibitem{Venkatesan2004b} 
    Venkatesan M, Fitzgerald C B and Coey J M D
    2004 {\it Nature} {\bf 430} 630
\bibitem{Hong2006} 
    Hong N H, Sakai J, Poirot N and Briz\'{e} V
    2006 {\it Phys. Rev. B} {\bf 73} 132404
\bibitem{Dietl2006} 
    Dietl T 2006 {\it Nature Mater.} {\bf 5} 673
\bibitem{Kuroda2007a} 
    Kuroda S, Nichizawa N, Takita K, Mitome M, Bando Y, Osuch K and Dietl T
    2007 {\it Nature Mater.} {\bf 6} 440 (2007)
\bibitem{Wi2004} 
    Wi S C, Kang J-S, Kim J H, Cho S-B, Kim B J, Yoon S, Suh B J, Han S W, Kim K H, Kim K J, Kim B S, Song H J, Shin H J, Shim J H and Min B I
    2004 {\it Appl. Phys. Lett.} {\bf 84} 4233
\bibitem{Zhou2006} 
    Zhou S, Potzger K, Zhang G, Eichhorn F, Skorupa W, Helm M and Fassbender J
    2006 {\it J. Appl. Phys.} {\bf 100} 114304
\bibitem{Li2007} 
    Li X Z, Zhang J and Sellmyer D J
    2007 {\it Solid State Commun.} {\bf 141} 398
\bibitem{Sudakar2007} 
    Sudakar C, Thakur J S, Lawes G, Naik R and Naik V M
    2007 {\it Phys. Rev. B} {\bf 75} 054423
\bibitem{Wei2009} 
    Wei H, Yao T, Pan Z, Mai C, Sun Z, Wu Z, Hu F, Jiang Y and Yan W
    2009 {\it J. Appl. Phys.} {\bf 105} 043903
\bibitem{Ahlers2006}
    Ahlers S, Bougeard D, Sircar N, Abstreiter G, Trampert A, Opel M and Gross R
    2006 {\it Phys. Rev.} B {\bf 74} 214411
\bibitem{Jaeger2006}
    Jaeger C, Bihler C, Vallaitis T, Goennenwein S T B, Opel M, Gross R and Brandt M S
    2006 {\it Phys. Rev.} B {\bf 74} 045330
\bibitem{Ney2008} 
    Ney A, Kammermeier T, Ney V, Ye S, Ollefs K, Manuel E, Dhar S, Ploog K H, Arenholz E, Wilhelm F and Rogalev A
    2008 {\it Phys. Rev. B} {\bf 77} 233308
\bibitem{Venkatesan2007} 
    Venkatesan M, Stamenov P, Dorneles L S, Gunning R D, Bernoux B and Coey J M D
    2007 {\it Appl. Phys. Lett.} {\bf 90} 242508
\bibitem{Kaspar2008b} 
    Kaspar T C, Droubay T, Heald S M, Engelhard M H, Nachimuthu P and Chambers S A
    2008 {\it Phys. Rev.} B {\bf 77} 201303
\bibitem{Jedrecy2009} 
    Jedrecy N, von Bardeleben H J and Demaille D
    2009 {\it Phys. Rev.} B {\bf 80} 205204
\bibitem{Cullity2001}
    Cullity B D and Stock S R
    2001 {\it Elements of X-ray diffraction}, 3rd edition, (New Jersey: Prentice Hall), p. 167
\bibitem{Schutz1987} 
    Sch\"{u}tz G, Wagner W, Wilhelm W, Kienle P, Zeller R, Frahm R and Materlik G
    1987 {\it Phys. Rev. Lett.} {\bf 58} 737
\bibitem{Thole1992} 
    Thole B T, Carra P, Sette F and van der Laan G
    1992 {\it Phys. Rev. Lett.} {\bf 68} 1943
\bibitem{Carra1993} 
    Carra P, Thole B T, Altarelli M and Wang X
    1993 {\it Phys. Rev. Lett.} {\bf 70} 694
\bibitem{Chen1995} 
    Chen C T, Idzerda Y U, Lin H-J, Smith N V, Meigs G, Chaban E, Ho G H, Pellegrin E and Sette F
    1995 {\it Phys. Rev. Lett.} {\bf 75} 152
\bibitem{Coey2006} 
    Coey J M D 2006 {\it Curr. Op. Sol. State Mater. Sci.} {\bf 10} 83
\bibitem{Kundaliya2004a}
    Kundaliya D C, Ogale S B, Lofland S E, Dhar S, Metting C J, Shinde S R, Ma Z, Varughese B, Ramanujachary K V, Salamanca-Riba L and Venkatesan T
    2004 {\it Nature Mater.} {\bf 3} 709
\bibitem{Garcia2005a} 
    Garcia M A, Ruiz-Gonz\'{a}lez M L, Quesada A, Costa-Kr\"{a}mer J L, Fern\'{a}ndez J F, Khatib S J, Wennberg A, Caballero A C, Mart\'{\i}n-Gonz\'{a}lez M S, Villegas M, Briones F, Gonz\'{a}lez-Calbet J M and Hernando A
    2005 {\it Phys. Rev. Lett.} {\bf 94} 217206



\bibitem{Kumar2006} 
    Kumar M, Kim T-H, Kim S-S, Lee B-T
    2006 {\it Appl. Phys. Lett.} {\bf 89} 112103
\bibitem{Wei2007} 
    Wei Z P, Lu Y M, Shen D Z, Zhang Z Z, Yao B, Li B H, Zhang J Y, Zhao D X, Fan X W, Tang Z K
    2007 {\it Appl. Phys. Lett.} {\bf 90} 042113
\bibitem{Senthil2010a} 
    Kumar E S, Venkatesh S, Rao M S R
    2010 {\it Appl. Phys. Lett.} {\bf 96} 232504
\bibitem{Walukiewicz1994} 
    Walukiewicz W 1994 {\it Phys. Rev.} B {\bf 50} 5221
\bibitem{vandeWalle1993} 
    van de Walle C G, Laks D B, Neumark G F, Pantelides S T
    1993 {\it Phys. Rev.} B {\bf 47} 9425
\bibitem{Park2002} 
    Park C H, Zhang S B and Wei S-H
    2002 {\it Phys. Rev.} B {\bf 66} 073202
\bibitem{Wessler2002}
    Wessler B, Steinecker A and Mader W
    2002 {\it J. Cryst. Growth} {\bf 242} 283
\bibitem{Senthil2010b}
    Kumar E S 2010 {\it priv.~communication}
\bibitem{Frenzel2010} 
    Frenzel H, Lajn A, von Wenckstern H, Lorenz M, Schein F, Zhang Z and Grundmann M
    2010 {\it Adv. Mater.} {\bf 22} 5332



\bibitem{Awschalom2002} 
    Awschalom  D D and Samarth N 2002
    {\it Semiconductor Spintronics and Quantum Computation}, eds D D  Awschalom, D  Loss and N Samarth (Berlin: Springer) pp 147-193
\bibitem{Ghosh2005} 
    Ghosh S, Sih V, Lau W H, Awschalom D D, Bae S-Y, Wang S, Vaidya S and Chapline G
    2005 {\it Appl. Phys. Lett.} {\bf 86} 232507
\bibitem{Liu2007} 
    Liu W K, Whitaker K M, Smith A L, Kittilstved K R, Robinson B H and Gamelin D R
    2007 {\it Phys. Rev. Lett.} {\bf 98} 186804
\bibitem{Ghosh2008} 
    Ghosh S, Steuerman D W, Maertz B, Ohtani K, Xu H, Ohno H and Awschalom D D
    2008 {\it Appl. Phys. Lett.} {\bf 92} 162109
\bibitem{Bratschitsch2008} 
    Jan{\ss}en N, Whitaker K M, Gamelin D R and Bratschitsch R
    2008 {\it Nano Lett.} {\bf 8} 1991
\bibitem{Schwark2011}
    Schwark C \textit{et al.}
    2011 {\it to be published}
\bibitem{Meyer2004} 
    Meyer B K, Alves H, Hofmann D M, Kriegseis W, Forster D, Bertram F, Christen J, Hoffmann A, Stra{\ss}burg M, Dworzak M, Haboeck U and Rodina A V
    2004 {\it Phys. Status Solidi} B {\bf 241} 231
\bibitem{Schmidt2000}
    Schmidt G, Ferrand D, Molenkamp L W, Filip A T and van Wees B J
    2000 {\it Phys. Rev.} B {\bf 62} R4790;
    2005 {\it J. Phys. D: Appl. Phys.} {\bf 38} R107
\bibitem{Hanbicki2003}
    Hanbicki A T, van´t Erve O M J, Magno R, Kioseoglou G, Li C H, Jonker B T, Itskos G, Mallory R, Yasar M and Petrou A
    2003 {\it Appl. Phys. Lett.} {\bf 82} 4092
\bibitem{Kennedy1999} 
    Kennedy R J and Stampe P A
    1999 {\it J. Magn. Magn. Mater.} {\bf 195} 284;
    1999 {\it J. Phys. D: Appl. Phys.} {\bf 32} 16
\bibitem{Lu2004} 
    Lu Y X, Claydon J S, Xu Y B, Thompson S M, Wilson K and van der Laan G
    2004 {\it Phys. Rev. B} {\bf 70} 233304;
    2004 {\it J. Appl. Phys.} {\bf 95} 7228
\bibitem{Lu2005} 
    Lu Y X, Claydon J S, Ahmad E, Xu Y, Thompson S M, Wilson K and van der Laan G
    2005 {\it IEEE Trans. Magn.} {\bf 41} 2808;
    2005 {\it J. Appl. Phys.} {\bf 97} 10C313
\bibitem{Watts2004} 
    Watts S M, Nakajima K, van Dijken S and Coey J M D
    2004 {\it J. Appl. Phys.} {\bf 95} 7465;
    2005 {\it Appl. Phys. Lett.} {\bf 86} 212108
\bibitem{Boothman2007} 
    Boothman C, S\'{a}nchez A M and van Dijken S
    2007 {\it J. Appl. Phys.} {\bf 101} 123903
\bibitem{Paul2009} 
    Paul M, M\"{u}ller A, Ruff A, Schmid B, Berner G, Mertin M, Sing M and Claessen R
    2009 {\it Phys. Rev. B} {\bf 79} 233101
\bibitem{Wong2010} 
    Wong P K J, Zhang W, Cui X G, Xu Y B, Wu J, Tao Z K, Li X, Xie Z L, Zhang R and van der Laan G
    2010 {\it Phys. Rev. B} {\bf 81} 035419
\bibitem{Nielsen2008}
    Nielsen A, Brandlmaier A, Althammer M, Kaiser W, Opel M, Simon J, Mader W, Goennenwein S T B and Gross R 2008
    {\it Appl. Phys. Lett.} {\bf 93} 162510
\bibitem{Beschoten2011} 
    Beschoten B 2011 {\it priv.~communication}



\bibitem{Schmid1994} 
    Schmid H 1994 {\it Ferroelectrics} {\bf 162} 317
\bibitem{Bea2008b} 
    B\'{e}a H, Gajek M, Bibes M and Barth\'{e}l\'{e}my A
    2008 {\it J. Phys.: Condens. Matter} {\bf 20} 434221
\bibitem{Eerenstein2006}
    Eerenstein W, Mathur N D and Scott J F
    2006 {\it Nature} {\bf 442} 759
\bibitem{Ramesh2007}
    Ramesh R and Spaldin N A
    2007 {\it Nature Mater.} {\bf 6} 21
\bibitem{Spaldin2010} 
    Spaldin N A, Cheong S-W and Ramesh R
    2010 {\it Physics Today} {\bf 63(10)} 38
\bibitem{Hill2000}
    Hill N A 2000 {\it J. Phys. Chem.} B {\bf 104} 6694
\bibitem{Kimura2003a} 
    Kimura T, Kawamoto S, Yamada I, Azuma M, Takano M and Tokura Y
    2003 {\it Phys. Rev.} B {\bf 67} 180401
\bibitem{Lawes2011} 
   Lawes G and Srinivasan G 2011 {\it J. Phys. D: Appl. Phys.} {\bf 44} 243001



\bibitem{vanAken2004} 
    van Aken B B, Palstra T T M, Filippetti A and Spaldin N A
    2004 {\it Nature Mater.} {\bf 3} 164
\bibitem{Fennie2005} 
    Fennie C J and Rabe K M
    2005 {\it Phys. Rev.} B {\bf 72} 100103
\bibitem{Kimura2003b} 
    Kimura T, Goto T, Shintani H, Ishizaka K, Arima T and Tokura Y
    2003 {\it Nature} {\bf 426} 55
\bibitem{Ikeda2005} 
    Ikeda N, Ohsumi H, Ohwada K, Ishii K, Inami T, Kakurai K,
    Murakami Y, Yoshii K, Mori S, Horibe Y and Kit\^{o} H
    2005 {\it Nature} {\bf 436} 1136
\bibitem{Subramanian2006} 
    Subramanian M A, He T, Chen J, Rogado N S, Calvarese T G and Sleight A W
    2006 {\it Adv. Mater.} {\bf 18} 1737
\bibitem{Kiselev1963}
    Kiselev S V, Ozerov R P and Zhdanov G S
    1963 {\it Sov. Phys. Dokl.} {\bf 7} 742
\bibitem{Smolenskii1961}
    Smolenskii G A, Isupov V A, Agranovskaya A I and Krainik N N
    1961 {\it Sov. Phys. Solid State} {\bf 2} 2651
\bibitem{Wang2003} 
    Wang J, Neaton J B, Zheng H, Nagarajan V, Ogale S B, Liu B, Viehland D, Vaithynathan V
    Schlom D G, Waghmare U V, Spaldin N A, Rabe K M, Wuttig M and Ramesh R
    {\it Science} {\bf 299} 1719
\bibitem{Eerenstein2005} 
    Eerenstein W, Morrison F D, Dho J, Blamire M G, Scott J F and Mathur N D
    2005 {\it Science} {\bf 307} 1203a
\bibitem{Wang2005} 
    Wang J, Scholl A, Zheng H, Ogale S B, Viehland D, Schlom D G, Spaldin N A,
    Rabe K M, Wuttig M, Mohaddes L, Neaton J, Waghmare U, Zhao T and Ramesh R
    2005 {\it Science} {\bf 307} 1203b
\bibitem{Bea2005} 
    B\'{e}a H, Bibes M, Barth\'{e}l\'{e}my A, Bouzehouane K, Jacquet E, Khodan A, Contour J-P,
    Fusil S, Wyczisk F, Forget A, Lebeugle D, Colson D and Viret M
    {\it Appl. Phys. Lett.} {\bf 87} 072508
\bibitem{Bea2006b} 
    B\'{e}a H, Bibes M, Fusil S, Bouzehouane K, Jacquet E, Rode K, Bencok P and Barth\'{e}l\'{e}my A
    2006 {\it Phys. Rev.} B {\bf 74} 020101
\bibitem{Bea2009a} 
    B\'{e}a H, Dup\'{e} B, Fusil S, Mattana R, Jacquet E, Warot-Fonrose B, Wilhelm F, Rogalev A,
    Petit S, Cros V, Anane A, Petroff F, Bouzehouane K, Geneste G, Dkhil B, Lisenkov S,
    Ponomareva I, Bellaiche L, Bibes M and Barth\'{e}l\'{e}my A
    {\it Phys. Rev. Lett.} {\bf 102} 217603
\bibitem{Geprags2007}
    Gepr\"{a}gs S, Opel M, Goennenwein S T B and Gross R 2007 {\it Philos. Mag. Lett.} {\bf 87} 141
\bibitem{Ederer2005a} 
    Ederer C and Spaldin N A 2005 {\it Phys. Rev.} B {\bf 71} 060401
\bibitem{Lu2010} 
    Lu J, Schmidt M, Lunkenheimer P, Pimenov A, Mukhin A A, Travkin V D and Loidl A
    2010 {\it J. Phys.: Conf. Ser.} {\bf 200} 012106
\bibitem{Catalan2009} 
    Catalan G and Scott J F 2009 {\it Adv. Mater.} {\bf 21} 2463
\bibitem{Zhao2006} 
    Zhao T, Scholl A, Zavaliche F, Lee K, Barry M, Doran A, Cruz M P, Chu Y-H,
    Ederer C, Spaldin N A, Das R R, Kim D M, Baek S H, Eom C B and Ramesh R
    2006 {\it Nature Mater.} {\bf 5} 823
\bibitem{Bea2006a} 
    B\'{e}a H, Bibes M, Cherifi S, Nolting F, Warot-Fonrose B, Fusil S,
    Herranz G, Deranlot C, Jacquet E, Bouzehouane K and Barth\'{e}l\'{e}my A
    2006 {\it Appl. Phys. Lett.} {\bf 89} 242114
\bibitem{Bea2008a} 
    B\'{e}a H, Bibes M, Ott F, Dup\'{e} B, Zhu X-H, Petit S, Fusil S
    Deranlot C, Bouzehouane K and Barth\'{e}l\'{e}my A
    2008 {\it Phys. Rev. Lett.} {\bf 100} 017204
\bibitem{Lebeugle2010} 
    Lebeugle D, Mougin A, Viret M, Colson D, Allibe J, B\'{e}a H,
    Jacquet E, Deranlot C, Bibes M and Barth\'{e}l\'{e}my A
    2010 {\it Phys. Rev.} B {\bf 81} 134411
\bibitem{Bibes2008} 
    Bibes M and Barth\'{e}l\'{e}my A 2008 {\it Nature Mater.} {\bf 7} 425
\bibitem{Chu2008} 
    Chu Y-H, Martin L W, Holcomb M B, Gajek M, Han S-J, He Q, Balke N, Yang C-H, Lee D, Hu W,
    Zhan Q, Yang P-L, Fraile-Rodr\'{\i}guez A, Scholl A, Wang S X and Ramesh R
    2008 {\it Nature Mater.} {\bf 7} 478
\bibitem{Bea2009b} 
    B\'{e}a H and Paruch P 2009 {\it Nature Mater.} {\bf 8} 168
\bibitem{Seidel2009} 
    Seidel J, Martin L W, He Q, Zhan Q, Chu Y-H, Rother A, Hawkridge M E, Maksymovych P,
    Yu P, Gajek M, Balke N, Kalinin S V, Gemming S, Wang F, Catalan G, Scott J F, Spaldin N A,
    Orenstein J and Ramesh R
    2009 {\it Nature Mater.} {\bf 8} 229
\bibitem{Lubk2009} 
    Lubk A, Gemming S and Spaldin N A
    2009 {\it Phys. Rev.} B {\bf 80} 104110
\bibitem{Aird1998} 
    Aird A and Salje E K H
    1998 {\it J. Phys.: Condens. Matter} {\bf 10} L377
\bibitem{Salje2010} 
    Salje E K H
    2010 {\it ChemPhysChem} {\bf 11} 940
\bibitem{Sugawara1968} 
    Sugawara F, Iiida S, Syono Y and Akimoto S I
    1968 {\it J. Phys. Soc. Jpn.} {\bf 25} 1553
\bibitem{Murakami2006} 
    Murakami M, Fujino S, Lim S H, Long C J, Salamanca-Riba L G,
    Wuttig M, Takeuchi I, Nagarajan V and Varatharajan A
    2006 {\it Appl. Phys. Lett.} {\bf 88} 152902
\bibitem{Kim2006} 
    Kim D H, Lee H N, Varela M and Christen H M
    2006 {\it Appl. Phys. Lett.} {\bf 89} 162904
\bibitem{Ederer2005b} 
    Ederer C and Spaldin N A 2005 {\it Phys. Rev.} B {\bf 71} 224103
\bibitem{Atou1999}
    Atou T, Chiba H, Ohoyama K, Yamagauchi Y and Syono Y
    1999 {\it J. Solid State Chem.} {\bf 145} 639
\bibitem{Moreira2002a} 
    Moreira dos Santos A F, Cheetham A K, Atou T, Syono Y, Yamaguchi Y, Ohoyama K, Chiba H and Rao C N R
    2002 {\it Phys. Rev.} B {\bf 66} 064425
\bibitem{Moreira2002b} 
    Moreira dos Santos A F, Parashar S, Raju A R, Zhao Y S, Cheethan A K and Rao C N R
    2002 {\it Solid State Commun.} {\bf 122} 49
\bibitem{Seshadri2001}
    Seshadri R and Hill N A 2001 {\it Chem. Mater.} {\bf 13} 2892
\bibitem{Chiba1997}
    Chiba H, Atou T and Syono Y 1997 {\it J. Solid State Chem.} {\bf 132} 139
\bibitem{Zhong2004} 
    Zhong C, Fang J and Jiang Q
    2004 {\it J. Phys.: Condens. Matter} {\bf 16} 9059
\bibitem{Baettig2005} 
    Baettig P and Spaldin N A 2005 {\it Appl. Phys. Lett.} {\bf 86} 012505
\bibitem{Nechache2009} 
    Nechache R, Harnagea C, Carignan L-P, Gautreau O, Pintilie L, Singh M P, M\'{e}nard D, Fournier P, Alexe M and Pignolet A
    2009 {\it J. Appl. Phys.} {\bf 105} 061621
\bibitem{Sakai2007} 
    Sakai M, Msauno A, Kan D, Hashisaka M, Takata K, Azume M, Takano M and Shimakawa Y
    2007 {\it Appl. Phys. Lett.} {\bf 90} 072903
\bibitem{Langenberg2009} 
    Langenberg E, Varela M, Garc\'{\i}a-Cuenca M V, Ferrater C, Polo M C, Fina I, F\`{a}brega L, S\'{a}nchez F and Fontcuberta J
    2009 {\it J. Magn. Magn. Mater.} {\bf 321} 1748



\bibitem{Gonnenwein2010} 
    Goennenwein S T B 2010 {\it Europhysics News} {\bf 41} 14
\bibitem{Ma2011} 
    Ma J, Hu J, Li Z and Nan C-W
    2011 {\it Adv. Mater.} {\bf 23} 1062
\bibitem{Boomgaard1974}
    van den Boomgaard J, Terrell D R and Born R A J
    1974 {\it J. Mater. Sci.} {\bf 9} 1705
\bibitem{Avellaneda1994}
    Avellaneda M and Harshe G 1994 {\it J. Intel. Mat. Syst. Str.} {\bf 5} 501
\bibitem{Ryu2001a}
    Ryu J, Priya S, Carazo A V, Uchino K and Kim H
    2001 {\it J. Am. Ceram. Soc.} {\bf 84} 2905
\bibitem{Ryu2007} 
    Ryu S, Park J H and Jang H M
    2007 {\it Appl. Phys. Lett.} {\bf 91} 142910
\bibitem{Ryu2001b}
    Ryu J, Carazo A V, Uchino K and Kim H
    2001 {\it J. Electroceram.} {\bf 7} 17
\bibitem{Srinivasan2001} 
    Srinivasan G, Rasmussen E T, Gallegos J and Srinivasan R
    2001 {\it Phys. Rev.} B {\bf 64} 214408; 2002 {\it Phys. Rev.} B {\bf 66} 029902
\bibitem{Srinivasan2002} 
    Srinivasan G, Rasmussen E T, Levin B J and Hayes R
    2002 {\it Phys. Rev.} B {\bf 65} 134402
\bibitem{Srinivasan2004} 
    Srinivasan G, Rasmussen E T, Bush A A and Kamentsev K E
    2004 {\it Appl. Phys.} A {\bf 78} 721
\bibitem{Brandlmaier2008} 
    Brandlmaier A, Gepr\"{a}gs S, Weiler M, Boger A, Opel M, Huebl H, Bihler C, Brandt M S, Botters B, Grundler D,
    Gross R and Goennenwein S T B 2008 {\it Phys. Rev.} B {\bf 77} 104445
\bibitem{Ryu2002}
    Ryu J, Priya S, Uchino K and Kim H E
    2002 {\it J. Electroceram.} {\bf 8} 107
\bibitem{Lee2000} 
    Lee M K, Nath T K, Eom C B, Smoak M C and Tsui F
    2000 {\it Appl. Phys. Lett.} {\bf 77} 3547
\bibitem{Dale2003} 
    Dale D, Fleet A, Brock J D and Suzuki Y
    2003 {\it Appl. Phys. Lett.} {\bf 82} 3725
\bibitem{Murugavel2005} 
    Murugavel P, Singh M P, Prellier W, Mercey B, Simon C and Raveau B
    2005 {\it J. Appl. Phys.} {\bf 97} 103914
\bibitem{Chopdekar2006} 
    Chopdekar R V and Suzuki Y 2006 {\it Appl. Phys. Lett.} {\bf 89} 182506
\bibitem{Eerenstein2007} 
    Eerenstein W, Wiora M, Prieto J L, Scott J F and Mathur N D
    2007 {\it Nature Mater.} {\bf 6} 348
\bibitem{Tian2008} 
    Tian H F, Qu T L, Luo L B, Yang J J, Guo S M, Zhang H Y, Zhao Y G and Li J Q
    2008 {\it Appl. Phys. Lett.} {\bf 92} 063507
\bibitem{Vaz2009} 
    Vaz C A F, Hoffman J, Posadas A-B and Ahn C H
    2009 {\it Appl. Phys. Lett.} {\bf 94} 022504
\bibitem{Czeschka2009} 
    Czeschka F D, Gepr\"{a}gs S, Opel M, Goennenwein S T B and Gross R
    2009 {\it Appl. Phys. Lett.} {\bf 95} 062508
\bibitem{Geprags2010} 
    Gepr\"{a}gs S, Brandlmaier A, Opel M, Gross R and Goennenwein S T B
    2010 {\it Appl. Phys. Lett.} {\bf 96} 142509
\bibitem{Martin2008} 
    Martin L W, Crane S P, Chu Y-H, Holcomb M B, Gajek M, Huijben M, Yang C-H, Balke N and Ramesh R
    2008 {\it J. Phys.: Condens. Matter} {\bf 20} 434220
\bibitem{Bichurin2003} 
    Bichurin M I, Petrov V M and Srinivasan G
    2003 {\it Phys. Rev.} B {\bf 68} 054402
\bibitem{Shayegan2003}
    Shayegan M, Karrai K, Shkolnikov Y P, Vakili K, Poortere E P D and Manus S
    2003 {\it Appl. Phys. Lett.} {\bf 83} 5235
\bibitem{Botters2006}
    Botters B, Giesen F, Podbielski J, Bach P, Schmidt G, Molenkamp L W and Grundler D
    2006 {\it Appl. Phys. Lett.} {\bf 89} 242505
\bibitem{Gonnenwein2008} 
    Goennenwein S T B, Althammer M, Bihler C, Brandlmaier A, Gepr\"{a}gs S, Opel M, Gross R, Schoch W, Limmer W, Huebl H and
    Brandt M S 2008 {\it phys. stat. sol. (RRL)} {\bf 2} 96
\bibitem{Weiler2009} 
    Weiler M, Brandlmaier A, Gepr\"{a}gs S, Althammer M, Opel M, Bihler C, Huebl H, Brandt M S, Gross R and
    Goennenwein S T B 2009 {\it New J. Phys.} {\bf 11} 013021
\bibitem{Shebanov1981}
    Shebanov L A 1981 {\it Phys. Status Solidi} A {\bf 65} 321
\bibitem{Kay1949}
    Kay H F and Vousden P 1949 {\it Philos. Mag.} {\bf 40} 1019



\bibitem{Velev2009} 
    Velev J, Duan C, Burton J, Smogunov A, Niranjan M, Tosatti E, Jaswal S and Tsymbal E
    2009 {\it Nanolett.} {\bf 9} 427
\bibitem{Gajek2007} 
    Gajek M, Bibes M, Fusil S, Bouzehouane K, Fontcuberta J, Barth\'{e}l\'{e}my A and Fert A
    2007 {\it Nature} {\bf 6} 296
\bibitem{Thomas2010} 
    Thomas R, Scott J F, Bose D N and Katiyar R S
    2010 {\it J. Phys.: Condens: Matter} {\bf 22} 423201
\bibitem{Rogers1989} 
    Rogers C T, Inam A, Hegde M S, Dutta B, Wu X D and Venkatesan T
    1989 {\it Appl. Phys. Lett.} {\bf 55} 2032
\bibitem{Locquet1998} 
    Locquet J-P, Perret J, Fompeyrine J, M\"{a}chler E, Seo J W and van Tendeloo G
    1998 {\it Nature} {\bf 394} 453
\bibitem{Ohtomo2004} 
    Ohtomo A and Hwang H Y
    2004 {\it Nature} {\bf 427} 423
\bibitem{Reyren2007} 
    Reyren N, Thiel S, Caviglia A D, Fitting Kourkoutis L, Hammerl G, Richter C, Schneider C W, Kopp T,
    R\"{u}etschi A-S, Jaccard D, Gabay M, Muller D A, Triscone J-M and Mannhart J
    2007 {\it Science} {\bf 317} 1196
\bibitem{Caviglia2008} 
    Caviglia A D, Gariglio S, Reyren N, Jaccard D, Schneider T, Gabay M, Thiel S, Hammerl G, Mannhart J and Triscone J-M
    2008 {\it Nature} {\bf 456} 624
\bibitem{Heber2009} 
    Heber J 2009 {\it Nature} {\bf 459} 28


\end{thebibliography}
\end{document}